\documentclass[a4paper,10pt,twocolumn,twoside]{article}

\usepackage{graphicx,natbib,aas_macros}
\usepackage{fancyhdr}

\tt

\title{Decoding quasars: gravitationally redshifted spectral lines! }

\author{Nimisha G. Kantharia \\
National Centre for Radio Astrophysics, \\ Tata Institute of Fundamental
Research, \\ Post Bag 3, Ganeshkhind, Pune-411007 \\nkprasadnetra@gmail.com}

\date{September 2016}

\begin{document}
\maketitle

\noindent
{\bf This version includes errata and comments in the 9 pages appended to the original paper.}

\thispagestyle{empty}

\begin{abstract}
{\small
Further investigation of data on quasars, especially in the ultraviolet band, 
yields an amazingly coherent narrative 
which we present in this paper.  Quasars are characterised by strong continuum emission and 
redshifted emission and absorption lines which includes the famous
Lyman $\alpha$ forest.  We present irrefutable evidence in support of (1) the entire line spectrum 
arising in matter located inside the quasar system,  (2) the range of 
redshifts shown by the lines being due to the
variable contribution of the gravitational redshift in the observed line velocity,  
(3) existence of rotating black holes and of matter inside its ergosphere, 
(4) quasars located within cosmological redshifts $\sim 3$, 
(5) $\gamma$ ray bursts being explosive events in a quasar.  
These results are significant and a game-changer when we realise that the absorbing gas has been
postulated to exist along the line-of-sight to the quasar and observations 
have accordingly been interpreted. 

In light of these definitive results which uniquely constrain the quasar structure,
we need to drastically revise our understanding of the
universe built on the faulty assumptions of observed redshifts of quasars
having an entirely cosmological origin and the absorption lines arising in the intervening medium. 
}
\end{abstract}

\section*{\small Keywords}
quasars: general, absorption lines, emission lines, 
supermassive black holes; gamma ray burst: general;
galaxies: high redshift

\section{Introduction}
Quasars are a well-observed and well-studied subgroup of objects generally
known as active nuclei - we refer to quasi-stellar objects
as quasars in this paper.  However, they famously remain one of the least understood objects. 
The study which identified quasars \citep{1963Natur.197.1040S} reported wide emission lines ($\sim 50$A)
on a strong blue continuum in 3C 273.  The lines could only be idenfied if a redshifted velocity
component was included although the object had a star-like appearance.
Quasars have spurned extensive research but continue to intrigue even after 50+ years of discovery.  
Quasars show rich observational
signatures which includes strong thermal emission in the ultraviolet (and optical) 
and a power law continuum emission 
which dominates the optical to infrared bands and continuing in the radio when detected. 
Their line spectra show the presence of wide emission lines and numerous absorption features
spanning a range of velocities and widths.
Some of the observed properties of active nuclei, in particular quasars, can be summarised to be:  
\begin{itemize}
\item Flat or power law continuum emission from radio to ultraviolet wavelengths. 
Enhanced ultraviolet emission referred to as the
`blue bump' or 'uv upturn'.  Many active nuclei also detected in X-rays.  
\item Quasar spectra are characterised by broad emission lines and narrow/broad absorption
lines.  BL Lac objects show a featureless continuum with only a few showing spectral lines.  
Seyfert 1 galaxies show broad emission lines while Seyfert 2 galaxies show narrow emission lines. 
\item High ionization lines such as doublets of
C IV (1548.188A 1550.762A), S IV (1393.755A,1402.770A), 
N V (1238.808A, 1242.796A), O VI (1031.928A, 1037.619A) are detected from
many active nuclei especially quasars.
Low ionization lines such as C II (1335A), Fe II (2383A,2586A), Si II (1260A),
Mg II (2795.528A, 2802.704A) are also detected in the spectra of many active nuclei.
These lines are detected either in emission and/or absorption in the quasar spectrum.
\item Quasar spectra show a host of redshifts with the emission line redshifts
being the largest.  Absorption lines span a range of redshifts. 
\item Many active nuclei show variability especially quasars, blazars and Seyfert 1 galaxies.
\end{itemize}

While there is general agreement that a  supermassive black hole is the central
object in all active nuclei, rest of the details remain perplexing at best. 
It is instructive to glimpse the exciting research that quasars sparked
due to their exotic nature as captured by their observations. 
Soon after quasars were identified \citep{1963Natur.197.1040S}, the debate on whether these
objects were extragalactic and very distant or whether these were 
Galactic or in the neighbourhood has been going on.  
\citet{1963Natur.197.1040S} and \citet{1964ApJ...140....1G} concluded that 
quasars were distant extragalactic objects.  
The controversy arose since the large observed redshifted
velocities of the spectral lines, if interpreted to indicate Hubble expansion,
would make quasars very distant objects.  This, then, led to 
the observed magnitudes translating to very high luminosities for the quasars
which had, hitherto, not been observed in any extragalactic object.  
%Quasars did not follow the redshift- magnitude relation.  
However, there was a group of astrophysicists who were convinced observations
indicated that quasars were local and the redshifts were intrinsic.
\citet{1967ApJ...148..321A} suggested that radio sources were associated with nearby peculiar galaxies 
\citep{1966ApJS...14....1A} or bright galaxies \citep[e.g.][]{1974IAUS...58..199A}. In fact
a few such pairs were also found to be physically connected by a bridge 
\citep[e.g. NGC 4319 and Mrk 205;][]{1971ApL.....9....1A}. 
However the local origin did not find favour with most astronomers.  
One major problem with the local origin and association of quasars with nearby galaxies
were the distinct redshifts noted for the quasar (high) and the nearby galaxy (low).
Since one of the explanations was that quasars are ejected from nuclei of galaxies
\citep[e.g.][]{1967ApJ...148..321A}, the quasars can show large redshifts.  However 
in this scenario, the quasars would show both redshifted and blueshifted velocities wrt to the nearby
galaxy whereas the quasars always showed a redshift wrt to the nearby galaxy. This essentially
ruled out the ejection origin for the redshift and an hitherto unknown non-velocity 
intrinsic origin for the redshifts of quasars
was postulated.  \citet{1974IAUS...58..199A}, \citet{2007ARA&A..45....1B} and others continued
to advocate the scenario of quasars being local objects and the observed redshifts 
having an intrinsic origin.  In this paper, we revisit the quasar redshifts and
present evidence for a sizeable intrinsic redshift component in quasar spectra.

Another perplexing observational result noted around the same time
was the arrangement of galaxies in the Coma cluster
along bands in the redshift-magnitude diagram which was especially significant when
the nuclear redshifts and magnitudes were plotted \citep{1972ApJ...175..613T,1973ApJ...179...29T}.  
This result, in addition to a periodicity observed in the velocity differences 
in pairs of galaxies \citep{1980ApJ...236...70T} suggested that the velocities
were quantised in factors or multiples of $\sim 72$ kms$^{-1}$ \citep[e.g.][]{1980ApJ...236...70T}.  
Although these results are not yet understood, 
its interesting that this value is close to the currently accepted value of the Hubble constant.
Since the Hubble constant gives the velocity difference between two galaxies
separated by 1 Mpc, the
%whose currently estimated value is $\sim 72$ kms$^{-1}$ - 
Hubble law can be understood as quantifying the radial velocity 
distribution of galaxies in space or in other words  `redshift quantisation'.  
In fact, it is interesting that Tifft had estimated a value for
the Hubble constant without realising it.  Obviously, observational astronomy was throwing up
several puzzling results which were difficult to understand.
%and this seems to have also led to the development of exotic theoretical explanations. 
We present a possible explanation for the redshift-magnitude bands in the paper. 

Most of the astronomical community has currently accepted the cosmological origin of redshifts of 
quasars and observations have been examined with this implicit assumption. 
Observations which could not be explained like Tifft's bands and Arp's quasar/galaxy
associations were considered to be faulty or spurious which is alarming since 
even if one did not agree with their interpretation, these were observational results
by solid astronomers and needed to be scientifically investigated. 
A careful study clearly brings out the fantastic nature of 
quasars and one senses that the unique properties of quasars
are likely extreme signatures of an active nucleus. 
In this paper, we present our study which provides strong evidence
to a sizeable internal contribution to the observed redshifts of quasars.  We show that
one of the physical processes considered for the origin of the observed high redshifts 
by \citet{1963Natur.197.1040S,1964ApJ...140....1G} is indeed the most plausible explanation.  

We start with a study of the origin of the large observed velocity shifts of the
lines in a quasar spectrum followed
by the ultraviolet continuum emission.    Then we suggest a structure for the quasar
which can explain most ultraviolet observations within the framework of known physics,
discuss other active nuclei and variability and end with some concluding remarks.  

\section{Quasars}
Quasars have been observed at redshifts ranging from $\sim 0.1$ to $\sim 7$. 
Since the identification of quasars from emission lines detected at high redshifts, with the first being
3C 273 \citep{1963Natur.197.1040S}, their study has been intense and in several
wavebands.   Prior to this, these objects were found to
be radio sources whose optical appearance was star-like instead of galaxy-like - the first such
source to be identified was 3C 48 \citep{1961S&T...21..148M} followed by 3C 196 
and 3C 286 \citep{1962PASP...74R.406M,1963ApJ...138...30M}.
Subsequent to this, absorption lines were detected in the ultraviolet
in the quasar spectrum with the first extensive study being of 3C191 
\citep{1967ApJ...147..388B}.  For most quasars, the observed absorption
lines could be identified if different lines were interpreted as showing a range of velocity shifts 
\citep[e.g.][]{1966ApJ...144..847B} as was demonstrated for the spectral lines from 
quasar 1116+12 \citep{1966ApJ...145..369B}.  Thus, the spectral
lines observed in absorption with redshifts ranging from close to the emission line
redshift to much lower redshifts along the same sightline to a quasar
were identified with different species, ionization levels and locations along the line-of-sight
to quasars.  Intriguingly, several absorption lines could be identified at
a common redshift of 1.95 in the spectra of a few quasars \citep{1967ApJ...148L.107B}.
In fact, \citet{1971A&A....13..333K} suggested that the
absorption line redshifts are discretised at an interval of $\Delta ln (1+z) = 0.089$
ie detected at redshifts of 0.061, 0.3, 0.6, 0.96, 1.41, 1.96 etc and the effect
was stronger in quasars located close to bright galaxies in the sky plane \citep{1990A&A...239...50K}
- arguing for the spectra to arise within the quasar.

Several early quasar studies suggested that absorption lines were detected in the
high redshift quasars especially those with emission line redshifts $\ge 2$ and that
variability was observed in the continuum emission. 
However as more data accumulated the first inference has not been confirmed whereas the
second has been modified to include variability in both continuum and lines. 
Opinion has been divided  between astrophysicists regarding the origin of
the absorption spectrum - those who believed that it was intrinsic to the quasar
\citep[e.g.][]{1973ApJ...183..767S,1976A&A....53..275C} and those who believed that it
was extrinsic to the quasar ie arising along the sightline 
\citep[e.g.][]{1969ApJ...156L..63B}.  How fast the field was evolving and how strong the
divided opinion was is
evident in comparing the basic premise in a 1976 review article by \citet{1976ARA&A..14..307S}
where they favoured an intrinsic origin for the redshifts whereas
in a 1981 review article by \citet{1981ARA&A..19...41W} an extrinsic ie cosmological origin
for the redshifts was favoured.  A hybrid inference was derived by \citet{1967Natur.216..351B} who 
on finding a comparable number of quasars and radio galaxies in the 3CR catalogue
concluded that the two types of objects occupied the same
volume of space and hence the high redshifts of quasars were mostly gravitational in origin whereas
the low redshifts were cosmological in nature.  \citet{1969ARA&A...7..527S} comments
on this hypothesis as being too arbitrary and hence not acceptable. 
Thus, we find that due to a lack of suitable explanation for the intrinsic cause of redshifts
and a disbelief in the gravitational origin kept astronomers from favouring the internal cause. 
It appeared easier to accept that the range of redshifted lines indicated the different locations
of the absorbing gas along the line of sight. 

It is now widely accepted by astronomers that the
highest redshift emission and absorption lines are intrinsic to quasars whereas
the lower redshift absorption lines arise in intervening galaxies/gas.  
The intervening absorption line systems are believed to be located at velocities offset by 
$-3000$ to $-5000$ kms$^{-1}$ from the emission line velocity.
%However, there still remain staunch believers in the intrinsic origin of all the absorption lines.  
While most astronomers have accepted that quasars are high
redshift objects there have also been suggestions that they are local objects,
probably ejected from our Galaxy or nearby galaxies and located within 10-100 Mpc 
\citep[e.g.][]{1966Sci...154.1281T,1966ApJ...144..534H,1974IAUS...58..199A}.  
\citet{1964Sci...145..918T} had also inferred, based on the observed
variability in quasars, that the emitting region might not be bigger than a few light days in size.
It is clear that quasars continue to defy a coherent consistent physical
explanation which can encompass its range of observational characteristics.
We summarise some of these questions which are explored in this work and explain
them in the summary section:
\begin{itemize}
\item The spectral lines in absorption in ultraviolet: intrinsic or extrinsic?
\item Quasars: local or cosmological?
\item The origin of the ultraviolet continuum emission and variability.
\item Quasars: galaxies or isolated black holes ? 
\end{itemize}

\begin{figure}
\includegraphics[width=8cm]{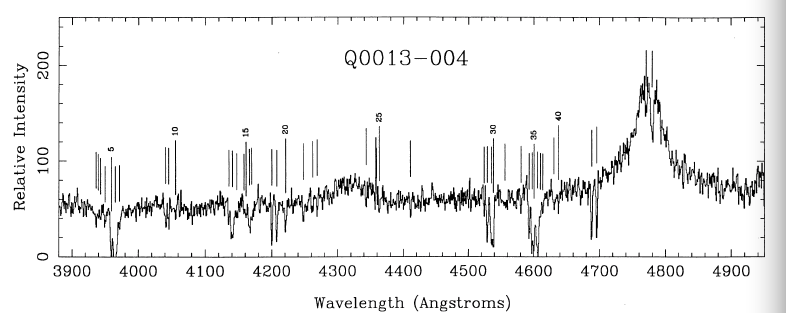}
\includegraphics[width=8cm]{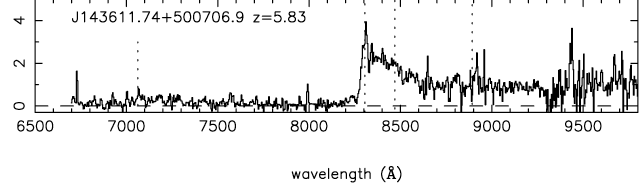}
\caption{\small Ultraviolet spectra of quasars Q0013-004 at $z_{em}=2.086$ (top)
from \citet{1988ApJS...68..539S} and of J143611.74+500706.9 at $z_{em}=5.83$ 
(bottom) from \citet{2006AJ....131.1203F}. 
The strong feature in both is the redshifted Lyman $\alpha$ emission line ($\lambda_{rest}=$1215.67A) 
and the absorption features are detected bluewards of the Lyman $\alpha$ emission line. }
\label{fig1}
\end{figure}

\subsection{Emission and absorption line spectra}
The observed line spectra towards quasars are complex with high redshift broad/narrow emission lines
and several absorption lines detected shortwards of the redshifted Lyman $\alpha$ emission line
(see Figure \ref{fig1}). 
We refer to the observed emission line redshift by $z_{em}$ and absorption line redshift by $z_{abs}$. 
Since it is widely accepted that the quasar redshifts are
cosmological in origin, the emission lines which are detected at the highest
redshift in the spectrum are used to estimate the 
cosmological redshift i.e. $z_c=z_{em}$.  The redshifts of active nuclei are preferably 
estimated from the narrow forbidden
emission lines like [OIII] 5007A, [SII]4068.6A \citep[e.g.][]{2008ApJ...680..926K}
but since forbidden lines are not detected in all  
quasars, the broad resonance lines of Mg II in emission are often used. 
It has been noted that $z_{em}$ estimated from the broad resonance lines show a velocity
offset wrt [OIII] or [SII] lines.   For example, high ionization lines like the broad resonance
line of C IV shows a blueshifted velocity of $-564$ kms$^{-1}$ whereas low
ionization lines like Mg II show a redshifted velocity of 161 kms$^{-1}$ 
wrt the [OIII] velocity \citep{2001AJ....122..549V}.
To calibrate this difference, systems where both the forbidden [OIII] lines and
broad resonance lines are detected have been used to estimate the velocity offset and 
the resonance lines in emission have then been used to estimate $z_c$ for the 
active nuclei where narrow forbidden lines are not detected 
\citep[e.g.][]{1992ApJS...79....1T, 2001AJ....122..549V}. 

Most quasars show a multitude of absorption lines bluewards of the redshifted
Lyman $\alpha$ ($\lambda_{rest}=1215.67$A)  emission line such that $z_{abs}<z_{em}$.  
These absorption lines, many of them doublets, could be identified only if they 
were detected with different velocity offsets wrt to the quasar redshift.  This,
then, led to the obvious question of whether these velocity shifts which ranged from
a few thousand kms$^{-1}$ to several tens of thousands of kms$^{-1}$ were intrinsic
to the quasar or was due to absorption in the gas between us and the quasar.  
Observations could be interpreted to support either origin but the widely favoured
origin is the intervening one probably because no physical reason 
for such large redshifts arising in the
quasar system was convincing enough.   
The absorption lines which are commonly detected at multiple redshifts in the quasar spectra are 
doublets of C IV  (1548.188A, 1550.762A), Si IV (1393.755A, 1402.770A), N V (1238.808A, 1242.796A) 
Mg II (2795.528A, 2802.704A), triplet of Fe II (2383A, 2586A, 2600A) and  
the singlets of HI (1215A), C II (1335A), Si II (1260A) and Al II (1671A). 
Observations indicate that the C IV doublets are detected in emission and absorption
at relatively high redshifts close to $z_{em}$ whereas the Mg II doublet is 
identified in emission close to $z_{em}$ and in absorption
at much lower redshifts.  Neither show significant evolution with
redshift  \citep[e.g.][]{1988ApJ...334...22S, 1988ApJS...68..539S, 2013ApJ...779..161S,
2013ApJ...763...37C, 2015ApJS..218....7B}.  Mg II lines can arise in
gas with a large range in hydrogen column densities - $10^{17}$ to $10^{22}$ cm$^{-2}$
\citep{1986A&A...169....1B}.

\begin{table}[h]
\centering
\caption{\small Estimating the intrinsic redshift ($z_{in}$) from the emission line
redshifts ($z=z_{em}$) of quasars and redshift of the absorption lines of Mg II ($z_c=z_{MgII}$) using
Eqn \ref{eqn1}.  The sample of 27 quasars is taken from \citet{1988ApJ...334...22S}. }
\begin{tabular}{l|c|c|c}
\hline
{\bf Quasar} & {$\bf z=z_{em}$} & {$\bf z_c=z_{MgII}$} &  {$\bf z_{in,em}$}  \\
\hline
Q0013-004 & 2.086 & 0.4466 & 1.133   \\
Q0014+818 & 3.377 & 1.1109 & 1.073   \\
Q0058+019 & 1.959 & 0.6128 & 0.835  \\
Q0119-046 & 1.937 & 0.6577 & 0.772  \\
Q0150-203 & 2.147 & 0.3892 & 1.265   \\
Q0207-003 & 2.849 & 1.0435 & 0.883  \\
Q0229+131 & 2.067 & 0.3722 & 1.235 \\
Q0348+061 & 2.060 & 0.3997 & 1.186 \\
Q0440-168 & 2.679 & 1.0067 & 0.833  \\
Q0450-132 & 2.253 & 0.4940 & 1.177  \\
Q0528-250 & 2.765 & 0.9441 & 0.937  \\
Q0837+109 & 3.326 & 1.4634 & 0.756  \\
Q0848+163 & 1.925 & 0.5862 & 0.844  \\
Q0852+197 & 2.221 & 0.4151 & 1.276  \\
Q0958+551 & 1.751 & 0.2413 & 1.216  \\
Q1222+228 & 2.040 & 0.6681 & 0.822 \\
Q1329+412 & 1.935 & 0.5009 & 0.955  \\
Q1331+170 & 2.084 & 0.7443 & 0.768 \\
Q1517+239 & 1.898 & 0.7382 & 0.667  \\
Q1548+093 & 2.749 & 0.7703 & 1.118 \\
Q1623+269 & 2.526 & 0.8876 & 0.868  \\
Q1715+535 & 1.929 & 0.3673 & 1.142  \\
Q2206-199 & 2.559 & 0.7520 & 1.031  \\
Q2342+089 & 2.784 & 0.7233 & 1.196  \\
Q2343+125 & 2.515 & 0.7313 & 1.03 \\
Q2344+125 & 2.763 & 1.0465 & 0.839 \\
Q2145+067 & 0.990 & 0.7908 & 0.112 \\
\hline
\end{tabular}
\label{tab1}
\end{table}

\begin{figure*}
\centering
\includegraphics[width=8.5cm]{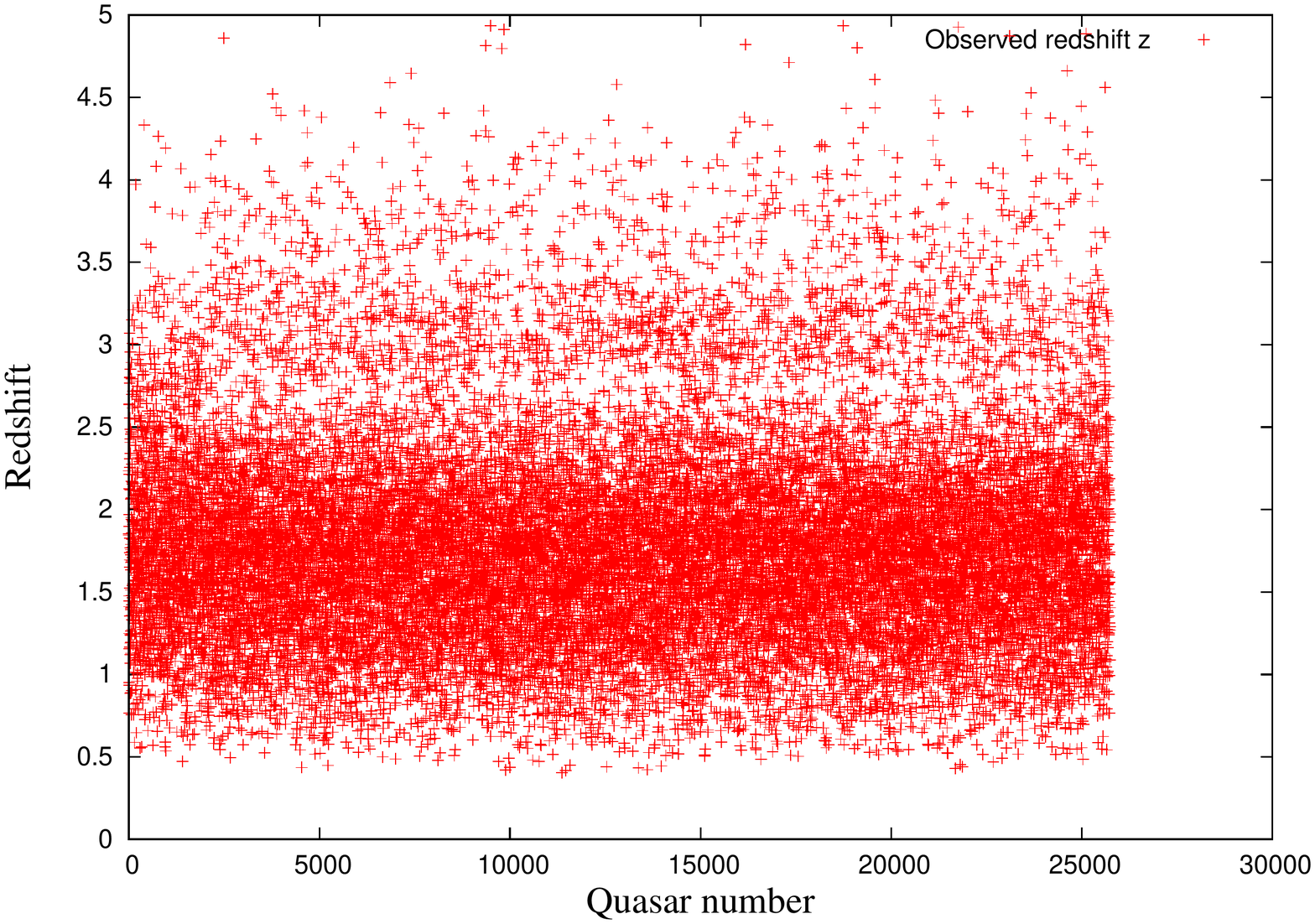}
\includegraphics[width=8.5cm]{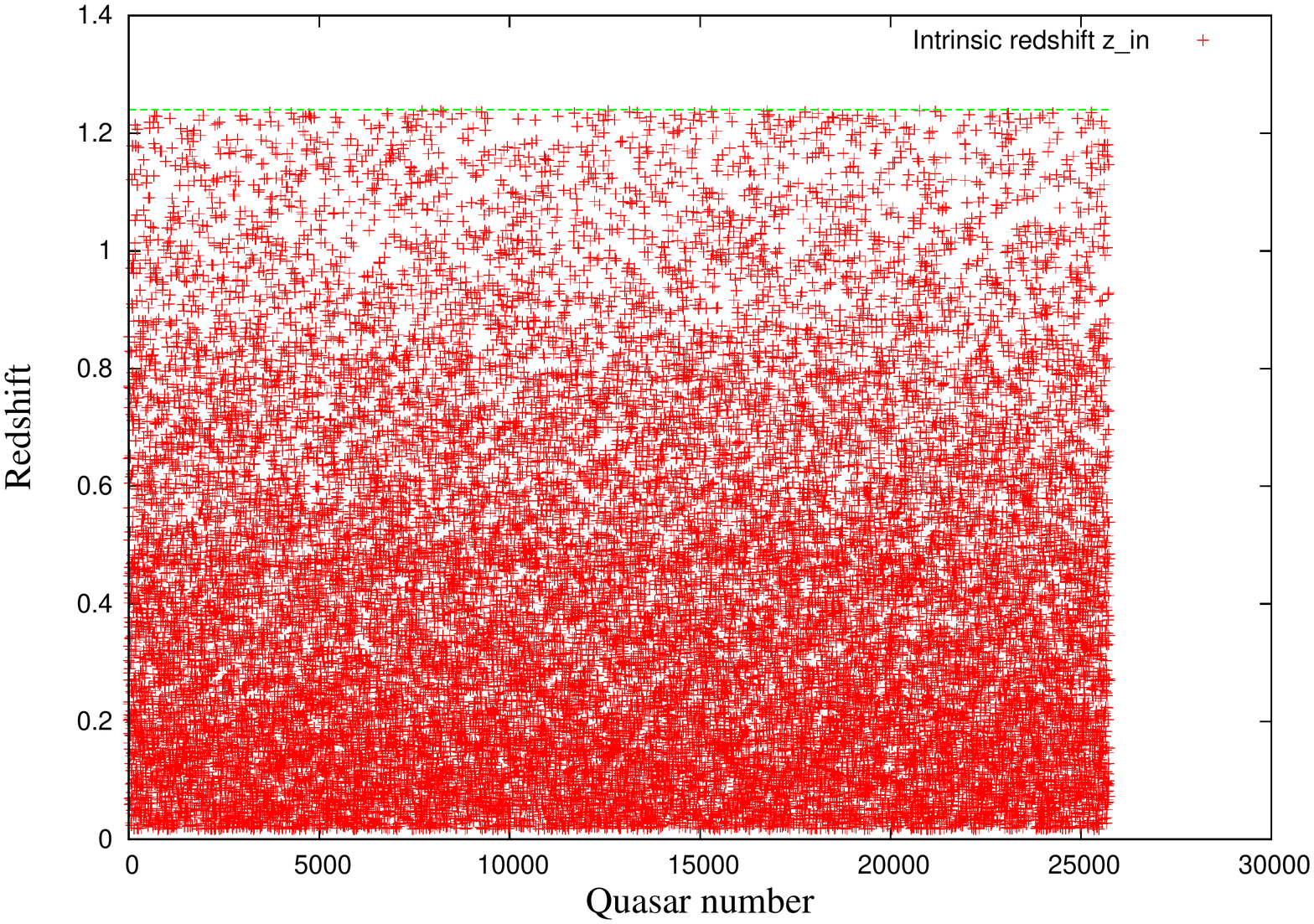}
\caption{\small Left panel shows the quasar redshift $z=z_{em}$ and right panel
shows the intrinsic (difference) redshift ($z_{in}$) of the emission line 
estimated assuming the Mg II absorption
detected at the lowest redshift in the quasar spectrum is $z_c$ using Eqn \ref{eqn1}.
Each point represents a quasar.  
The horizontal line in right panel shows the observed cutoff at z=1.25.  The data are taken
from the catalog presented by \citet{2013ApJ...779..161S}.}
\label{fig2}
\end{figure*}

We recall that the observed redshift of the quasar $z$, if due to multiple reasons,
can be written as follows:
\begin{equation}
(1+z) = (1+z_c)(1+z_{in});~~ z_{in} = \frac{1+z}{1+z_c} - 1
\label{eqn1}
\end{equation}
\begin{equation}
(1+z_{in}) = (1+z_D)~(1+z_{unknown})
\label{eqn2}
\end{equation}
where $z_c$ is the cosmological redshift
and $z_{in}$ is the intrinsic redshift which also consists of two components
namely a Doppler shift ($z_D$) and a hitherto unknown ($z_{unknown}$) component
as given in Eqn. \ref{eqn2}.  In the currently favoured
scenario, the intrinsic component consists of only local Doppler shifts which are
negligible in comparison to $z_c$.
Since the cosmological origin of the redshift has been widely accepted but
still remains unable to consistently explain several observables, we decided to examine
the possibility and origin of the $z_{unknown}$ in Eqn \ref{eqn2}. 

We assume that all the deduced redshifts from an absorption line spectrum of a quasar, except the lowest
redshift, include an intrinsic component. In this case, the lowest detected redshift
is equal to $z_c$ of the quasar.  
The absorption lines of C IV and Mg II are commonly detected in quasar
spectra and are generally the highest and lowest redshifted absorption lines respectively. 
We, thus, started with the assumption that the lowest redshift at which a Mg II absorption 
line ($z_{MgII}$) 
is detected in a quasar spectrum indicates the cosmological redshift $z_c$ of the quasar 
i.e. $z_c = z_{MgII}$.  Using $z = z_{em}$ and $z_c = z_{MgII}$ in Eqn \ref{eqn1}, we can 
estimate $z_{in}$ for the emission line.  If we use $z=z_{abs}$ then we can estimate the $z_{in}$ for
the concerned absorption line (ie other than $z_{MgII}$).  However since $z_{abs}<z_{em}$,
$z_{in,abs} < z_{in,em}$. 
Thus we use $z=z_{em}$ to estimate the largest values of $z_{in}$ in the quasar. 

We used this method on the sample of 27 quasars in \citet{1988ApJ...334...22S} and estimated
$z_{in}$ which are listed in Table \ref{tab1}.  
If multiple Mg II absorption line redshifts were listed for a quasar, then we used the lowest 
redshift as $z_c$.  
$z_{em}$ ranges from 1.751 to 3.377 for 26/27 quasars and $z_{Mg II}$ 
ranges from 0.2413 to 1.46 while the estimated $z_{in}$ ranges from 0.667 to 1.276.
To show that $z_{in}$ is not always largest for the quasar with the largest observed redshift -
we draw attention to quasars Q0013-004 and Q0014+818 in Table \ref{tab1} -
Q0013-004 has $z_{em} = 2.086$ and $z_{in}=1.133$ while Q0014+818 has
$z_{em}=3.377$ while $z_{in}=1.073$. 
We were dumbfounded by the remarkable result that $0.6 < z_{in} < 1.28$ for 26/27 quasars
and were convinced that this result demanded further investigation. 

\begin{figure}[h]
\centering
\includegraphics[width=8cm]{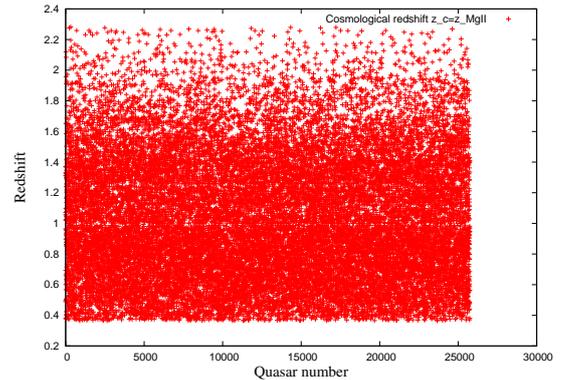}
\caption{\small The lowest $z_{MgII}$ detected towards a quasar for the same sample
shown in left panel of Figure \ref{fig2}.  $z_c = z_{MgII}$ as we find from our study.
Data taken from the \citet{2013ApJ...779..161S} sample.  } 
\label{fig3}
\end{figure}

We, then, used the sample of 35629 Mg II absorption
lines given in  \citet{2013ApJ...779..161S} which included multiple
redshifted Mg II absorption lines detected towards a quasar.
Since we required the lowest redshift at which a Mg II absorption line is detected towards a given quasar, 
we filtered the samples appropriately and derived a sample of about 25500 quasars for which
$z_{em}$ ranges from 0.5 to 5 (Figure \ref{fig2} left) and the range of $z_{MgII}$ which is
primararily determined by the observing band is from 0.36 to 2.29 
(Figure \ref{fig3}).   We then repeated the above exercise ie using $z = z_{em}$ and $z_c=z_{MgII}$ in
Eqn \ref{eqn1} to estimate $z_{in}$. 
{\it The result (Figure \ref{fig2} right) confirms that $z_{in}$ is always $<1.25$.} 
It indicates that the observed limit has to be due to some physical process inside the quasar. 
The median value of the distribution is $z_{in} \sim 0.333$. 
If there was no connection between 
$z_{em}$ and $z_{MgII}$ i.e. if the former was $z_c$ of the quasar and latter
was an intervening system, as is generally assumed, then there should have been no limit on 
$z_{in}$.  This result, we believe, provides irrefutable evidence to a non-trivial 
intrinsic component in the observed redshift of quasars and the physical process
which gives rise to this component has to be constrained by the result that
the maximum value of $z_{in}$ is 1.25.  This value includes contributions from
a Doppler effect and an unknown cause as given in Eqn. \ref{eqn2}.  

From the above results, our reasoning leading to the assumption that 
$z_c=z_{MgII}$ where $z_{MgII}$ refers to the lowest redshift detected in a quasar spectrum
is validated.    This, then, has very important implications: 
\begin{enumerate}
\item All the multi-redshifted spectral lines arise inside the quasar system.
\item $z_c \ne z_{em}$ for a quasar.  Instead $z_c \sim z_{MgII}$ and Figure \ref{fig3} shows
the distribution of $z_c = z_{MgII}$ of the quasars.  The median value of this $z_c$ 
distribution is 0.977.   
\item The observed range of redshifts in a quasar spectrum are due to an intrinsic physical
process such that $z_{in} < 1.25$ for all lines and the median value of $z_{in}$ is 0.333.
\end{enumerate}

\begin{figure}[t]
\centering
\includegraphics[width=8.0cm]{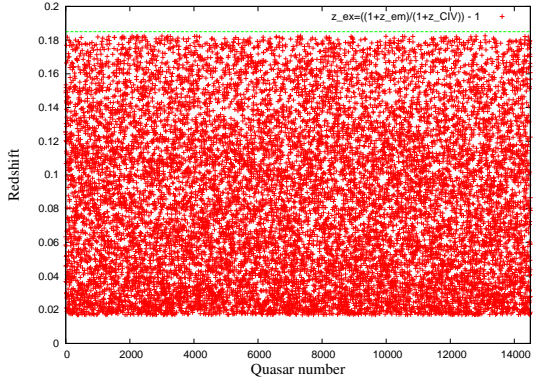}
\caption{\small The difference redshift of the emission line and C IV absorption lines towards quasars
from \citet{2013ApJ...763...37C}. 
The difference redshift  $(z_{em}+1)/(z_{CIV}+1) - 1$ is always less than 0.185 
shown by the horizontal line.}
\label{fig4}
\end{figure}

We, then, used the C IV absorption line catalogue in
\citet{2013ApJ...763...37C} and estimated the difference between $z_{em}$ and $z_{abs,CIV}$
for all C IV line detections along a quasar.  The quasar sample is same as the one
shown in Figure \ref{fig2} (left panel) but only those with $z_{em} > 1.7$ are used 
in the C IV line analysis. 
The results are shown in Figure \ref{fig4}.  The difference redshifts are $< 0.185$.  
Thus, $z_{abs,CIV}$ are close to $z_{em}$ and will also contain a large contribution from $z_{in}$
which is distinct from $z_{in,em}$.
This result supports the origin of the C IV lines in the
quasar itself otherwise the difference between $z_{abs,CIV}$ and $z_{em}$ should have been
larger and random.  
Being entirely convinced that the observed redshifts of lines in quasars definitely contain a
varying intrinsic component responsible for the multiply-redshifted absorption lines, we
examined other observational results in literature with this new perspective.
We present some results which are trivially explained with an intrinsic
redshift component but often require a contrived explanation for a cosmological redshift:

\begin{figure*}
\centering
%\mbox{\includegraphics[width=8.0cm]{qso-MgII-redshift-SDSS.png}
%\includegraphics[width=8.0cm]{qso-CIV-redshift-SDSS.png}}
\mbox{\includegraphics[width=7.5cm]{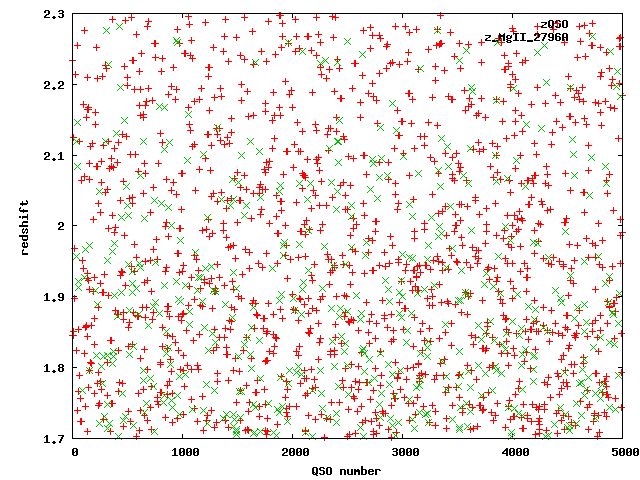}
\includegraphics[width=7.5cm]{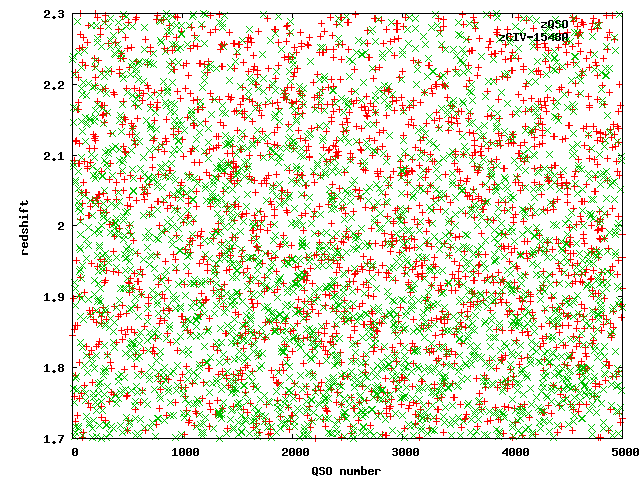}}
\caption{\small Redshifts of absorption lines of Mg II in green (x; left) and C IV lines in green 
(x; right) and the emission line redshifts in red (+) in the 
common redshift range of 1.7 to 2.3.  Note the significantly
larger number of C IV systems as compared to the Mg II systems at these
redshifts.  Data on Mg II
redshifts from \citet{2013ApJ...779..161S} and C IV redshifts from 
\citet{2013ApJ...763...37C} are plotted.}
\label{fig5}
\end{figure*}

\begin{enumerate}

\item While emission lines are observed in all types of active nuclei, absorption
lines at blue-shifted velocities wrt to the emission line are mainly
detected towards quasars, a few blazars and Seyfert 1 galaxies.  This is a strong argument
against the intervening origin for the absorption lines since then the spectra of
the nuclei of all distant active galaxies should have detected these.  However if the origin of
the absorption lines is intrinsic to the quasar then
it indicates a difference in the attributes of quasars and other active nuclei. 

\item The redshift-magnitude correlation is not followed by quasars
unlike active galaxies.  This point is explained in the cosmological
redshift model as indicating a large spread in the intrinsic luminosities of quasars
as compared to other galaxies.  We suggest that this correlation is corrupted by the
intrinsic redshift component in quasars and should be examined after removing the
intrinsic contribution as outlined above.

\item It is believed that the emission lines and the highest redshifted absorption lines in a quasar
spectrum are intrinsic to the quasar whereas the rest of the absorption lines arise in the 
intervening medium.  While it is a reasonable assumption, it appears to be fairly adhoc 
due to lack of better understanding. 

\item In a quasar spectrum,
the high ionization absorption lines such as C IV and Si IV always appear at relatively higher
redshifts compared to the low ionization absorption lines such as Mg II and Fe II.
Some of this could be a bias due to the optical observing band. 
To check the influence of this bias, we plotted both the lines detected in the common
redshift range of 1.7 to 2.3 (Figure \ref{fig5}). 
As seen in Figure \ref{fig5}, a significantly larger number of C IV lines (right panel) are detected
compared to Mg II lines (left panel).  In fact it appears that C IV lines in this redshift range
are detected from most quasars unlike Mg II lines.  This result, in the intervening gas model would mean
that high ionization systems are more distant than 
low ionization systems.  On the other hand, taking recourse to our inference
above wherein all absorption lines arise in the quasar and $z_{MgII} \sim z_c$, this
result indicates that 
$z_c$ of quasars  are fairly modest and $z_{em}$, $z_{abs}$ appear large due to the contribution of $z_{in}$. 
Mg II lines contain a smaller contribution of $z_{in}$ compared to the 
C IV lines so that the latter appear at higher redshifts.  

\item \citet{2015ApJS..218....7B} presented an extensive study of absorption spectra towards
nine quasars.  Since they believed that the lines arose in the intervening
medium whose evolution they were studying, they combined the absorption
line data towards all quasars based on the observed redshift.  They infer several interesting points 
based on their comprehensive analysis and we present a few here.  Firstly, they find no 
evolution in any parameter of the CIV systems with redshift although they are detected at the highest
redshifts towards quasars.  Secondly, they find that the absorption lines with
redshift $\le 3000$ kms$^{-1}$ offset from the quasar redshift have comparatively harder radiation
impinging on them.  Thirdly and we think very significantly, \citet{2015ApJS..218....7B}
find that the column density ratios Si IV/C IV and Si II/C II do not
change with redshift whereas the ratios  C II/C IV and Si II/Si IV continuously
vary with redshift.  This is difficult to understand if these lines arise the intervening gas.  
In an intrinsic origin, this can be understood as different contribution of
$z_{in}$ on the high ionization and low ionization lines so that the ratios of high or low ionization
lines will show no evolution with redshift but ratios of high and low ionization species will show
a dependence on $z_{in}$ ie $z$.   We suggest that
to study the cosmological variation in these ratios, the $z_{in}$ component in $z$ needs to be eliminated. 

\item Recently \citet{2016arXiv160203894S} have studied the dependence of the velocity offset between
emission lines of two species and differing ionization 
on quasar luminosity.  Figure 1 in their paper shows their results.  
%We examine Figure 1 in their paper with a different perspective by using the velocity
%at which the lines appear to locate the line forming region.
The velocity offsets between
the narrow forbidden emission lines is small and so is the scatter.  This could 
mean that these lines arise in the same region around the ionizing source.  
On the other hand, the velocity offsets and scatter between the wide lines
are large but similar for He II, C IV and Si IV i.e. these appear to arise
in the same region.  This is indicative of the different
contribution by $z_{in}$ on the forbidden line velocities and on the wider lines of
He II, C IV and Si IV. 

\end{enumerate}

Thus, we conclude that both $z_c$ and $z_{in}$ make a significant contribution to the
observed line velocities in a quasar spectrum and hence influence their overall
behaviour.  In the following section,
we explore the dominant physical process which contributes to the intrinsic redshifts. 

\subsection{$\bf z_{in} \sim z_{unknown} = z_{gravitational} (z_g)$ }

In this section, we search for a physical cause of the
intrinsic redshifts in a quasar spectrum which can explain the following: 
\begin{itemize}
\item $z_{in} < 1.25$. 
\item Large, varying redshifts and line widths of the emission lines and absorption lines.
\item No contribution of $z_{in}$ in the lowest redshift absorption feature due to Mg II.
\item The difference between spectra of quasars and other active nuclei
especially the absence of absorption lines.
\end{itemize}

Soon after quasars were discovered, \citet{1964ApJ...140....1G} examined
the high redshift emission lines detected in the spectra of 3C 48 ($z_{em}=0.3675$) and 
3C 273 ($z_{em}=0.1581$).  Two possible origin scenarios for the high redshift lines
were considered \citep{1963Natur.197.1040S,1964ApJ...140....1G}:
(1) cosmological redshifts which was favoured by them and 
has come to be widely accepted by the astronomical community;  
(2) gravitational redshifts which was ruled out by them which we recount here.
In the gravitational redshift origin suggested by
\citet{1963Natur.197.1040S}, the lines had to arise in a star with a radius of $\sim 10$ km.  
On the other hand, the study by \citet{1964ApJ...140....1G} found that 
the gravitational redshift can explain the emission line redshifts if the lines arose in a thin
shell of thickness $10^{-4}$ pc (i.e. $3\times10^{14}$ cm) within a radius of 
0.01 pc (i.e. $3\times10^{16}$ cm) around a massive object of mass $\sim 10^{11}$ M$_\odot$.
They ruled out this origin since it was not clear if such massive 
compact systems could be stable.
Interestingly, the authors stated `If stable, massive configurations exist, we must re-examine this possibility.'
Moreover, they concluded that if the redshifts were gravitational then the object had to be extragalactic.
\citet{1967Natur.213..373H} put forward a different model which according to them could give rise
to large gravitationally redshifted lines.  They suggested that the quasar consisted of
a central hot gas cloud which emitted the continuum and spectral lines and which 
was surrounded by a large number of compact objects like neutrons stars.
The compact objects then gravitationally redshifted the emission lines from the central gas cloud
as the radiation left the system.  Curiously, they could obtain arbitrarily large values of 
gravitational redshifts in their model which explained the entire $z$ of a line. 
The absorption lines were postulated by them to arise in floating clouds of ions that surrounded the
central emitting cloud.  The viability of this model has been examined in 
detail in \citet{1975MNRAS.171...87D, 1979MNRAS.186....1D, 1968ApJ...153L.163Z}. 
The gravitational redshift origin has indeed been considered by several astronomers 
as a possible cause for the range of absorption line redshifts.
It also appears that many of these studies were trying to explain the entire observed redshifts
of quasars as being due to a gravitational redshift and hence being stumped. 
Many observational studies which were not convinced of the gravitational
origin, still concluded that the observed redshift
contained an intrinsic component \citep[e.g.][] {1969Natur.222..735B} 
but appear to have been unable to find a cause for this as against the cosmological origin.  
We believe that with much more data on quasars now available for inspection and 
interpretation, we should
be able to conclusively arrive at a physical solution and pave way for future research. 
We begin by examining gravitational redshift as the $z_{unknown}$.

Gravitationally redshifted spectral lines are a natural outcome of the general theory of relativity.
As Robert Lawson's English translation of Einstein's book \citep{1920Rel.book.....E} 
notes 'An atom absorbs or emits light of a 
frequency which is dependent on the potential of the gravitational field in which it is situated.'
The shifted frequency of a line emitted in a strong gravitational field was quantified by Einstein to be:
\begin{equation}
\nu = \nu_0 (1+\frac{\phi}{c^2})
\label{eqn3}
\end{equation}
where $\nu$ is the observed frequency, $\nu_0$ is the rest frequency of
the spectral line and  $\phi = -G M /R$ is the gravitational
potential where M is the mass of the massive object and R is the distance of the line 
forming medium from the massive object.  
The above equation then gives the expected shifts in the frequency and wavelength
of the spectral line arising in the gravitational potential $\phi$:
\begin{equation}
\frac{\nu_0 - \nu}{\nu_0} = \frac{G M}{R c^2} ~~ or ~~ \frac{\lambda - \lambda_0}{\lambda_0} = \frac{G M}{R c^2}
\label{eqn4}
\end{equation}
These are the gravitational redshifts ($z_g$). 
Einstein noted at that time that it was an open question if such an effect existed.  
Soon after, \citet{1925PNAS...11..382A} seems to have measured a displacement of about 23 kms$^{-1}$
in the velocity of the spectral lines from the white dwarf Sirius B which
was attributed to a gravitational redshift.   This provided the first evidence of a frequency
shift in a line due to the effect of a gravitational potential. 
The frequency shift caused by the earth's gravitational potential was demonstrated
in a neat experiment by \citet{1960PhRvL...4..337P}
and \citet{1964PhRvL..13..539P}.  More evidence for the existence and
detectability of this effect came from the gravitationally redshifted spectral lines 
detected from several white dwarfs \citep{1967ApJ...149..283G} and   
absorption lines in the X-ray band from neutron stars showing a gravitational redshift of $0.35$ in
EXO 0748-676 \citep{2002Natur.420...51C} and a redshift of
0.12-0.23 in 1E 1207.4-5209 \citep{2002AAS...200.8001P}.
Thus, it is clear that gravitational redshifts are detectable in lines formed near compact objects in
our Galaxy and it is reasonable to examine their effect, if any, in the line
spectra observed from quasars whose central engine is a supermassive compact object. 

Here we revisit the original suggestion in \citet{1964ApJ...140....1G} wherein
the emission line redshifts of 3C 48 and 3C 273  are due to
gravitational redshifts $z_g$, if the emission arose in a very thin dense shell located 
very close to a supermassive black hole in extragalactic systems.  
We use Eqn \ref{eqn1} to define the observed redshift of the quasar and 
examine results from the previous section for the possibility of $z_{g}$ being the 
component $z_{unknown}$ in $z_{in}$ (Eqn. \ref{eqn2}).

We start with some physical background on black holes which
is relevant for this discussion before we go to the analysis of the redshifted lines.   
We note that the Schwarzchild radius of an object of mass M is given by:
\begin{equation}
\rm
R_s = \frac{2GM}{c^2}
\label{eqn5}
\end{equation}

We modify the labels in Eqn \ref{eqn4} and relate the gravitational redshift to
the observed velocity shift of the spectral lines $v$ for connecting to observations:
\begin{equation}
\rm
z_{g}(R) = \frac{(\lambda_{obs} - \lambda_{rest})}{\lambda_{rest}}
= \frac{GM}{Rc^{2}}  = \frac{v}{c}
\label{eqn6}
\end{equation}

Combining Eqns \ref{eqn5},\ref{eqn6},  we note that a line emitted from $R \sim R_s$ will show
$z_g\sim0.5$ when detected by us and R can be expressed in terms of the Schwarzschild radius $R_s$
for a given gravitational redshift i.e.
\begin{equation}
z_g(R) = \frac{R_s}{2 R} 
\label{eqn7}
\end{equation}
We note that the expression for the gravitational redshift which should be used
when $z_g(R) \rightarrow 1$ is
\begin{equation}
z_g(R) = \frac{1}{\sqrt{1 - \frac{2 G M}{R c^2}}}  - 1
\label{eqn8}
\end{equation}
and the expression for the cosmological redshift which should be used when $z_c$ approaches one is 
\begin{equation}
z_c  = \sqrt{\frac{c+v}{c-v}}  - 1
\label{eqn9}
\end{equation}
However, for simplicity since we are trying to establish if $z_{unknown}=z_g$, 
we use the approximations given in Eqn \ref{eqn6} when required. 

\begin{figure}
\centering
\includegraphics[width=8cm]{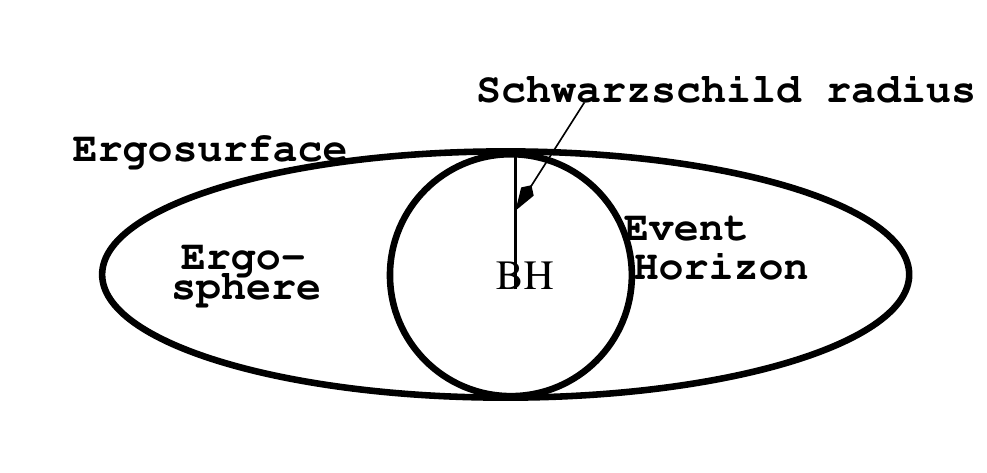}
\caption{\small The event horizon and ergosphere of a rotating black hole.  For non-rotating
black holes, the event horizon and the ergosurface coincide ie a spherical surface with
radius $R_s$.}
\label{fig6}
\end{figure}

The event horizon of a black hole is the effective radius
of a zone around it from which no information which includes photons can be extracted.
Thus, the event horizon defines the black hole extent. 
In case of non-rotating black holes also known as Schwarzchild black holes, the extent of the event
horizon is defined by the Schwarzchild radius $R_s$ (see Figure \ref{fig6}).  
Thus, if the spectral lines are formed
just outside the event horizon, they will be detected with a gravitational redshift of 0.5 and this   
is the maximum redshift that the gravitational potential of a non-rotating black hole
can introduce in the spectral lines.  In case of a rotating black hole also known as a Kerr black hole,
the event horizon is smaller than the Schwarzchild radius in the non-polar regions.
It is $R_s/2$ at the equator for a maximally rotating black hole.  
There exists a region between the event horizon and the Schwarzschild
radius in rotating black holes which is referred to as the ergosphere. 
The ergosphere has a peculiar egg-shaped structure (see Figure \ref{fig6}) 
since it arises due to a modification in the gravitational field of the black hole due to its rotation.
The rotation is maximum in the equatorial zones so the extent
of the ergosphere is largest there and rotation is zero at the poles where the event horizon 
coincides with $R_s$ and there is no ergosphere (see Figure \ref{fig6}).  
In such a black hole, a spectral line formed near the event horizon in the equatorial regions 
can be detected with a maximum gravitational redshift close to one whereas a spectral line
formed in the polar regions can only suffer a maximum $z_g$ of 0.5.  
To summarise, a non-rotating Schwarzschild black hole can explain upto $ z_g \sim 0.5$
while a rotating Kerr black hole can explain upto $z_g \sim 1$ i.e. the maximum allowed $z_g$
for any black hole configuration is one. 

\begin{figure}
\centering
\includegraphics[width=7cm]{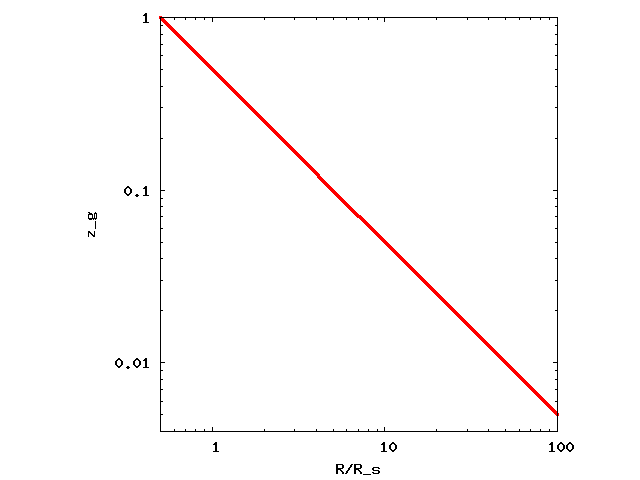}
\includegraphics[width=7cm]{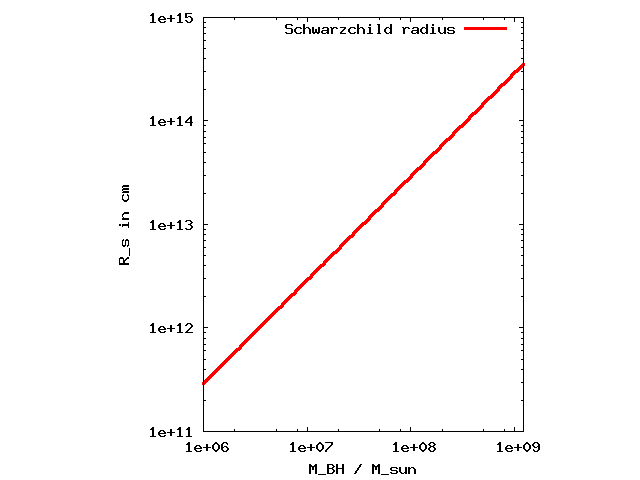}
\caption{\small Top panel shows the gravitational redshift that a line will suffer
when formed at different distances from the black hole.  The second panel shows the 
Schwarzchild radius in physical units for different black hole masses. }
\label{fig7}
\end{figure}

We show the gravitational redshifts suffered by  
spectral lines arising at different distances from the black hole
in the top panel of Figure \ref{fig7}.  In the lower panel, we show 
$R_s$ for black holes of different masses.  Thus, the Schwarzchild radius of a black hole of
mass $10^{9}$ M$_\odot$ is $3\times10^{14}$ cm while that of mass $10^6$ M$_\odot$ is
$3\times10^{11}$ cm.  However as evident from Figure \ref{fig7} (top), the
effect of the gravitational potential of the black hole felt by the
line forming gas, if located near $R_s$ will be the same in both cases.   In other words,
the gravitational redshift can be used to estimate the ratio $M_{BH}/R$ but other observational
results are required to estimate $M_{BH}$. 

Equipped with this background on black holes, we find that the result we obtained in the 
previous section that $z_{in} < 1.25$ for all quasars assuming $z_c=z_{MgII}$ 
is extremely significant and conclusive when considered with two other details - (1)
the largest value of $z_g$ is one and (2) $z_{in}$ consists of $z_D$ and $z_{unknown}$  (Eqn \ref{eqn2}). 
If we assume that $z_{unknown}=z_g$ then substituting the maximum value of $z_g$ and $z_{in}=1.25$ 
in Eqn \ref{eqn2} gives $z_D = 0.12$ i.e. the contribution of
Doppler velocity component is $\le 36000$ kms$^{-1}$.
This component will be different for different quasars but will be similar for the spectral
lines detected from a quasar unlike $z_g(R)$ which changes for lines which form at different
$R$ from the black hole in the same quasar. 
$z_D$ can be due to incorrect $z_{MgII}$ which introduces an offset in $z_c$ and/or 
due to outflow/inflow in the line forming region.  $z_D$ makes a small contribution
to $z$ and we are not concerned with it here.   
The spectral lines from quasars in a 25500 large sample include contribution from $z_{unknown} = z_g \le 1$
and a small component $z_D < 0.12$. 
{\it We thus conclude that the limit $z_{in}<1.25$ is mainly dictated by the limit $z_g \le 1$ imposed
by black hole physics.}  Since $z_g(R)$ varies as a function of R, this explanation trivially explains 
the range of redshifted lines observed in a quasar spectrum as lines that arise at different R.
{\it This, then, leads to the inevitable and clear conclusion that the large observed spectral 
line redshifts and the large range in absorption line redshifts in a quasar spectrum are 
a result of the varying gravitational potential experienced by the line photons arising
in gas located at different distances from the black hole.}

\begin{table}
\centering
\caption{\small The estimated values of cosmological component of redshift ($z_c$)
for different observed redshifts ($z$) after removing $z_{in} \sim z_g =1$  in Eqn \ref{eqn1} are listed here.}
\begin{tabular}{c|c}
\hline
{$\bf z=z_{emission}$} & {$ \bf z_{cosmological}$} \\
\hline
%8  & 3.5 \\
7 & 3 \\
6 & 2.5 \\
5 & 2 \\
4 & 1.5 \\
3 & 1 \\
2 & 0.5 \\
\hline
\end{tabular}
\label{tab2}
\end{table}

This discussion then leads us to the following important inferences:
\begin{itemize}
\item For quasars, $z_{unknown}=z_g$ and $z_c \le 3$.  The large observed redshifts $z$
are because of the contribution of $z_g$.  The $z_c$ estimated after removing $z_g$ from
z are listed in Table \ref{tab2}. 
\item $z_g \le 1$ is the major component of $z_{in}$ and hence leads to the limit 
$z_{in} < 1.25$ in quasars. 
%implying that $z_D \le 0.12$.
\item Now that we know that lines are shifted by $1 < z_g < 0.5$, 
this proves that rotating black holes exist and contain matter within their ergosphere. 
\item The emitting zone is located closer whereas absorption zones
are further from the black hole and hence lines appear at different redshifts.
\end{itemize}

We note that the explanation
presented above violates no black hole physics nor does it demand any `new' physics.
In fact, it is heartening to note that all the required physics has been known for a long time.
%stressing the universality and constancy of laws of physics. 
There are obviously several questions which need to be addressed now that it
is clear that the quasar spectrum is intrinsic and the range of redshifts are due to
the gravitational potential of the black hole. 

\begin{table}
\centering
\caption{\small The gravitational redshifts estimated for the multiple absorption lines of
C IV 1548A detected at different redshifts ($z_{CIV,abs}$) in the spectra of six quasars.}
\begin{tabular}{c|c}
\hline
$\bf z_{CIV,abs}$ & $\bf z_{in} \sim z_{g,CIV,abs}$  \\ 
\hline
\multicolumn{2}{c}{Q0848+163$^1$}\\
\multicolumn{2}{c}{$z_{em}=1.925,z_{MgII}=0.5862,z_{g,em}=0.844$} \\
\hline
1.4575  &  0.5493 \\
1.4684	&  0.5562 \\
1.4704  &  0.5574 \\
1.9159  &  0.8383 \\
\hline
\multicolumn{2}{c}{Q0014+818$^1$} \\
\multicolumn{2}{c}{$z_{em}=3.377,z_{MgII}=1.1109,z_{g,em}=1.073$} \\
\hline
2.4932 & 0.6548 \\
2.7980 & 0.7992 \\
2.8004 & 0.8004 \\
3.2265 & 1.0022 \\
\hline
\multicolumn{2}{c}{Q0837+109$^1$} \\
\multicolumn{2}{c}{$z_{em}=3.326,z_{MgII}=1.4634,z_{g,em}=0.7561$} \\
\hline
2.4165 & 0.3869 \\
2.9558 & 0.6058 \\
3.1428 & 0.6817 \\
\hline
\multicolumn{2}{c}{54452-2824-554$^2$} \\
\multicolumn{2}{c}{$z_{em}=2.5566,z_{MgII}=2.08431,z_{g,em}=0.1531$} \\
\hline
2.08788 &       0.0011  \\
2.14618 &       0.0200  \\
2.21361 &       0.0419  \\
2.38185 &       0.0965  \\
\hline
\multicolumn{2}{c}{52178-0702-503$^2$} \\
\multicolumn{2}{c}{$z_{em}=2.6929,z_{MgII}=0.83151,z_{g,em}=1.0163$} \\
\hline
2.16934&       0.7305  \\
2.18427&       0.7386  \\
2.22892&       0.7629  \\
2.52017&       0.9220  \\
\hline
\multicolumn{2}{c}{52618-1059-146$^2$} \\
\multicolumn{2}{c}{$z_{em}=4.1541,z_{MgII}=2.07092,z_{g,em}=0.6783$} \\
\hline
3.42263 & 0.4402 \\
3.53558 & 0.4769 \\
3.73894  & 0.5432 \\
\hline
\end{tabular}

$^1$ {\small $z_{MgII}$ from \citet{1988ApJ...334...22S} and $z_{CIV}$ from \citet{1988ApJS...68..539S}}

$^2$ {\small $z_{MgII}$ from \citet{2013ApJ...779..161S} and $z_{CIV}$ from \citet{2013ApJ...763...37C}. }
\label{tab3}
\end{table}

\begin{figure}[t]
\centering
\includegraphics[width=8.0cm]{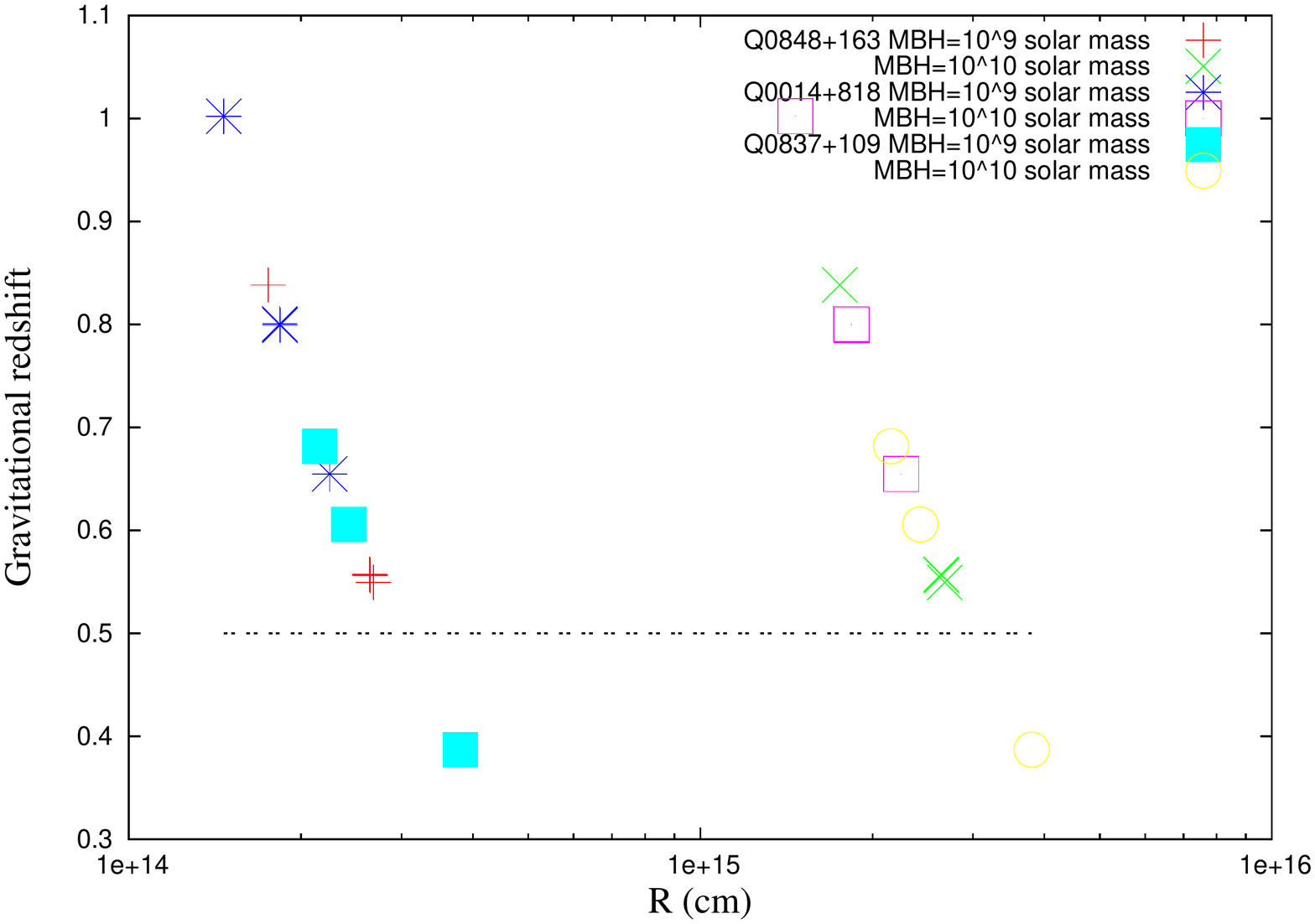}
\includegraphics[width=8.0cm]{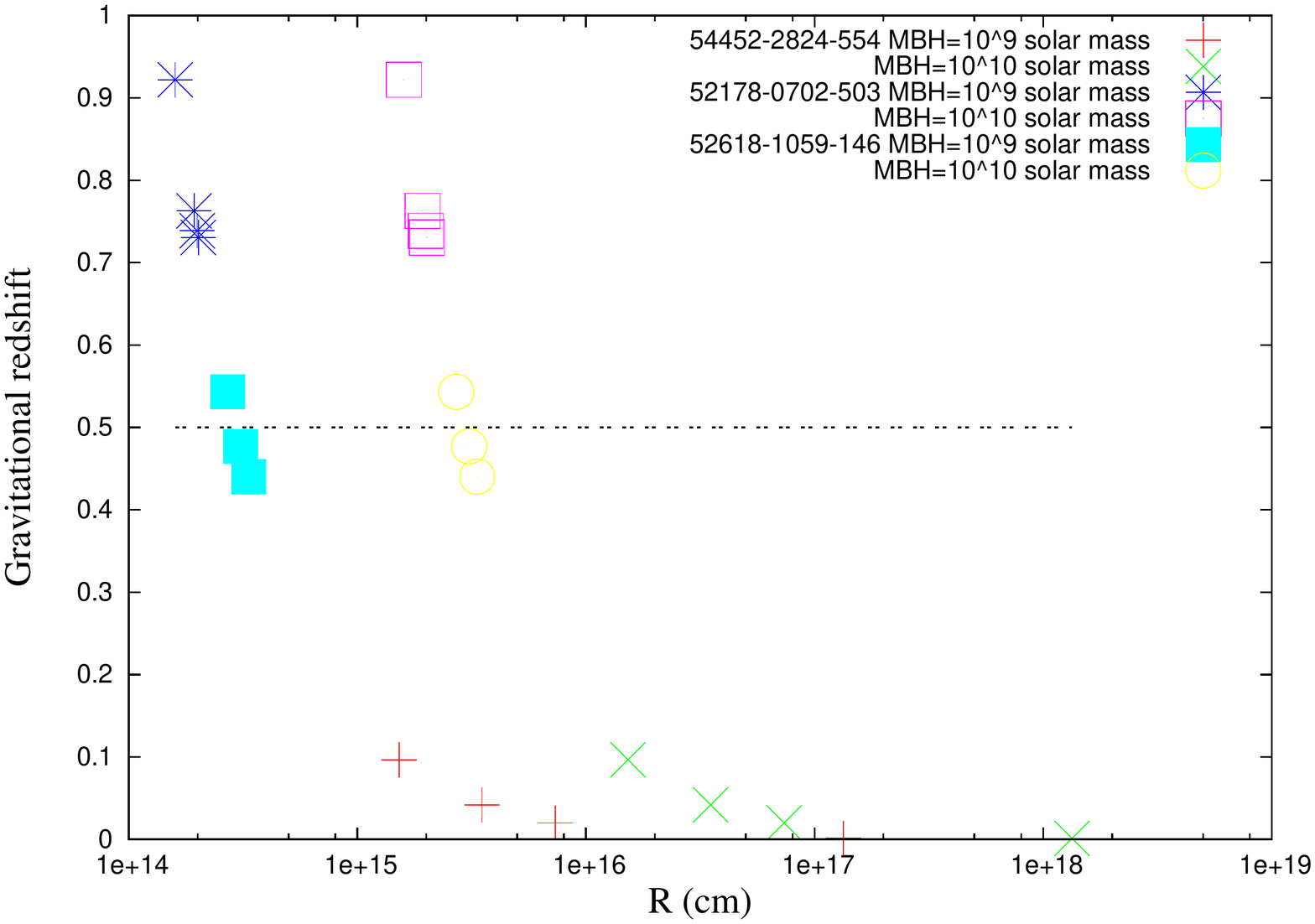}
\caption{\small The  estimated $z_g$ for C IV absorption lines listed in Table \ref{tab3}
and the implied distances of the line forming region from black holes of
masses $10^8$ and $10^9$ M$_\odot$ are shown in both the panels.  The dashed line 
is drawn at $z_g=0.5$ and the
intersection with the data points indicate the Schwarzchild radius of that black hole mass. All
points above this line require the absorbing gas to be located inside the ergosphere. }
%In the top panel the $z_{MgII}$ is taken from
%\citet{1988ApJ...334...22S} and the $z_{CIV}$ from \citet{1988ApJS...68..539S} whereas in the bottom panel
%$z_{MgII}$ is from \citet{2013ApJ...779..161S} and $z_{CIV}$ is from \citet{2013ApJ...763...37C}.  }
\label{fig8}
\end{figure}

After finding a satisfactory explanation for the redshifted quasar spectra, 
we decided to estimate the $z_g(R)$ of the absorption lines for a few quasars. 
In the remaining discussion, we refer to $z_{in}$ as $z_g$ but with
the clear understanding that $z_g$ can only take values upto one and if $z_{in}$ is greater than one then
it definitely includes a contribution from $z_D$.  By the same token, we are aware that
for $z_{in}< 1$ there will be a contribution from $z_D$ but this is difficult
to estimate here and hence, for simplicity, we assume that $z_{in} = z_g$ when $z_{in} < 1$. 
The $z_g(R)$ of the multiple C IV absorption lines detected towards six quasars were estimated
using the same method we used for emission lines ie $z=z_{abs}$ and
$z_c=z_{MgII}$ in Eqn \ref{eqn1}.  These are listed in Table \ref{tab3}.  
Three quasars which had data on both Mg II and several C IV lines 
were selected from \citet{1988ApJ...334...22S,1988ApJS...68..539S} and another
three from the catalogues by \citet{2013ApJ...779..161S} and \citet{2013ApJ...763...37C}.
As seen in Table \ref{tab3}, C IV absorption lines are detected at a number of
redshifts towards the same quasar due to different $z_g(R)$ 
estimated in the second column of Table \ref{tab3}.  
In other words, the absorption lines arise in multiple zones located 
at different radial separations from the event horizon.  For example, 
the result on Q0014+818 with $z_{em}=3.377$ where
the $z_g(R)$ for the four different C IV absorption features vary from 1.0022 to 0.6548 
can be interpreted as lines arising in
multiple absorbing zones distributed from very close to the event horizon i.e. $R_s/2$ to 
$R_s$ in a rotating black hole as shown in top panel of Figure \ref{fig8}.
On the other hand, in quasar 52618-1059-146 with $z_{em}=4.1541$, the $z_g(R)$ of all the C IV absorption
features are close to 0.5 indicating absorbing zones near $R_s$ as shown in bottom panel
of Figure \ref{fig8}. These examples 
demonstrate that the quasar population at all cosmological redshifts show a large range in
$z_g(R)$.  
%We then used the estimates values of $z_g(R)$ in Table \ref{tab3} to
%estimate the radial separation R of the absorbing zones from black holes of  masses 
%$10^9$ and $10^{10}$ M$_\odot$ using Eqn \ref{eqn6}.  These are shown in Figure \ref{fig8}.
%We can also plot this in terms of $R_s$ but to
%quantify the small linear extent that is under discussion we show these in physical units.
As seen in Figure \ref{fig8},
if $M_{BH}=10^9$ M$_\odot$ then the C IV absorbing zone is located between $1.5\times10^{14}$ cm
to $4\times10^{14}$ cm in the quasars plotted in the top panel and between $1.5\times10^{14}$ cm and
$2\times 10^{17}$ cm in the quasars plotted in the lower panel.  Note that
$R_s = 3\times10^{14}$ cm for a black hole of mass $10^9$ M$_\odot$.  

\begin{figure}
\centering
\includegraphics[width=8cm]{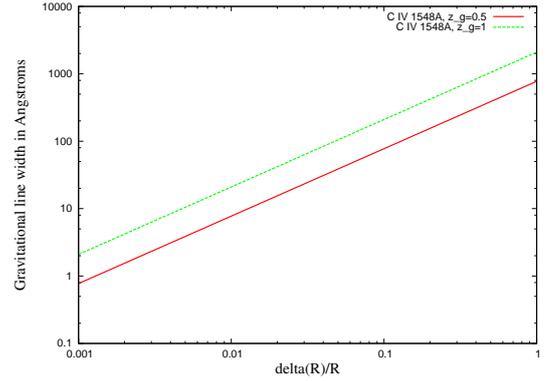}
\caption{\small The variation in the observed
line width of the C IV 1548A line as a function of $\Delta R/R$ for $z_g=0.5$ (red, solid) and 
$z_g= 1$ (green, dashed) are shown.  }
\label{fig9}
\end{figure}

%Since the observed spectral lines from quasars show large velocity shifts
%due the deep gravitational potential of the black hole, we note that 
The widths of the spectral lines arising so close to the black hole
will be dominated by the varying gravitational potential in the line forming zone
and hence cannot be used to estimate $M_{BH}$,  Instead, it 
can be used to estimate the fractional line-of-sight 
extent $\Delta R/R$ of the line forming region as given by \citet{1964ApJ...140....1G}:
\begin{equation}
\rm
\frac{\Delta R}{R} = \frac{\Delta \omega}{\lambda_{obs} - \lambda_{rest}}
\label{eqn10}
\end{equation}
where R is the radial separation of the region from the black hole and  
$\Delta \omega$ is the observed line width. 
For example, the C IV 1548A line which shows a gravitational redshift of 0.5 
will arise at a distance of $3\times10^{14}$ cm if we assume the black hole
has a mass of $10^9$ M$_\odot$.  If the observed linewidth is 50 A, then
it can be surmised that the region has an extent of $0.2\times10^{14}$ cm. 
This C IV line will appear at a wavelength of 2322 A.   We show how the observed 
C IV 1548 A line width gives input into the fractional thickness of the
line forming region $\Delta R/R$ for gravitational redshifts of 0.5 
and 1 in Figure \ref{fig9}.  As the observed line width increases, the fractional
thickness of the line forming region increases.  Thus we find that 
$z_g$ is determined from the observed line wavelength, M/R is determined from $z_g$
and $\Delta R/R$ can be determined from the  observed line width.  
However further independent information
is required to determine $M_{BH}$ and hence R and $\Delta R$. 

In the previous section,
we have already demonstrated how an intrinsic origin of the line spectrum in quasars
can explain several observational results.   Here we show how $z_{in} \sim z_g$ 
further simplifies the explanations.  We start with the six points listed in section 2.1: 

\begin{itemize}

\item Point 1 was already explained by an intrinsic origin. 
In the gravitational origin, it indicates differences in the distribution of 
matter around the black hole and their physical conditions. 
%is easily explained as intrinsic variation in the distribution of matter
%around the central black hole.  
Point 2 - the observed quasar redshift as concluded here, has 
a non-cosmological gravitational component which needs to be removed before the 
magnitude-redshift correlation can be studied. 
Point 3 is no longer an issue since all the spectral lines arise in the quasar system and 
no adhoc classification is needed. 

\item Point 4 - the high ionization lines appear at higher redshifts 
for all quasars indicating they are formed closer to the black hole
where the gravitational potential is deeper and the radiation field is stronger.  
The low ionization lines arise farther away
from the black hole at lower redshifts (ie potential) and in a weaker softer radiation field.

\item The explanation for Point 5,6 is similar to point 4.   
The dependence of column density ratios on redshifts in a quasar spectrum is actually a dependence on
the gravitational component of the redshift.  
If the high ionization lines of C IV and Si IV arise
in the same region closer to the black hole then the ratio C IV/Si IV will not 
show any dependence on redshift \citep{2015ApJS..218....7B}  
since both lines have same contribution from $z_g$.  The same argument holds for
C II and Si II.  However when the ratios of differing ionization species are
considered ie C II/C IV and Si II/Si IV, then since C II (and Si II) and C IV (and Si IV) 
arise in different absorbing zones around the black hole, they are shifted by different $z_g$ and hence appear to
show a variation with $z$.  In fact, results presented in \citet{2016arXiv160203894S}
also give evidence to the forbidden lines arising in a common
region and showing the same $z_g$ and He II, C IV and Si IV arising in another common zone 
and showing the same $z_g$.  This then indicates
segregation of line forming zones around the quasar based on densities and incident radiation field and can 
be used to understand the quasar structure.

\item Quasars at $z\sim6$.  It is observed that the Gunn-Peterson trough \citep{1965ApJ...142.1633G}
becomes more prominent with increasing redshift 
of a quasar, so that most quasars near $z\sim6$ show the trough.  However significant differences
are noted in spectra of quasars with the same observed redshift 
\citep[e.g.][]{2006AJ....131.1203F} making it difficult for cosmological models to explain the result.
In the gravitational redshift
model, the highest redshift quasars contain large contributions from both $z_g$ and $z_c$.
So if $z_{g}\sim1$ for a $z=6$ quasar, then its cosmological redshift is only $z_c=2.5$ as
listed in Table \ref{tab2}! 
It would be interesting to confirm this by searching for Mg II absorption near redshift 2.5 
along the sightlines to quasars with redshift near 6.  In this case, the 
Gunn-Peterson trough is due to the joint effect of
broad overlapping lines and coarse spectral resolution and has no 
connection to the universe at $z_c\sim6$.  

\item 
It has been observed that the number of absorption features detected in spectra of quasars 
at high redshifts are not an extrapolation from lower redshifts \citep[e.g.][]{1991AJ....102..837S}.  
Such a correlation is expected in the intervening origin for the absorbing
gas and the lack of correlation has baffled astronomers.  However,
we note that no such correlation is expected in a gravitational origin. 
Instead the absorption features depend on the distribution of matter around the black hole.  

\end{itemize}

%On a general note, any correlation which is expected with cosmological redshift but which is not found 
%to be followed by quasars gives clues to the contribution of a gravitational redshift
%component to the intrinsic redshift. 
Thus it is clear that the contribution of $z_g \le 1$ is responsible for several interesting
redshift characteristics observed in a quasar spectrum. 

\subsection{Ultraviolet continuum}
\begin{figure}
\centering
\includegraphics[width=7cm]{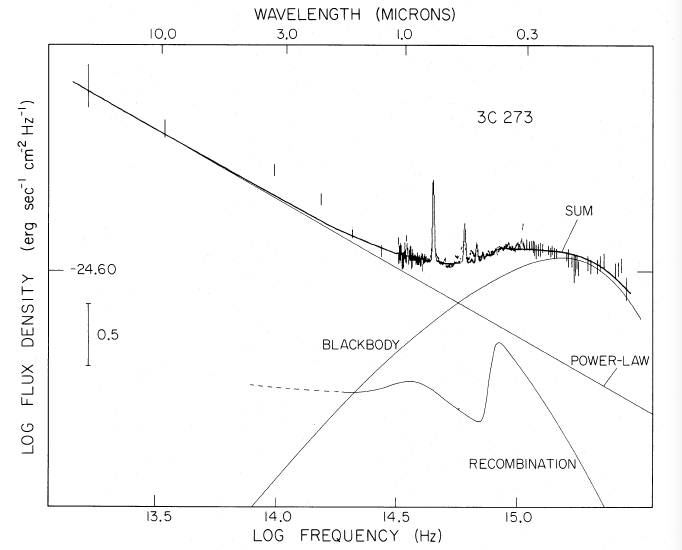}
\caption{\small The continuum emission of quasar 3C273 at $z_{em}=0.158$ is shown 
from ultraviolet to infrared.  Similar spectra are observed from other active nuclei also.
Note the blue 'bump' in the ultraviolet and the three components which are
required to fit the spectrum.  Figure is copied from \citet{1982ApJ...254...22M}.}
\label{fig10}
\end{figure}

Quasars generally show strong ultraviolet continuum emission (see Figure \ref{fig10}) often referred
to as the `blue' bump and which appears to be similar to the uv upturn (part of the blue bump)
widely observed in the nuclei of elliptical galaxies and spiral bulges.  
%The study of the ultraviolet continuum from quasars is non-trivial
%due to its rich line spectra which modifies its nature.  
IUE data suggested that the continuum emission of quasars showed excess in the ultraviolet 
\citep[e.g.][]{1983MNRAS.205.1053B}. 
The thermal component of the continuum emission is believed to arise
in the accretion disk around the black hole, modelled to fit the observed spectra of
quasars and Seyfert galaxies \citep[e.g.][]{1989ApJ...346...68S}. 
Since the quasar ultraviolet continuum is often corrupted by the presence of numerous spectral lines, 
we also examined studies of the upturned ultraviolet spectra of the nuclei of elliptical galaxies.  The 
assumption is that the basic physics behind the quasar continuum in the ultraviolet should be
applicable to all active nuclei which is reasonable since most seem to show the blue excess.  
Thus, in the following we discuss the ultraviolet continuum from quasars and nuclei of galaxies.

\begin{figure}
\centering
\includegraphics[width=7cm]{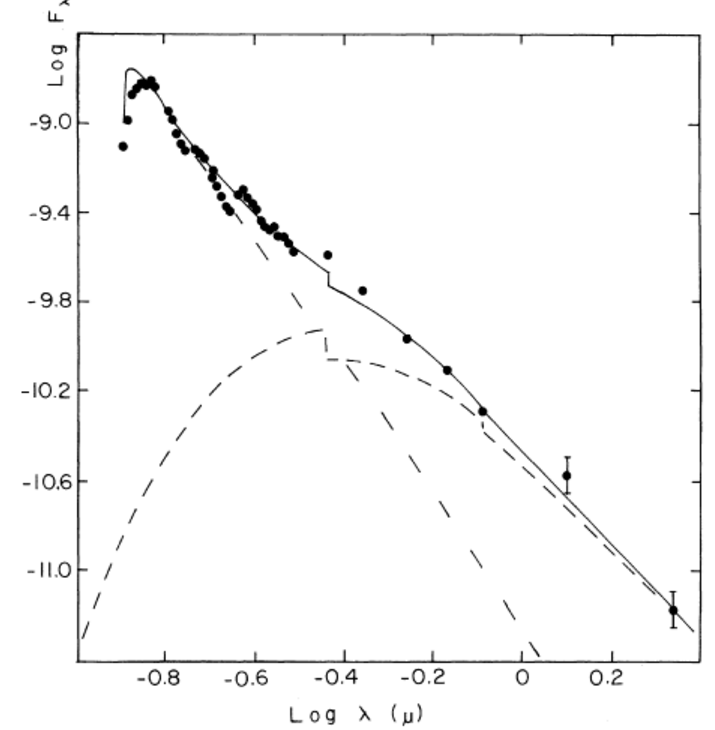}
\caption{\small The continuum emission spectrum of the white dwarf VW Hyi from ultraviolet
to near-infrared. The long dashed line shows the spectrum of a 20000 K stellar atmosphere and
short dashed line represents an accretion disk.  Figure is copied from \citet{1984AJ.....89..863M}.}
\label{fig11}
\end{figure}

The spectra taken by the International Ultraviolet Explorer (IUE), launched in 1978, 
of quiescent and active population of early type galaxies 
%in a sample of 50 galaxies 
showed a rise  in the blue/ultraviolet emission
shortwards of 1550 A which appeared to be absent in star forming galaxies \citep{1987IAUS..127..439B}.
Interestingly, it was found that this ultraviolet upturn was a
nuclear phenomenon which got diluted when the signal of the entire galaxy was included
\citep{1996A&A...313..405B}.  Moreover the nuclear spectra of some of these galaxies showed 
dips near 1400 A and 1600 A \citep{1996A&A...313..405B}.  The ultraviolet upturn 
was intriguingly similar to the excess recorded in the spectra of subdwarf stars with
temperatures 110000-150000K or of hot DO type white dwarfs with temperature $>75000$ K
or of nuclei of planetary nebulae and the
the absorption dips near 1400A and 1600A were similar to those seen in DA4 or DA5 white dwarfs
which have temperatures of 12000-20000 K \citep{1996A&A...313..405B}.   
This prompted \citet{1996A&A...313..405B}   
to conclude that the ultraviolet upturn could indicate the presence of such
stellar objects in the central region of the galaxies.
We recall that \citet{1963ApJ...138...30M} had noted a similarity between the 
colours of old novae/white dwarfs and quasars (3C 48, 3C 196, 3C 286). This was supported by
\citet{1967ApJ...147.1219B} who commented on the similarities in the
absorption spectra of the white dwarf Hz 29 \citep{1957ApJ...126...14G} and quasars. 
\citet{1982ApJ...254...22M} found that the observed continuum spectra from infrared to ultraviolet
of several quasars and Seyfert 1 galaxies could be explained by the combination spectrum
of (1) a power law of index $-1.1$ (2) optically thin
Balmer continuum emission  (3) single temperature black body emission from gas
with temperatures between 20000 and 30000 K as shown in Figure \ref{fig10}.  
For comparison, we show the spectrum of the white dwarf VW Hyi 
from near-infrared to ultraviolet wavelengths \citep{1984AJ.....89..863M} in Figure \ref{fig11}.  
The similarity is unmistakeable in that the continuum starts rising
at the near-ultraviolet wavelengths (note that the scales are different) and a separate
thermal component has to be included in the model to explain the observed ultraviolet emission. 
This indicates that both white dwarfs and quasars have a black body 
component which dominates the continuum emission at ultraviolet wavelengths.  

Another similarity to quasars is observed in the absorption features detected in the ultraviolet 
spectra of white dwarfs \citep[e.g.][]{1957ApJ...126...14G,1983ApJ...275L..71D, 2003MNRAS.341..477B}. 
IUE detected narrow absorption lines of highly ionized ions such as C IV, Si IV 
with a range of systemic velocities
in the white dwarf spectra \citep[e.g.][]{1983ApJ...275L..71D} indicating that such  
highly ionized species exist in the immediate vicinity of the white dwarf. 
\citet{2003MNRAS.341..477B} studied 23 hot white dwarfs and detected absorption lines of
CIV and also found that these lines were blue-shifted from the photospheric
line velocities.  \citet{1998ApJS..119..207H} studied 55 white dwarfs and
separated the absorption lines forming in the photosphere, circumstellar material
and in the interstellar medium and inferred that several lines were blue shifted
from the photospheric velocities.  These lines, interestingly, were always detected at a
blue shifted velocity wrt to the photospheric velocities and distinct from the interstellar medium
velocities.  A possible origin for the blue-shifts wrt to the photospheric velocities could be the reducing
effect of the gravitational potential of the white dwarf as the line forms
at an increasing distance from the white dwarf.  
The velocity shifts due to the gravitational potential have been estimated to be
$25-80$ kms$^{-1}$ depending on the mass of the white dwarf
and its composition \citep[][Figure 3 in the paper]{1967ApJ...149..283G}. 
\citet{1984AJ.....89..863M} reported  a wide Lyman $\alpha$ absorption feature 
in the spectrum of the white dwarf VW Hyi.
\citet{1984ApJ...277..700P} also reported the detection of a broad Lyman $\alpha$ absorption with
wings (similar to a damped Lyman $\alpha$ feature in a quasar spectrum)
and  Mg II lines in emission in the spectrum of the dwarf nova U Gem. 
We recall that a classical nova is formed in a binary system consisting of a white dwarf and 
a main sequence or red giant
companion star when accreted mass on the white dwarf ignites in an explosive thermonuclear
reaction once it crosses a certain density and temperature.  The nova explosion is
detectable at wavelengths ranging from $\gamma$ rays to radio.  
%Thus,  white dwarf spectra show the presence of gravitationally redshifted absorption lines due to 
%C IV, Si IV and Lyman $\alpha$ as seen in quasar spectra. 

The observational similarities in the ultraviolet between white dwarfs and quasars
are many.  They share a similar continuum emission spectral shape.
They both show gravitationally redshifted absorption lines of highly ionized species and a
broad Lyman $\alpha$ absorption.  All these 
suggest that quasars share some physical properties with white dwarfs.
In the next subsection we use these important clues to gain further insight into quasars. 

\subsubsection{Hot degenerate matter surface}
We infer from the above discussion that the ultraviolet continuum in quasars can be explained by
radiation from a black body with temperatures ranging from 10000 to 150000 K.   

We recall that once a low mass star exhausts the fuel in its core, thermonuclear
reactions stop and the star starts collapsing under gravity.  It is estimated that
when the densities are around
$10^5-10^8$ gm cm$^{-3}$, the electron degeneracy pressure can balance
the gravity of the low mass star and a white dwarf is formed.  If the star is massive but less
than about 8 M$_\odot$, then the phase of contraction after the fuel in its core is exhausted will stop
only when the densities
exceed the nuclear densities ie $>1.2\times10^7$ gm cm$^{-3}$ and neutron degeneracy
pressure halts the gravitational collapse.  In fact, the densities in neutron stars can get as
high as $10^{16} - 10^{17}$ gm cm$^{-3}$ \citep{1983bhwd.book.....S}.

Keeping this in mind, we shift the discussion to quasars.   
It is accepted that the central object in a quasar is a supermassive black hole.
%and that the emitting and absorbing zones are arranged around it.   
The black hole will accrete matter and it is reasonable to expect the density of the 
matter to increase as it falls towards the black hole. 
%radially inwards as the gravitational influence of the black hole increases. 
The infalling matter, at some distance from the black hole, can accumulate with 
densities $> 10^8$ gm cm$^{-3}$ leading to degenerate neutron matter 
collecting around the black hole and at some point, being
supported against further gravitational collapse by neutron degeneracy pressure.  
Thus, it appears reasonable to postulate that a degenerate neutron
surface forms and exists in equilibrium around the black hole.  The immense gravity
of the black hole can heat up this surface to high temperatures and it can possibly
emit in X-rays.   Outside this dengerate neutron surface, the matter will
have lower densities but once matter accumulates at densities $10^5-10^8$ gm cm$^{-3}$ 
then the electron degeneracy pressure can counter
the strong gravity of the black hole and the degenerate matter can remain in
equilibrium.  This, then, forms a degenerate electron shell
outwards of the neutron shell.  Again the immense gravitational energy
of the black hole can cause this surface to be heated up to high temperatures and
radiate ultraviolet continuum like a white dwarf surface.  
Outwards of this could be lower density non-degenerate matter which contributes to the line spectrum. 
In fact, this appears to be the most logical setup of matter around any black hole.

Thus, the ultraviolet continuum observed in quasars will be emission
from the degenerate matter surface around the black hole.  Since there are quasars
which are X-ray bright and others which are ultraviolet-bright: it needs
to be investigated if these can be interpreted as evidence to the type of
degenerate matter surface which dominates the continuum emission of a particular
quasar.   While it appears most reasonable to expect both a degenerate neutron and 
electron surface around a black hole - there might exist a variety which 
needs to be examined from observations and physical understanding of degenerate matter. 
 
\section{Proposed structure of a quasar}
\begin{figure}
\centering
\includegraphics[width=8cm]{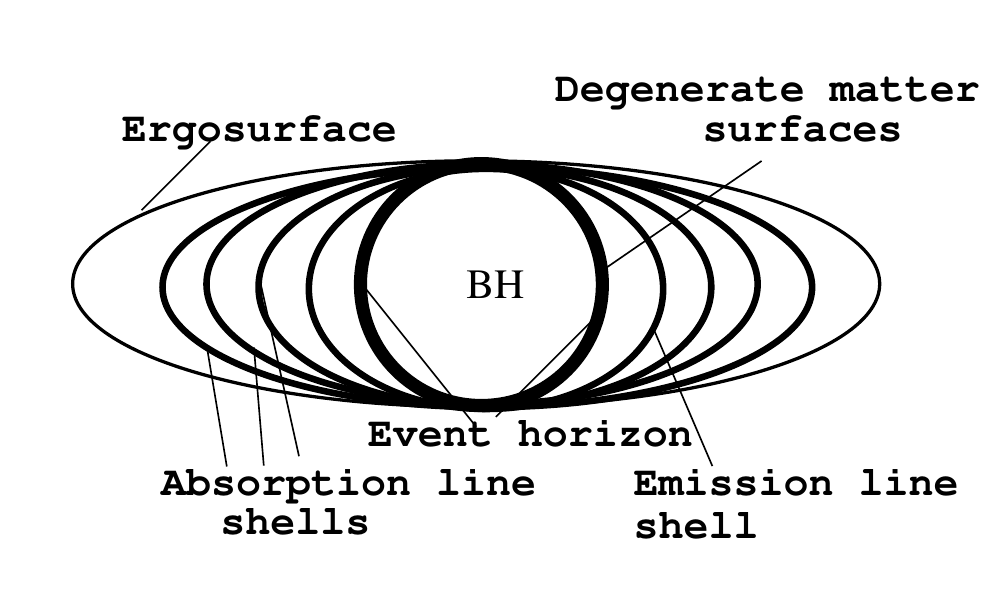}
\caption{\small Proposed structure of a quasar.  The dense degenerate matter shells, line 
emitting shell and absorbing shells are all located inside the ergosphere of a rotating
black hole.  The spectral lines arising in matter inside the ergosphere will 
suffer a redshift ranging from $\sim 1$ (close to event horizon) to 0.5
(close to the ergosurface) due to the strong gravitational potential of the black hole (BH).  }
\label{fig12}
\end{figure}

Amazingly, it appears that the discussion so far has
uniquely determined the structure of a quasar which is schematically shown in Figure \ref{fig12}
and described below. 

A supermassive black hole is the central object.  Matter is being accreted by the black hole
and compressed to progressively larger densities as it approaches the black hole.  
At appropriate matter densities, the degeneracy pressure of neutrons and electrons will 
be sufficient to support the matter from further
gravitational collapse.  Thus, in the quasar structure, the 
black hole is surrounded by a shell of matter supported by degenerate neutron pressure
and outside that a shell supported by degenerate electron pressure.
As mentioned in the previous section,
the degenerate neutron shell can have densities upto $10^{17}$ gm cm$^{-3}$ while the
degenerate electron shell can have densities upto $10^8$ gm cm$^{-3}$. 
The matter outside the degenerate shells will be non-degenerate and 
will accumulate in dense shells with densities $\le 10^5$ gm cm$^{-3}$.  
The hot degenerate electron shell will give
rise to an intense ultraviolet continuum which bears close resemblance to the
white dwarf continuum.  Its hard radiation field will ionize
and heat a Stromgren shell \citep{1939ApJ....89..526S} 
around it where several elements including hydrogen will emit spectral lines.
Since some quasars show the existence of forbidden emission lines which will 
collisionally deexcite if densities $>10^8$ cm$^{-3}$, 
we suggest that there should exist a relatively lower density ($<10^8$ cm$^{-3}$) 
region around the degenerate 
matter shells in some quasars where the forbidden lines can arise and 
which could have an origin similar to the stellar cavities seen around stars. 
Surrounding this `cavity' would then be the dense ionized shell which emits Lyman $\alpha$ 
and resonance lines of highly ionized
species such as C IV, Si IV.   We note that the forbidden lines can also arise in small
lower density pockets located inside the Stromgren shell. 
Since the emission lines arise in the shell located closest to the black hole and are formed
under the influence of the strongest possible gravitational potential,
their velocities contain the largest contribution from $z_g(R)$ in the spectrum. 
Outwards of this ionized Stromgren shell, matter
will accumulate to form cooler dense shells which will absorb the ultraviolet
continuum from the hot degenerate matter surface and give rise to the absorption line spectrum.  
The matter at different distances from the black hole will absorb different wavelengths
and give rise to the observed absorption line spectrum of the quasar. 
Matter is highly ionized in the shells closest to the emission line shell
as can be surmised from the highly ionized species such as 
C IV and Si IV which are detected in absorption at redshifts which are closest to the
emission line redshifts (see Figure \ref{fig4}).  
Such absorbing shells will continue forming as matter is
accreted.  The spectral lines will keep appearing at decreasing redshift as compared to
the emission lines as they arise in radially distant shells as long as they lie within detectable
influence of the gravitational potential of the black hole.  
The absorption lines which arise far from the black hole and hence are not shifted by
the gravitational potential, can then give an estimate of the $z_c$ of the quasar
as we had assumed in $z_c = z_{MgII}$.
The emission and absorption lines which show $z_g>0.5$, have to arise from matter
inside the ergosphere of a rotating black hole. 
The distinct geometry of the ergosphere (see Figure \ref{fig12}) can be assumed by the equipotential
shells which can lend the appearance of a thick accretion disk to 
the continuum emitting and line forming regions.  Any emitting/absorbing
shells in the polar regions of the black hole will necessarily have to be outside the ergosphere since
the ergosurface and the event horizon coincide in the polar regions.  
Thus, in this structure, lines formed in the equatorial plane of a rotating black hole 
can suffer a maximum gravitational redshift
of 1 whereas those formed in the polar regions can only suffer a maximum gravitational
redshift of 0.5 which is also the maximum gravitational redshift that can
result from a non-rotating black hole.   

This structure,  we note, can satisfactorily explain the ultraviolet observations of quasars -
both continuum and spectral lines within the ambit of black hole physics.   It also
explains the similarities noted between the ultraviolet properties of quasars and white dwarfs such
as the blue bump and gravitationally redshifted absorption lines of highly ionized ions C IV and Si IV. 
The large gravitational field of the black hole can
amplify the luminosity of the quasar since the emission from the event horizion of a black hole 
is expected to be boosted by a staggering factor of $\sim 10^{27}$ \citep{1983bhwd.book.....S}.  The
details of this needs to be studied but since the degenerate matter shells are
expected to be located just outside the event horizon, an obvious inference would be that
such boosting is responsible for the large luminosities of quasars.  
This structure for a quasar naturally emerges from the observational results and hence  
successfully accounts for quasar observables making it the most plausible model of a quasar. 

\subsection{Other active nuclei}
In this section, we discuss blazars and Seyfert 1 nuclei and comment on their structure.
A few blazars and Seyfert 1 galaxies show the presence of
absorption features detected bluewards of the emission lines in their spectra while broad emission lines
are observed in Seyfert 1 and some blazars and radio galaxies. 

\subsubsection{Blazars}
Blazars are classified into BL Lac objects and flat spectrum radio quasars (FSRQs).  
BL Lacs occasionally show line features superimposed on a mostly featureless flat continuum spectrum.
BL Lacs are radio-loud, highly variable and show a stellar-like appearance in the optical. 
In fact, BL Lacs are believed to be radio quasars observed along the radio jet or minor axis.   
The FSRQs are generally more energetic and show wide emission lines in their spectra.
Here we estimate the intrinsic redshifts shown by the emission lines from BL Lac and FSRQ objects using
the method outlined earlier. 
%whereas only few BL Lacs show spectral lines in their spectra.

We use the sample of 23 BL Lac objects given in \citet{2011A&A...525A..51B} for which 
Mg II absorption lines have been recorded 
to estimate the maximum gravitational redshift in the spectrum  (see Table \ref{tab4}).
We notice that BL Lacs show a lower observed redshift range (0.875 to 1.522) 
compared to quasars although the
range of $z_{MgII}$ (0.2527 to 1.2847) seems comparable (also see Table \ref{tab1}).  
Since  $z_{MgII}$ is a proxy for $z_c$, this
means that quasars and BL Lacs have similar cosmological redshift distribution and 
the difference in the observed redshifts is due to the different contributions
of $z_g$ to the line velocities.
We find the remarkable result that 22/23 BL Lacs show $z_g< 0.5$ as listed in 
Table \ref{tab4} ($z_g$ ranges from 0.017 to 0.49).  
{\it Clearly the spectral lines in BL Lacs, when detectable,
arise in a lower gravitational potential compared to quasars. } 
This, we believe, is highly significant and should be verified on a larger sample. 
It gives independent support to the existing model of a polar sightline 
defining a BL Lac.  Since at the poles the ergosphere coincides with the event 
horizon for all black holes, $ z_g \le 0.5$ for any line arising there.
The other explanation can be that the black hole
in  BL Lacs are non-rotating and hence $z_g\le 0.5$. 
We recall that BL Lacs generally show a flat spectrum with few spectral lines. 
This can be explained if the equipotential degenerate matter and line
forming shells are located inside the ergosphere and hence not detectable when viewed from the poles. 
If BL Lacs contained a non-rotating black hole then the degenerate matter and
line forming shells can be arranged as spherical equipotential shells around the event horizon and we 
should have detected numerous spectral lines in their spectrum since orientation would play no role
in the nature of the observed spectra.
We end by inferring that BL Lacs are quasars with a rotating black hole viewed polewards
and hence the maximum $z_g$ shown by spectral lines is $ 0.5$.  We note that this supports
the inference derived from radio observations.

\begin{figure}
\includegraphics[width=7cm]{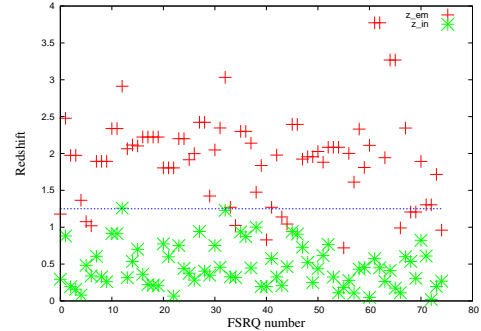}
\caption{\small The observed redshift $z$ (in red +) and intrinsic redshift $z_{in} \sim z_g$ (in green *) estimated from
$z_{MgII}$ for FSRQ are shown.  The horizontal line is drawn at redshift=1.25. The data has 
been taken from \citet{2012ApJ...754...38C}.}
\label{fig13}
\end{figure}

We also used the same method on a sample of 75 FSRQs with available Mg II absorption line
redshifts \citep{2012ApJ...754...38C}. 
The estimated $z_{in} \sim z_g$ are larger than BL Lacs but similar to quasars i.e. $z_{in} < 1.25$ 
and these are summarised in Figure \ref{fig13}. 
FSRQs have to be observed through the ergosphere of a rotating black hole ie away from
the poles, to explain the deduced distribution of $z_g$ and which also explains the
presence of wide emission lines in FSRQs.  

These results then unequivocally support the suggestion
in literature that blazars are quasars and possess the structure shown in Figure \ref{fig12}. 
BL Lacs are quasars viewed from the pole.   

\begin{table}[h]
\centering
\caption{\small Estimating the $z_{in} \sim z_{g}$ component in the observed
redshift of a sample of BL Lac  objects for
which Mg II absorption line redshifts have been recorded.  
Data is taken from \citet{2011A&A...525A..51B}.} 
\begin{tabular}{l|c|c|c}
\hline
{\bf BL Lac objects} & {$\bf z_{em}$} & {$\bf z_{MgII}=z_c$} &  {$\bf z_{g}\sim z_{in}$}  \\
\hline
0100-337    &    0.875 &   0.6810   & 0.115   \\
0238+1636   &    0.940  &  0.5245   & 0.272   \\
0241+0043   &    0.989   & 0.6310   & 0.219   \\
0334-4008   &    1.351    &1.0791   & 0.130   \\
0423-0120   &    0.915  &  0.6338   & 0.172   \\
0428-3756   &    1.110   & 0.5592   & 0.353   \\
0457-2324   &    1.003 &   0.8922   & 0.057  \\
0538-4405   &    0.890  &  0.6725   & 0.130   \\
0745-0044   &    0.994   & 0.7979   & 0.109   \\
0909+0121   &    1.022  &  0.5369   & 0.316   \\
0942-0047   &    1.362   & 0.8182   & 0.299   \\
0948+0839   &    1.489 &   1.0763   & 0.199   \\
1147-3812   &    1.049  &  0.3750   & 0.49   \\
1408-0752   &    1.5     & 1.2753   & 0.099   \\
1410+0203   &    1.253 &   1.1123   & 0.067   \\
1427-4206   &    1.522 &   1.0432   & 0.234   \\
1522-2730   &    1.294  &  1.2847   & 0.004   \\
1743-0350   &    1.054  &  0.2527   & 0.639   \\
1956-3225   &    1.242 &   0.6236   & 0.381   \\
2031+1219   &    1.215  &  1.1158   & 0.047   \\
2134-0153   &    1.285 &   1.2458   & 0.017   \\
2225-0457   &    1.404 &   0.8458   & 0.302   \\
0221+3556   &    0.944  &  0.6850   & 0.154   \\
\hline  
\end{tabular}
\label{tab4}
\end{table}

\subsubsection{Seyfert 1 galaxies}
Spiral galaxies with a bright compact nucleus which showed emission lines 
similar to the lines detected in planetary nebulae were first identified by
\citet{1943ApJ....97...28S}.  These galaxies are now known as Seyfert galaxies.
The nuclear emission lines were much broader ($> 30$ A)
than those found in emission nebulae such as HII
regions in our Galaxy or in neighbouring galaxies \citep{1943ApJ....97...28S}
thus indicating their distinct nature.  It was also inferred that the maximum observed 
width of the hydrogen emission lines
increased with absolute magnitude and with the ratio of light in the nucleus to the total
light in the nebula \citep{1943ApJ....97...28S}.   Since the central object
in a Seyfert galaxy is also a black hole, we note that the large observed widths
of the emission lines could be due to the varying gravitational potential across the emitting
region as given by Eqn \ref{eqn10} while the increase in absolute magnitude could indicate 
proximity to the event horizon of the black hole. 
Absorption lines were also detected by \citet{1943ApJ....97...28S} and were identified with 
the stellar spectrum from the galaxy.  
Seyfert galaxies were classified into two types based on their line spectra and referred to
as type 1 and type 2 \citep{1971BAAS....3R.237K}.
IUE detected absorption lines in the Seyfert 1 spectra which were blueshifted
wrt emission features by upto 2500 kms$^{-1}$ and these were inferred to
have a non-stellar origin \citep{1988MNRAS.230..121U}.
Further observations with the Hubble Space Telescope (HST) found that more than half the Seyfert 1 galaxies
showed both emission and absorption lines in the nuclear spectrum 
and that high ionization lines like C IV were generally detected whereas few Seyfert 1 galaxies showed
Mg II absorption lines \citep{1999ApJ...516..750C}. 
The C IV doublet was detected in absorption at multiple blueshifted velocities  
wrt to the emission line and the absorption features of widths 100-300 kms$^{-1}$
were resolved into several narrow kinematic components at high spectral resolution.
%Variability in the absorption profiles is also observed and 
Column densities of C IV ranging from $10^{13-15}$ cm$^{-2}$ are estimated \citep{1999ApJ...516..750C}.
\citet{1999ApJ...516..750C} also find an excellent correlation between the ultraviolet
absorption features and  X-ray 'warm absorbers' and infer that the two phenomena
are related.  

From the above discussion, it appears that quasars and Seyfert 1 
galaxies share a few common properties:  (1) compact bright nucleus (2) broad
emission lines (3) multiple blue-shifted absorption lines.   The main differences between
the two types of objects are  (1) quasars are observed at redshifts ranging from low
to high ($\sim 6$) whereas Seyferts are mostly
observed at low redshifts.  For example, all galaxies in a sample of 964 
Seyfert galaxies  \citep{1988SoSAO..55....5L} has a redshift $z < 1$ with the median redshift
being $<0.1$, (2) the range of absorption line redshifts are accordingly lower in Seyfert 1 galaxies, 
(3) it is difficult to locate a host 
galaxy around a quasar whereas Seyfert nuclei are hosted in spiral galaxies.  

Thus, we suggest that the basic structure of the nuclei of Seyfert 1 is same as
a quasar in that there is a black hole surrounded by degenerate matter surfaces
and line forming matter.  However the lower observed redshifts of Seyfert galaxies
and hence the small contribution of gravitational redshifts to the observed line velocities
indicates that the lines arise far from the event horizon.  It is clear that
matter is not located inside the ergosphere of a rotating black hole in Seyfert 1 nuclei.  
The multiple blue-shifted absorption lines detected in the spectra of Seyfert 1 nuclei 
indicate the varying separations of the absorbing zones from the black hole and hence the varying
$z_g$.  If we assume that a C IV 1548A line of half width 30A with  
$z_g=0.5$ (arises at $R=R_s$) is observed from a quasar then Eqn. \ref{eqn10} tells us that 
the emitting region will be sheet-like i.e. $\Delta R \sim 0.04 R = 0.04 R_s$.  However if the
same line with the same parameters was observed from a
low redshift Seyfert nucleus with $z_g=0.01$ (arises at $R=50 R_s$) then Eqn. \ref{eqn10} estimates
a thickness of $\Delta R \sim 2R = 100 R_s$ for the emitting layer around the black hole. 
Thus, observations and the inferred structure of quasars allows us to infer the following
about Seyfert 1 galaxies :  
(1) the hot degenerate matter surface is located close to the event horizon giving the nucleus
a compact bright morphology like quasars.
(2) the nuclear line spectrum arises over a large region ($100 R_s$) from the event
horizon as inferred from the low gravitational
redshifts suffered by the lines and large line widths due to the varying gravitational potential.
(3) the line forming shells are physically larger and tenuous than in quasars.

The above discussion can also be extended to other kind of active nuclei like
those hosted in elliptical and Seyfert 2 galaxies and its properties similarly studied. 
From our study of quasars, blazars and Seyfert 1 galaxies above, we are convinced that
the structure suggested for quasars in Figure \ref{fig12} holds for all active nuclei with
the differences in observed properties being due to (1) the separation between the 
degenerate matter surface and the
event horizon and (2) the separation between the emitting and absorbing shells from
the event horizon (3) rotation status of black holes.   

The gravitational redshift shown by the spectral lines from all non-quasar active nuclei 
appear to be $<<0.5$. However we caution against interpreting this result to mean that
all non-quasar objects host a non-rotating black hole and all quasars host a rotating black hole.
It might be likely that all active nuclei host a rotating black hole and the observational
differences are only due to the varying location of the line forming regions.  
We should further understand observations before arriving at any conclusions regarding
the spin of the black hole.  

{\it In summary, we conclude that quasars are the most extreme form of an active nucleus with
a structure shown in Figure \ref{fig12}.  In case of other active nuclei
the matter is located far from the event horizon and spread over large radial distances. }

\section{Variability in quasars}
Variability over several timescales ranging from minutes to years
is observed in several active nuclei especially in quasars
(includes blazars) and Seyfert 1 nuclei.  The variability is generally confined to the continuum 
emission although there are reports of variability in the line strengths \citep[e.g.][]{1999ApJ...516..750C}.
Here we restrict the discussion to variability in the continuum emission signal. 

We recall that the continuum spectrum of quasars from optical to radio bands can be
fitted by a power law likely to be synchrotron emission whereas 
thermal processes dominate the ultraviolet emission and also contribute to the optical emission.  
The location of the synchrotron emitting region wrt the black hole is not clear and
hence we do not discuss variability in the synchrotron emission.
However, since in the proposed structure, the
black body continuum arises on the hot degenerate matter surface,  it appears reasonable
to suppose that any variability detected in
the ultraviolet (and optical in many cases) will be due to an event on the degenerate matter surface. 
Thus, once the observed variability is established to be in the thermal emission,
it can constrain the physical size of the emitting surface using light travel arguments 
and since this occurs close to the event horizon,  the black hole mass can also
be constrained.  This can then complete the decoding of quasars.
Using wavelengths other than ultraviolet and optical (in some cases) might 
require disentangling the power law and thermal contributions in its emission. 
For example, \citet{1963ApJ...138...30M} found that while
they could extrapolate the optical flux
for 3C 48 and 3C 196 from the radio, it was not possible to do so for 3C 286 - one way to
understand this would be that the optical emission in 3C 48 and 3C 196 is dominated by
synchrotron emission whereas in 3C 286, its a combination of synchrotron and thermal emission.  
3C 48 showed variability $\le 15$m \citep{1963ApJ...138...30M}. 
If we assume this arises on the degenerate matter
surface (although it is likely to be synchrotron), then it indicates 
a physical size of $2.7\times10^{13}$ cm and a black hole mass 
of $10^8$ M$_\odot$.  However if the variability is in the synchrotron
emission and arises far from the black hole, then this  mass estimate will be wrong. 
For example, correlated variability is detected in radio, optical, ultraviolet, X-ray bands
in the BL Lac object PKS 2155-304 with X-rays leading the other bands by 2-3 hours
which indicates a synchrotron origin for the wideband variability  \citep{1995ApJ...438..120E}.  
Unless it is established that this synchrotron variability arises close to the black hole, it
cannot be used to estimate the black hole mass. 

A change of 45 \% in the radio flux density was detected over a timescale of 3 hours
in the blazar PKS 0537-441 \citep{1994A&A...288..731R}.
This indicates a physical size of $\sim 3.2\times10^{14}$ cm. 
Rapid variability in the optical emission over a timescale of 1.5h was first detected in
BL Lac objects \citep{1989Natur.337..627M}.  The light travel time argument
constrains the size of the emitting region to $\sim 1.5 \times 10^{14}$ cm.
If the origin of this variability lies on the degenerate matter surface
then it implies a black hole of mass $\sim 10^9$ M$_\odot$ as suggested by \citet{1989Natur.337..627M}.
Obviously some assumption or a priori information regarding the
location of the variability wrt to the black hole is needed to estimate the black hole mass,
if at all.  If it is localised to the degenerate matter surface then the timescales 
can be used to estimate the black hole mass.

In the next section we discuss one of the fastest and most energetic variable signal known i.e.
$\gamma$ ray bursts and show how these are related to quasars.

\subsection{$\gamma$ Ray Burst (GRB)}
GRBs were first reported by \citet{1973ApJ...182L..85K} as short bursts of duration $\sim 0.1$s to
$\sim 30$s and time-integrated flux densities of $10^{-5} - 2\times10^{-4}$ ergs cm$^{-2}$ in
the energy range of 0.2 to 1.5 MeV.  Soon after the
discovery of GRBs, the outbursts appeared to show two durations  - short 
($< 1$s) and long ($>10$s).  In the extensive data on the GRBs now available, 
a bimodal distribution in duration of the GRB with peaks centred
around 0.5s and 50s has been noted.  Such short timescale variability implies a compact emitting region 
of sizes $\sim 1.5\times10^{10}$ cm and $\sim 1.5\times10^{12}$ cm.  If this region is close
to the black hole then they imply masses of $10^5$ and $10^7$ M$_\odot$.
This, then suggests that low mass black holes can be responsible for GRBs.
However we note that several longer duration GRBs and recurrent GRBs have also now been
detected.  This, then, can indicate a range of properties for the host object.  
We are also aware that two possible origins have been put forward to explain GRBs.  
An 'afterglow' emission was
first detected in the X-ray \citep{1997Natur.387..783C} and has subsequently been detected in 
bands ranging from X-rays to radio. 

\begin{table}
\centering
\caption{\small Estimating the $z_{in} \sim z_{g}$ component in the observed
redshift of a sample of $\gamma$ ray bursts for
which Mg II absorption line redshifts have been recorded.
Data is taken from \citet{2006ApJ...648L..93P}.}
\begin{tabular}{l|c|c|c}
\hline
{\bf GRBs} & {$\bf z_{em}$} & {$\bf z_{MgII}=z_c$} &  {$\bf z_{g}\sim z_{in}$}  \\
\hline
GRB & & & \\
010222      &    1.477 &   0.927   & 0.285   \\
020405      &    0.695 &   0.472   & 0.151   \\
020813      &    1.255  &  1.224   & 0.014   \\
021004      &    2.328 &   1.38   &  0.398   \\
050505      &    4.275 &   1.695   & 0.957   \\
050820      &    2.6147 &  0.692   & 1.136   \\
050908      &    3.35  &   1.548   & 0.707   \\
051111      &    1.55  &   1.190   & 0.164   \\
060418      &    1.49  &   0.603   & 0.553   \\
030323$^1$  &    3.37  &   1.4092   & 0.814   \\
\hline
\end{tabular}

$^1$ From  \citet{2004A&A...419..927V}
\label{tab5}
\end{table}

Intriguingly, emission and absorption features have been detected in the 
afterglow spectrum of a GRB even at redshifts as high as $\sim 6$ \citep[e.g.][]{2001A&A...370..909J}
and which are similar to features in a quasar spectrum.  
Comparison of the incidence of absorption features due to Mg II
in a GRB afterglow spectrum and quasars have found that former shows four times
higher incidence of Mg II absorption as compared to quasars \citep{2006ApJ...648L..93P} and which has been
difficult to understand in an intervening origin for the absorption features.  
On the other hand, the similarity of
the absorption line spectra of quasars and GRBs have led to studies examining a possible 
connection between the two
\citep{2003ApJ...585..112B,2007RSPTA.365.1357B}.  These studies have concluded that
(1) one or more quasars are often found lying within a degree of a GRB,
(2) the deduced redshifts of the absorption lines in a GRB spectrum show a periodicity similar to
quasars 
(3) GRBs are local as also suggested for quasars
and are ejected from a galaxy and (4) the observed redshifts are
intrinsic.  Keeping all this in mind, 
we use the same method as before to estimate the maximum $z_g$ in a few GRBs.  We use
the data on 9 GRBs from \citet{2006ApJ...648L..93P} and the results are listed in
Table \ref{tab5}.  In this sample, the observed GRB redshifts range from 0.695 to 4.275; $z_{MgII}$ ranges
from 0.472 to 1.695 and $z_{in} \sim z_g$ ranges from 0.014 to 1.136.  
{\it The remarkable result that $z_{in} < 1.25$ is repeated for GRBs.}
It appears too contrived to not conclude a definite connection between quasars and GRBs - 
and we suggest that GRBs are transient events on quasars. 
Recalling the unique structure of a quasar that observations argue for (see Figure \ref{fig12}) 
- the most probable origin of GRBs is in an event on the degenerate matter surface in quasars
which subsequently leads to its brightening at multiple wavelengths in the afterglow.
This is supported by the detection of emission and absorption lines in the afterglow.
We argue that the afterglow is the brightening of the degenerate matter surface
which then illuminates the surrounding emitting and absorbing shells and provides a 
glimpse into the structure of the parent quasar till the afterglow lasts.  
The quasar might be `dark' and hence
not detectable except after an explosive GRB event.  The estimated values of $z_g$ clearly
indicate that matter is arranged inside the ergosphere of the rotating black hole
in the host quasar. 
We note that apparent luminosities of GRBs can be as high as $10^{54}$ erg s$^{-1}$.
We search for possible causes of such an energetic transient event on a quasar.

We refer to the relatively well-understood Galactic phenomena of novae which brighten
by $8-10$  magnitudes in the optical in a very short time and which also emit $\gamma$ rays. 
The most energetic novae are now well-established as being due to a cataclysmic thermonuclear
explosion in accreted matter on the surface of a white dwarf.
Since the proposed structure of a quasar has a degenerate matter surface surrounding
the black hole, we reason that matter will be accreted onto it. 
This matter will get denser and hotter as it accumulates on the degenerate matter
surface and can eventually ignite in a fast explosive
thermonuclear burst which can release $\gamma$ ray photons of energies of a few MeV.  
These low energy $\gamma$ rays can be gravitationally boosted to higher
energies due to the strong gravitational field of the black hole, before they leave the system.
The explosion is soon quenched as observed from the short duration of the burst which gives an
estimate of the size of the emitting region.  We note that
the luminosity of emission from the event horizion of a black hole can be boosted
by a staggering factor of $\sim 10^{27}$ and the black body temperature would appear to be boosted by
a factor of $\sim 4\times10^{13}$! \citep{1983bhwd.book.....S} which can easily explain the large 
luminosities of GRBs.
The huge energy release will also brighten the degenerate matter surface in other
wavebands and give rise to the multi-band afterglow of a GRB and the
line spectrum.  Once the afterglow becomes too faint, the spectrum also disppears.  
\citet{1998A&A...329..906P} reported that the quasar 4C49.29
was located in a 3' radius circle around GRB 960720.  We think many such 
correlations should exist otherwise it would mean that there exist `dark' quasars which
then contribute to dark matter and can only be detected gravitationally or through
GRBs.  Anyway, it is clear that a quasar should be present at the location of most GRBs
unless there exist multiple origin paths to GRBs as has been suggested.  

In summary, based on the observational results and the striking similarity of
the spectra and estimated $z_g$ with quasars,
we suggest that GRBs are energetic thermonuclear explosions on the degenerate matter surface
in a quasar similar to nova explosions on a white dwarf. 

\section{Other related topics}
\subsection{Nuclear redshift-magnitude bands in Coma cluster}
\begin{figure}
\centering
\includegraphics[width=7cm]{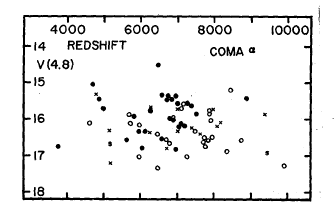}
\caption{\small Redshift-magnitude correlation for member galaxies of Coma cluster.   Filled
circles are ellipticals, open circles are S0s, crosses are SB0 and s are spiral.
Figure copied from \citet{1972ApJ...175..613T}. }
\label{fig14}
\end{figure}

In this section, we discuss how the explanation for the intriguing observational result of galaxies being arranged 
along bands in the nuclear redshift-magnitude diagram in the Coma cluster 
\citep{1972ApJ...175..613T,1973ApJ...179...29T} as shown in Figure \ref{fig14} 
emerges from what has been discussed so far in the paper. 
As Tifft had pointed out, 
there are three contributions to the observed velocity of a Coma galaxy namely orbital
motion in the cluster, Hubble expansion velocity of the cluster and an unknown origin.
\citet{1972ApJ...175..613T} concluded that the
velocity dispersion in the Coma cluster is $<220$ kms$^{-1}$ thus attributing most of the
observed velocity spread in the bands to an intrinsic origin.

We examine these results on Coma galaxies
in light of our result that gravitational redshifts suffered by spectral lines arising in material close to
the event horizon of a black hole can be as large as one. 
Thus the nuclear redshifts of galaxies which host a supermassive black hole
in the centre can be assumed to include a non-zero contribution from $z_g$.  While the
nucleus will be brighter if the degenerate matter surface is closer to the black hole, it will also
be obscured by the dense line forming material - the observed nuclear magnitude will
vary from galaxy to galaxy depending on these two opposing effects. 
The bands in the magnitude-redshift diagram are best discernible when the nuclear
magnitude is plotted against the nuclear redshift and are diluted once the signal from
the entire galaxy is included. 
In light of all this, we can explain the result in Figure \ref{fig14} as follows:

\begin{enumerate}

\item We suggest that the three bands are due to the orbital motion of the galaxies - the highest band
indicates the receding galaxies and the lowest band indicates the approaching galaxies.
The middle band indicates the galaxies which have no radial motion other than Hubble flow
of the cluster towards us.  

\item We suggest that the galaxies are located along a band depending on the
contribution of the gravitational redshift to its nuclear spectrum and obscuration of
its optical continuum from the degenerate matter surface by the line forming gas.  
Thus, the progressive shift of non-elliptical
galaxies to fainter high redshift end would then indicate that both the gravitational redshift
contribution and obscuration of degenerate matter surface is larger in 
non-ellipticals than in elliptical galaxies. 
This is as expected in the sense that non-ellipticals do have larger gas and dust fraction
as compared to ellipticals.  Moreover from the discussion till now, especially the presence
of wide lines in quasars and Seyfert 1 spectra, it appears that
the line forming gas in closest to the black hole in quasars followed by Seyfert 1 and then elliptical
galaxies.  

\end{enumerate}

Interestingly, the above explanation and the arrangement of all galaxies along bands
then indicates that all the galaxies in the Coma cluster host a central accreting black hole. 
Since the Coma cluster is virialised but not expected to be different, the result
can be extrapolated to support the existence of a central supermassive black hole
in all galaxies.

\subsection{Formation of supermassive black holes - primordial?}
Based on the observational results on quasars and active nuclei that are 
discussed in the paper and our inferences,
it appears reasonable to support a primordial formation of supermassive black holes
\citep{1998A&A...331L...1S} around
which matter arranged itself into galaxies.   Thus the supermassive black holes
could have formed first defining the location of galaxies.  It would be interesting
to examine the simple extrapolation that a disk galaxy formed around a rotating 
black hole whereas an elliptical
galaxy formed around a non-rotating black hole.  Quasars with their stellar-like compact
appearance do not seem to be a galaxy.  Instead they appear to consist of
the central supermassive black hole + dense gas shells surrounded by a spherical
halo of gas in which low ionisation lines like Mg II form - more compact than normal galaxies.  
The existence of a supermassive black hole, dense matter shells, 
lack of an extended galaxy but hosting galaxy-like mass make one wonder if quasars are failed
galaxies because all the matter rapidly fell in close to the central black hole during the phase
of galaxy formation.  With mounting evidence that all galaxies host a black hole at the centre including the
explanation of Figure \ref{fig14} presented in the previous section, the primordial origin
appears to be gaining support.  Interestingly,
there seem to be studies which suggest that 20-30 M$_\odot$ black holes
might also be primordial in nature \citep{2016ApJ...823L..25K}. 
While stellar mass black holes can be understood as end products of massive stars, 
it has always been difficult to understand the origin of supermassive non-stellar mass black holes.
Thus, the growing support for a primordial origin in which all galaxies should host a black hole
at its centre appears to be a possible explanation.  The kind of physical processes that can lead
to such large masses being concentrated in such small regions needs to be explored. 

\section{Summary, conclusions, future}
We have examined the origin of the continuum emission and the large redshifts shown by the 
emission and absorption lines in a quasar spectrum in the ultraviolet. 
The study has resulted in several important watertight inferences, all fitting
together like a jigsaw puzzle and uniquely constraining the quasar structure.  
The important results/inferences can be summarised to be: 

\begin{itemize}

\item All the  emission and absorption lines detected
in a quasar spectrum arise from matter inside the quasar system.  The observed redshift 
($z=z_{em}$) of a quasar is large due to contribution from both the cosmological ($z_c$) 
and intrinsic ($z_{in}$)
redshifts. $z_{in}$ comprises of a Doppler component ($z_D$) and a gravitational redshift ($z_g$).  We
show that $z_g$ going upto values of one comprises the dominant component of $z_{in}$. 

\item For quasars, we find $z_c<z_{em}$, $z_{in} < 1.25$ and $z_c\le3$. 

\item  The observations, particularly in the ultraviolet band,  determine a unique 
structure for a quasar.  The proposed quasar structure 
consists of a supermassive black hole surrounded by degenerate matter shells (where the neutron and electron
degeneracy pressure balance the black hole gravity)
emitting thermal continuum peaking in the ultraviolet which heats and ionizes matter around it
giving rise to a Stromgren shell where emission lines arise.  Around the emitting shells are
dense shells from where the absorption lines arise.  The entire structure is arranged just outside the
event horizon of a black hole in quasars (see Figure \ref{fig12}).  
The spectral lines are shifted by different $z_g$ depending on the
separation of the line forming region from the black hole i.e. gravitational potential.  
This structure explains the enhanced ultraviolet continuum of quasars 
and the multiple redshifted spectral lines. 

\item
We marvel at and believe that the existence of maximally gravitationally redshifted lines in 
quasar spectra are one of the most compelling 
proofs of Einstein's general theory of relativity \citep{1915SPAW.......844E, 1915SPAW...47..831E,
1915SPAW.......778E, 1915SPAW.......799E}.   There is no doubt that there exist
black holes with properties determined by the exact analytical solutions derived for
non-rotating black holes by \citet{1916SPAW.......189S} and for
rotating black holes by \citet{1963PhRvL..11..237K}.
Quasars strongly support $z_g$ upto 1 as expected for maximally 
rotating black holes and rule out $z_g >1$.

\item We suggest a method to separate $z_c$ and $z_{in}$ in a quasar spectrum. 
We assume that the lowest detected redshift of the Mg II absorption line $z_{MgII}$ 
in a quasar spectrum includes no contribution from $z_{in}$ and hence $z_c=z_{MgII}$.  
Using this, we find the remarkable result that the difference between the highest ($z_{em}$) and
lowest ($z_{MgII}$) redshifts deduced in a quasar spectrum are $ < 1.25$. This proves the existence
of a non-trivial intrinsic redshift contribution to the velocity of the lines. 

\item Spectral lines from quasars show a $z_g$ upto one. 
$z_g\sim0.5$ can be shown by lines arising in shells close to the Schwarzchild radius while $z_g\sim1$
is possible only from matter inside the ergosphere of a rotating black hole. 
This is the first time, to the best of our knowledge, that
$z_g$ between 0.5 and 1 has been inferred in astronomical spectral lines 
providing irrefutable evidence to
the existence of rotating black holes and matter inside its ergosphere. 

\item The gravitational redshift component can trivially explain several intriguing 
observational results on quasars.

\item We suggest that the peculiar morphology of the ergosphere is taken up 
by matter within as it arranges itself in equipotential shells around the black hole.
The matter inside the ergosphere will resemble a thick accretion 
disk which is often postulated to exist around active nuclei.

\item The emission and absorption lines detected
in a quasar spectrum are broadened due to the varying gravitational potential in
the line forming region and hence cannot be used to estimate the mass of the black hole.

\item We find that $z_g <0.5 $ for most BL Lac objects 
whereas $z_g$ upto one is shown by spectral lines from FSRQs - both comprising blazars. 
Thus blazars are indeed quasars with the structure shown in Figure \ref{fig12}. 

\item We show that GRBs are transient events in quasars consisting of 
$\gamma$ ray photons generated
in an explosive thermonuclear reaction on the hot degenerate matter surface of the quasar.  This explosion
illuminates the degenerate matter surface giving rise to the multiband afterglow emission
and a glimpse of the line spectrum of the quasar. 
It needs to be investigated whether this formation scenario explains all observed GRBs.

\item We suggest the observed variability in quasars, especially in the ultraviolet continuum emission,
can be unequivocally associated with energetic events on the degenerate matter surface -
GRBs being the most energetic.  Thus, the variability timescales alongwith
the decoded shell structure from the observed gravitational redshifted lines 
should be able to uniquely constrain the quasar - 
black hole mass and the physical properties of the surrounding degenerate matter and line formingshells.

\item The quasar model is applicable to other active nuclei 
with the variables being the separation between the black hole, degenerate matter surface and
the line forming zones.  

\item We explain the band structure in the nuclear redshift-magnitude diagram
shown by galaxies in the Coma cluster 
as being due to the effect of a gravitational redshift component and obscuration. 
This result strongly supports the presence of a supermassive black hole 
at the centres of all galaxies. 

\item Now that we have undeniable proof of the existence of matter inside the ergosphere 
of rotating black holes in quasars, 
it should be possible to further investigate how the energy of the black hole
is tapped \citep[e.g.][]{1969NCimR...1..252P,1977MNRAS.179..433B}.

\item We now revisit the four questions posed in Section 2 and answer them based
on the results: (1) the spectral lines
arise inside the quasar; (2) quasars exist in the same volume as other active galaxies;
(3) the ultraviolet continuum arises on the degenerate matter surface and a gravitational
instability on the same leads to the some of the observed variability; (4) quasars are isolated black holes
surrounded by matter with comparable masses but smaller physical sizes than galaxies. 

\item In light of the conclusive results on the presence of an intrinsic component in the quasar
redshifts presented here, the suggestion that nearby galaxies and quasars/active nuclei 
are related \citep{1967ApJ...148..321A,1974IAUS...58..199A} needs to be revisited.

\item Now that we know that observed redshifts of quasars
contain a sizeable redshift of non-cosmological origin, 
the existence and explanation of superluminal motions need to be revisited.          
For example, \citet{1971ApJ...170..207C} inferred that while the quasars 3C273 and 3C279 showed superluminal
expansion, the jets in the Seyfert galaxy NGC 1275 show non-superluminal expansion and in M87 show no
expansion.  This kind of gradation, if found to be widespread, could be due to
the wrong cosmological redshifts (and hence distance) which have been used for quasars (and to a lesser extent other active
nuclei) and needs to be examined.

%\item Quasars are likely to be strong emitters of gravitational waves.

\item The existence and implications of `dark' quasars should be examined. 

\item Possibility of a gravitational redshift contribution to the Hubble constant should be examined. 

\item This research then gives us new questions to ponder on - for example: How are
metals synthesized in quasars ?  What determines the separation between the
event horizon and the line forming zones ?  Where does the synchrotron emission arise
and is it due to a shock set up by the instability on the degenerate matter surface  ? 
What part of the structure of a quasar is common to the structure of any accreting black hole? 
Can we observe gravitationally redshifted lines from stellar mass black holes? 

\end{itemize}

\section*{Acknowledgements}
I am grateful to all the researchers who have taken meticulous observations
and drawn inferences from these data.  Many research papers are cited here but many more have been
referred to.  I acknowledge generous use of ADS abstracts, arXiv, gnuplot, Wikipedia, LaTeX and
Google search engines in this research.

%\bibliography{agn}
\bibliography{agn1.bib}

\onecolumn

\begin{center}

{\LARGE ERRATA \& COMMENTS on\\Decoding quasars: gravitationally redshifted spectral lines!  \\}

\vspace{0.5cm}

{\large Nimisha G. Kantharia \\
National Centre for Radio Astrophysics, \\ Tata Institute of Fundamental
Research, \\ Post Bag 3, Ganeshkhind, Pune-411007 \\ {\it Email: nkprasadnetra@gmail.com \\
URL: https://sites.google.com/view/ngkresearch/home \\} }

\vspace{0.2cm}
\date{\bf December 2017}
\end{center}

This appended document lists and corrects errors in the paper and includes comments relevant to the paper.  The
numbering of figures,tables and equations is independent of the paper.  The few references which
are referred to in this document are included in the reference list of the paper.  Further errors if and
when identified will be updated at https://sites.google.com/view/ngkresearch/home.

\setcounter{table}{0}

\begin{enumerate}

\item The wavelength part of Equations 4, 6 in the paper are wrong and these are corrected here.
The frequency formula is correct.
The correct form for Equation 4 is the following:

\setcounter{equation}{0}

\noindent
$ \rm \frac{\nu_0-\nu}{\nu_0} = \frac{GM}{Rc^2} ~~ or ~~ \frac{\lambda - \lambda_0}{\lambda} = \frac{GM}{Rc^2} $

\noindent
and the correct form for Equation 6 is: \\
$\rm z_g(R) = \frac{\lambda_{obs} - \lambda_{rest}}{\lambda_{obs}} = \frac{GM}{Rc^2} = \frac{v}{c}$

This, leads to the recognition of a major update that is required in the paper which is elaborated in
point 5 in this document.
{\it (September 2017)}

%Since the wavelength formula is not used anywhere else in the paper this change does not
%modify the results presented in the paper.
% Above two lines were included in the errata put out on my google website in Sep 2017.  I have commented
% November 2017 when point 5 has been included since they sound like famous last words which were so wrong!

\item Figure 6 in the paper gives a schematic structure of the event horizon, ergosphere and ergosurface
around a rotating black hole as is conventionally shown in literature.  However it fails
to represent the following important points for a rotating black hole: (1) The event horizon
of a maximally rotating black hole is prolate-shaped with its semi-major axis which is along the polar
axis being equal to the Schwarzchild radius $R_s$ and its semi-minor axis
which is in the equatorial plane being equal to half the
Schwarzchild radius.  For black holes with lower rotation speeds, the
semi-minor axis of the event horizon will lie between $R_s$ and $R_s / 2$.
(2) The ergosphere of a rotating black hole
is the region between the prolate-shaped event horizon and the pseudo-surface of a sphere of
radius $R_s$ known as the ergosurface.
The corrected schematic is shown in Figure \ref{BH1}.  We also include an extra schematic for
a non-rotating black hole in the figure.
{\it (September 2017)}

\item Figure 12 in the paper showing the proposed structure of a quasar
is modified to include the correct structure of a rotating black hole as
depicted in Figure 6.  The corrected figure is shown in Figure \ref{BH2}. There is no change
in the detailed structure of a quasar i.e the degenerate surfaces and line forming regions
(when a gravitational redshift $>0.5$ is inferred) continue to
be located inside the ergosphere and are shown.  The line-forming matter located beyond $R_s$
from the black hole will show a gravitational redshift $> 0.5$.
%The main change is in the structure of the event horizon and ergosphere for a rotating black hole
%so that the former is a prolate-shaped ellipse and the latter is a sphere.
{\it (September 2017)}

\setcounter{figure}{0}

\begin{figure}
\centering
\includegraphics[width=8.5cm]{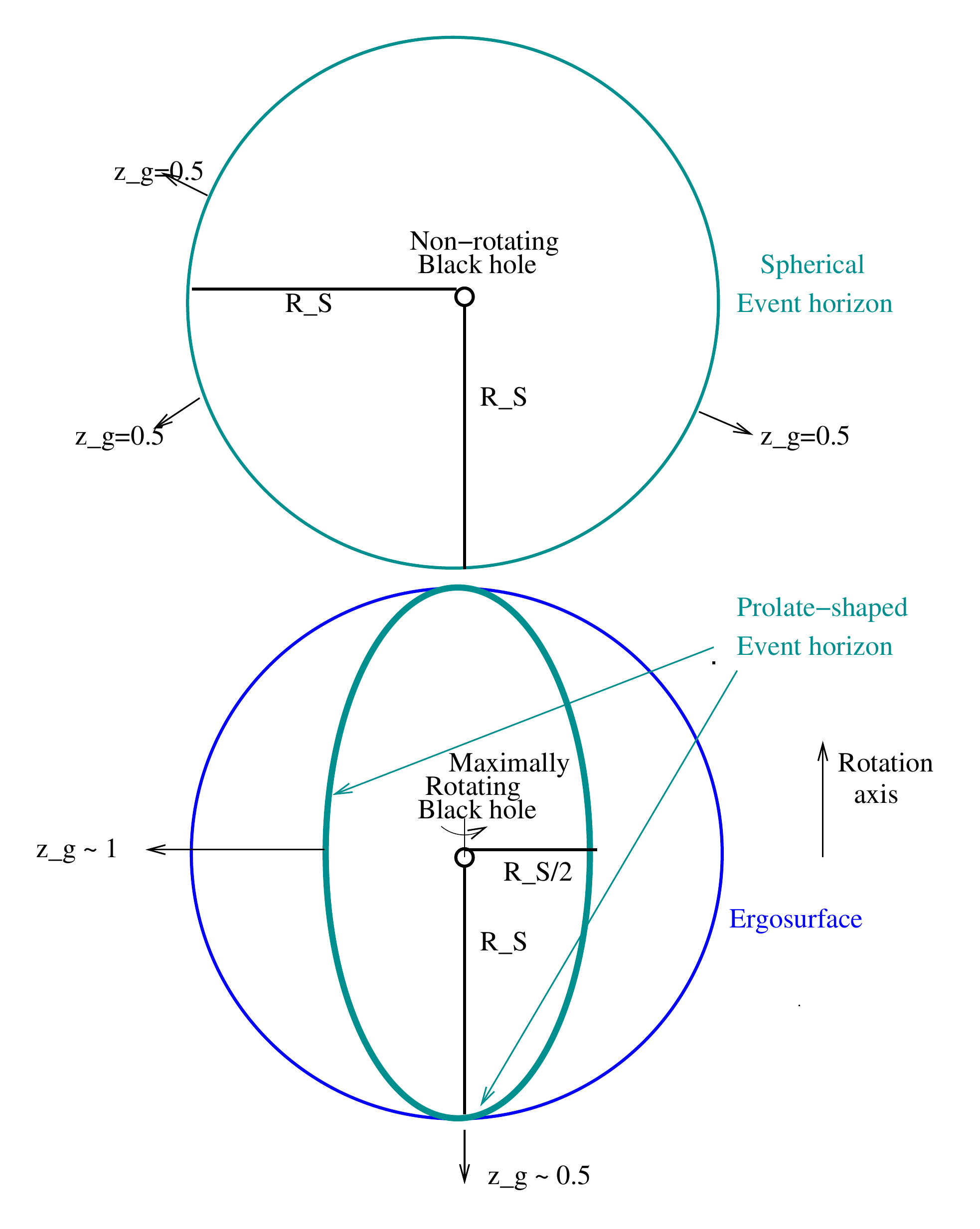}
\caption{ {\bf Figure 6 in the paper should be replaced by this figure.}
$R_s$ denotes Schwarzchild radius, $z_g$ denotes gravitational redshift.  The
edits in the schematic is in the shape of the event horizon and ergosphere and inclusion
of a schematic for a non-rotating black hole.  Rotation of the
black hole leads to the event horizon moving inwards in the non-polar regions and the
appearance of a
region known as the ergosphere which becomes visible.  For a non-rotating black hole,
the maximum $z_g < 0.5$ while for a rotating black hole, the maximum $z_g < 1$. }
\label{BH1}
\end{figure}

\begin{figure}
\centering
\includegraphics[width=11cm]{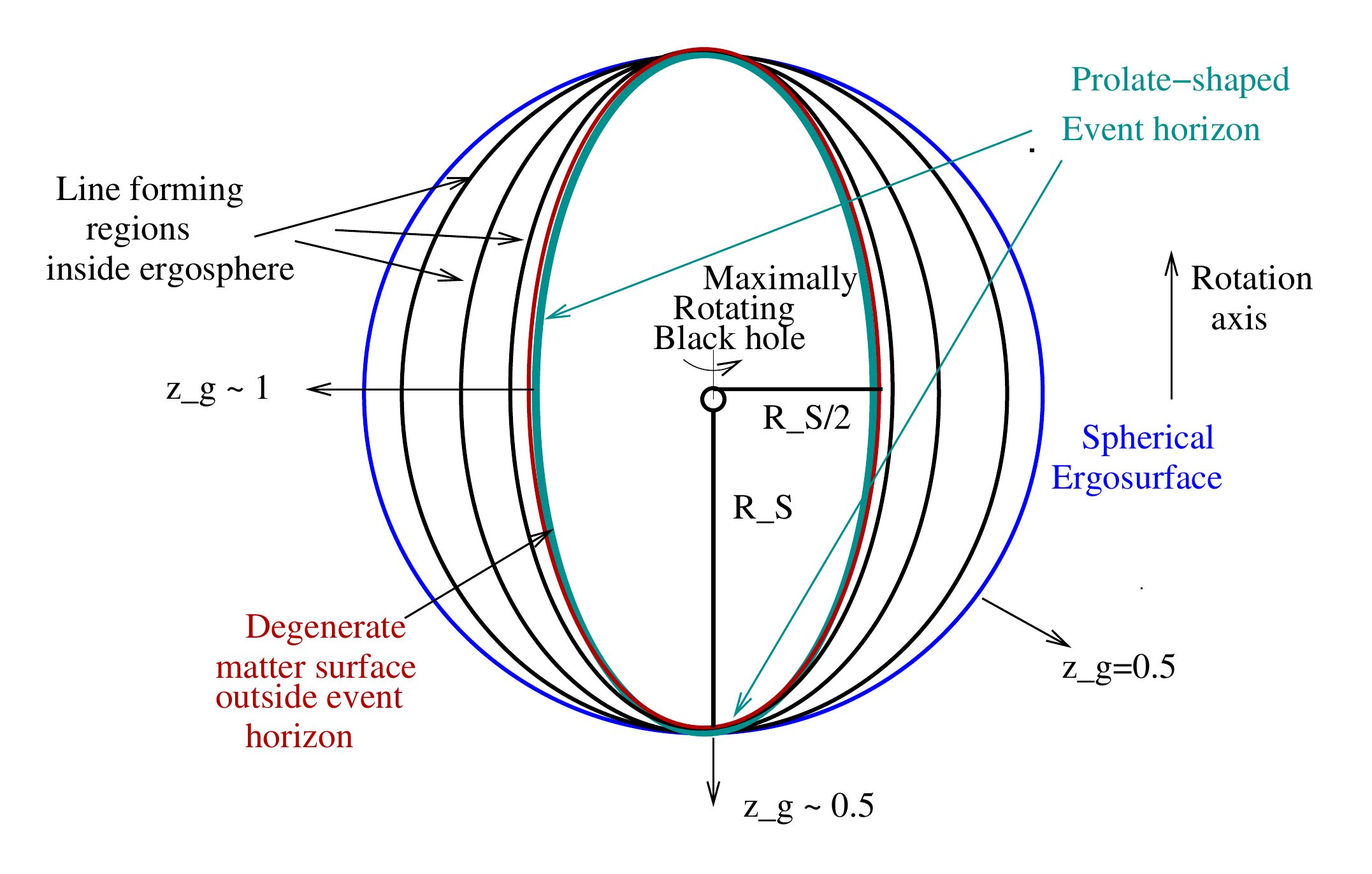}
\caption{{\bf Figure 12 in the paper should be replaced by this figure.}
$R_s$ denotes Schwarzchild radius, $z_g$ denotes gravitational redshift.  The edit
is in the shape of the event horizon (prolate-shaped ellipse) and ergosphere (spherical)
of the maximally rotating black hole.  The maximum gravitational redshift that
lines forming just outside the event horizon at the equator can suffer is
$< 1$ whereas those forming outside the event horizon at the poles can suffer
$z_g < 0.5$ in a rotating black hole.  The lines forming in matter outside the pseudo-surface of
radius $R_s$ in both types of black holes will show a shift $z_g < 0.5$.}
\label{BH2}
\end{figure}
\item
Alongwith the updated schematics, we include a few points here which we can surmise
from observations and black hole physics since there seems to prevail some confusion in literature:
\begin{itemize}
\item The event horizon is the fictitious surface around the black hole which defines
a surface of no-return for both light and matter.  As shown in the paper, the quasar
spectrum arises from matter located just outside this surface.  Observations support the
prolate-shaped structure for the event horizon in rotating black holes so that its radius varies from
$R_s/2$ to $R_s$ at the equator (depending on its rotation speed)  but is always $R_s$ at the poles.
\item  The event horizon is a pseudo-surface of a sphere of
radius $R_s$ around a non-rotating black hole while it is a prolate-shaped
pseudo-surface which is located inside the sphere of radius $R_s$
in a rotating black hole.   In a rotating black hole, the pseudo-surface of radius $R_s$ around the
black hole is referred to as the erogsurface.  The region between the event horizon and the ergosurface in
a rotating black hole is the ergosphere.  In a non-rotating black hole, the event
horizon and ergosurface are the same and there is no ergosphere.  The escape velocity
from the event horizon in a non-rotating black hole and ergosurface in a rotating black hole
is equal to velocity of light.  The escape velocity from inside the ergosphere has to be
$\ge$ velocity of light.
As quasar spectra demonstrate, electromagnetic radiation can escape
from within the ergosphere and indicates that the escape velocity continues
to be equal to velocity of light inside the ergosphere also.  We have no evidence of matter escaping
from the ergosphere and no matter should be able to escape from the ergosphere since
the escape velocity from the ergosphere is equal to the velocity of light.  So matter
once trapped inside the ergosphere can never escape but can only fall into the black hole.
This can also be understood as follows:
$ 1/2~ m  v_{esc}^2 = G m M / R^2$ so that $v_{esc} = \sqrt{(2GM/R)} $.
Schwarzchild radius is $R_s = 2 G M/c^2$ so that at this distance (event horizon or ergosurface),
$v_{esc}  = c$ (Schwarzchild radius is defined by this criterion).
We note that $z_g = GM/Rc^2 = R_s/2R$ so that the multi-redshifted spectral lines
in a quasar spectrum which show a
gravitational redshift $z_g\le 0.5$ will arise from the pseudo-surface of radius
$R_s$ or beyond it while if $z_g > 0.5$, then the lines have to arise from matter located within $R_s$
i.e. within the ergosphere.  Since we are able to observe these spectral lines means
that light is able to escape from the ergosphere and
indicates that the escape velocity continues to be $c$ within the ergosphere. \\
\item One often encounters mention of an inner and outer event horizon in literature.
It appears that these refer to the event horizon and the ergosurface in a rotating black hole.
The terminology might have come into practise since both will be
pseudo-surfaces of equal effective potential and different gravitational potential.
One can think of the ergosphere as
a `stretching' of the event horizon in rotating black holes which allows us an
electromagnetic glimpse into
a region around the black hole which would not have been possible in a non-rotating black hole.   Thus, while
observations unequivocally prove the existence of rotating black holes (from spectral lines
which show $z_g \ge 0.5$), it appears more difficult
to prove the existence of non-rotating black holes or rather disentangling the observational
signatures of non-rotating black holes from rotating black holes since the lines with
$z_g<0.5$ can arise from matter around both types of black holes.
Since most astrophysical systems show rotation, it might be that all black holes
are also rotating.  However it needs to be investigated using observational results.

\end{itemize}
{\it (September 2017)}

\item Reflecting more on the implications of the different wavelength-redshift conversion formula
for estimating cosmological and gravitational redshifts, it is found that a major update to the
paper is required.  This edit does not change the interpretation in that the results
continue to support the intrinsic origin of the entire quasar spectrum
as shown in the paper and in fact strengthens the inferences, nevertheless
this update is very critical and crucial to the correctness of the paper.  While it is still
advocated that the intrinsic redshift is a measure of the gravitational redshift, it is realised
that they cannot be equal.  A conversion
factor needs to be applied to the intrinsic redshift to convert it into
gravitational redshift.  This conversion has not been included in the paper
and is the fault that this update aims to correct.

The quasar redshift ($z$ or $z_{em}$) is estimated from the observed and rest wavelengths of the emission
line in the quasar spectrum as follows:
\begin{equation}
\rm
z = \frac{\lambda_{obs} - \lambda_{rest}}{\lambda_{rest}}
\label{eqn1}
\end{equation}
Equation 1 of the paper which was ($\rm (1 + z) = (1 + z_c) (1 + z_{in})$) indicates that
in the general case, the observed redshift of the emission line $z$ will be a combination of an
intrinsic redshift $z_{in}$ and an extrinsic i.e. cosmological redshift $z_c$.  The same will
be true for the absorption lines which are observed in the quasar spectrum.  While the
cosmological redshift will be the same for all the spectral lines which arise within the
quasar, it will differ if the spectral lines arise at different locations along the sightline to the quasar.
The spectral lines which arise within the quasar can suffer different
intrinsic redshifts due to some physical process in the quasar but will all have the same
cosmological redshift.  In the paper, this second scenario in which the observed redshifts contained
a constant contribution from cosmological redshift and a varying component due to an intrinsic process
was examined in detail.  A method for separating the two components was suggested and it was shown that
multiple redshifts of lines in the spectrum to a quasar were expected due to
the effect of a varying gravitational potential if the lines arose close to but
at different radial separations from the event horizon of the supermassive black hole in the quasar.
Thus, photons arising in the radially varying strong gravitational potential of the black hole and
hence gravitational redshifted was shown to be the intrinsic process which could
explain the observed multi-redshifted quasar spectrum and was a very plausible alternative to the
intervening matter scenario.   However as mentioned above, a mistake was made in the paper in assuming that
the gravitational redshift was the same as the intrinsic redshift estimated from $z$ and $z_c$.  We explain
this below and provide the corrected form for estimating gravitational redshifts.
The gravitational redshifts thus found for the emission/absorption lines of quasars
listed in tables in the paper are included here.  Two figures are also accordingly updated.
The y-axis label of Figure 8 in the paper should be changed to `Intrinsic redshift $z_{in}$'.

While both $z_c$ and $z_{in}$ are related to the shifted wavelength due to the respective
process i.e. $\lambda_{obs,c}$ and $\lambda_{obs,in}$ as given in Equation \ref{eqn1},
the gravitational redshift $z_g$ of a spectral line can be estimated from the wavelengths as:
\begin{equation}
\rm
z_g = \frac{\lambda_{obs,g} - \lambda_{rest}}{\lambda_{obs,g}}
\label{eqn2}
\end{equation}
where the subscript $g$ refers to the wavelength shift arising due to gravitational redshift.
The intrinsic redshift of the spectral lines in a quasar spectrum are shown
to be caused by the physical process which is responsible for gravitational redshifts and hence
$\lambda_{obs,g}=\lambda_{obs,in}$ (other possible intrinsic processes such as Doppler expansion
are ignored) i.e. the wavelength shift of a line due to intrinsic redshift and gravitational redshift
is the same.  However since the denominators in Equations \ref{eqn1} and \ref{eqn2} are different;
$z_{in} \ne z_g$.
{\bf So the major change required in the paper is to replace the heading of $z_g$ in the tables and
plotted quantity of $z_g$ in figures by $z_{in}$.   $z_{in}$ is a measure of $z_g$ but
$z_{in} \neq z_g$.   An extra factor is required for converting $z_{in}$ to $z_g$.}

\begin{figure}
\centering
\includegraphics[width=8cm]{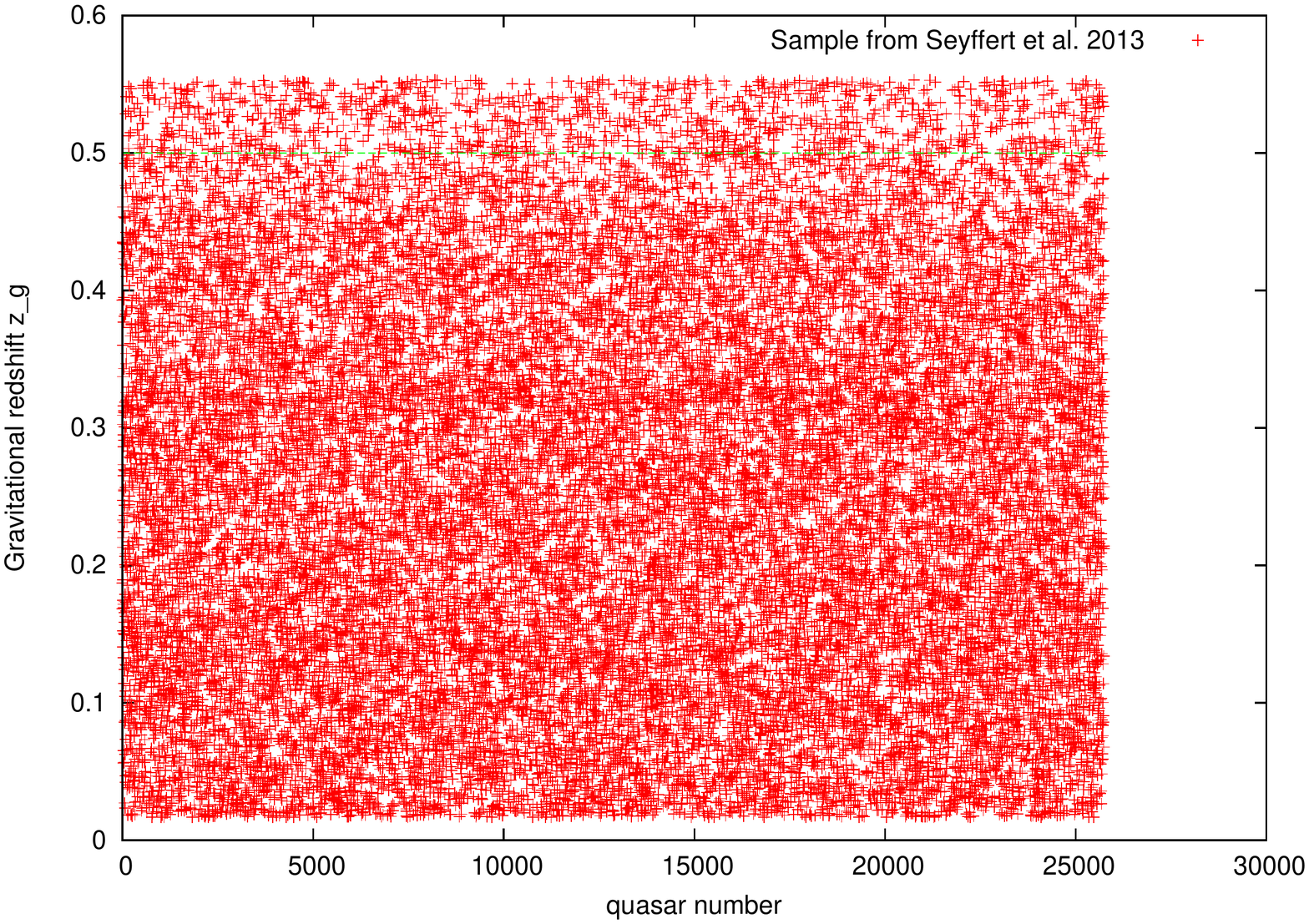}
\caption{The largest gravitational redshift detected in a quasar spectrum estimated using the above procedure
for the sample of absorption line quasars from the
\citet{2013ApJ...779..161S} sample.  The horizontal line is drawn at $z_g=0.5$.  The
emission lines which show $z_g$ above this line form in matter inside the ergosphere.  The median
$z_g=0.25$.}
\label{zg1}
\end{figure}

\begin{figure}[h]
\centering
\includegraphics[width=8cm]{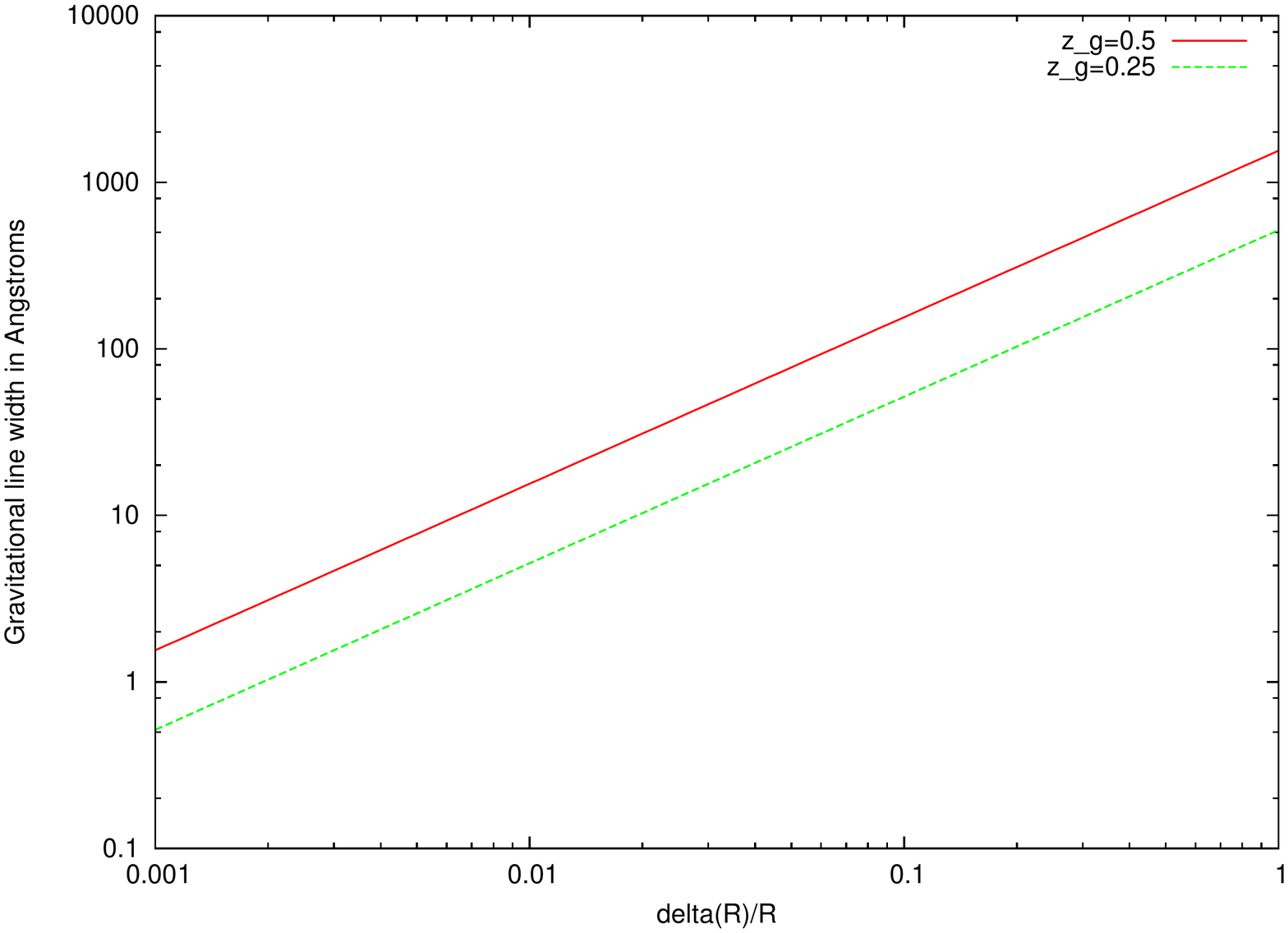}(a)
\includegraphics[width=8cm]{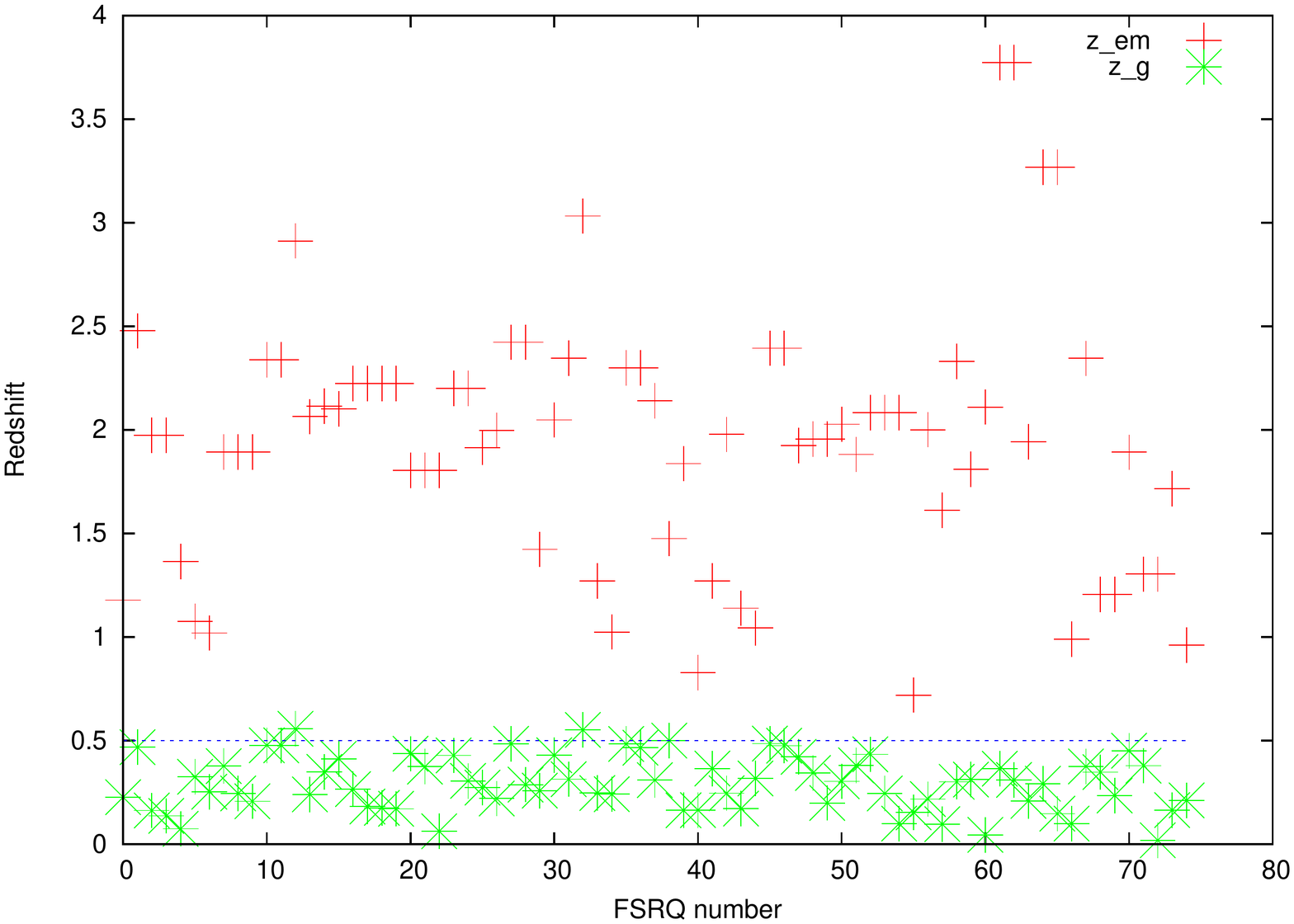}(b)
\caption{(a) Modified Figure 9 in the paper after using the correct equation for $z_g$ to
estimate the observed wavelength of the redshifted line of C IV 1548 A.  The plots are shown for
$z_g=0.5$ and $z_g=0.25$.
(b) An additional panel for Figure 13, is presented here.  In the paper, Figure 13 plotted the intrinsic
redshift distribution for a sample of flat spectrum radio quasars whereas here the
gravitational redshift distribution the same sample is shown.  The horizontal line is drawn at 0.5.}
\label{fsrq}
\end{figure}

To recall, we have determined the $z_{in}$ of the emission line from the difference between the quasar
redshift (emission line redshift i.e.
highest redshift in the quasar spectrum, $z$) and the lowest redshift at which Mg II absorption doublet
lines are detected (i.e. lowest redshift in the quasar
spectrum, $z_c$) by assuming that the lowest redshift in the quasar spectrum is the true
indicator of the distance to the quasar i.e. is the cosmological redshift.
%and the difference
%redshift is intrinsic to the quasar and is caused by the photon being emitted in the strong gravitational
%field of the black hole and hence redshifted as expected in the general theory of relativity.
The estimated $z_{in}$ is related to the observed and rest wavelength
as in Equation \ref{eqn1} and hence we can express $\lambda_{obs,in}$
in terms of $\lambda_{rest}$ as follows:
\begin{equation}
\rm
\lambda_{obs,in} =  (z_{in} + 1) \lambda_{rest}
\label{eqn3}
\end{equation}
Since $\lambda_{obs,in} = \lambda_{obs,g}$, we can substitute for $\lambda_{obs,g}$ in Equation \ref{eqn2}
from Equation \ref{eqn3}, so
that $z_g$ is expressed in terms of $z_{in}$ which we have already estimated:
\begin{equation}
\bf
z_g = \rm \frac{(z_{in} + 1)\lambda_{rest} - \lambda_{rest} }{(z_{in} + 1)\lambda_{rest}} =
\bf \frac{z_{in}}{(z_{in}+1)}
\label{zg}
\end{equation}

{\bf The gravitational redshift $z_g$ can be determined by dividing $z_{in}$ by the factor $z_{in} + 1$.}

The gravitational redshifts contributing to the quasar redshift has been estimated as above for
the sample of quasars from \citet{2013ApJ...779..161S}
that has been used in the paper and the resulting distribution of gravitational redshifts of the
emission lines is shown in Figure \ref{zg1}.  The median value of $z_g$ is 0.25.
%With this major edit regarding the conversion factor from $z_{in}$
%to $z_g$ which was not done in the paper,  the most important change required in the paper
%is to replace all labels which include $z_g$ in tables and figures with $z_{in}$.
%All the redshifts estimated in the paper use the form of Equation \ref{eqn1} and hence are intrinsic redshifts.
In Table \ref{tab1}, we include the emission line
gravitational redshifts for the quasars listed in Tables 1,3,4,5 in the paper.
This change should also lead to edits in the text of the paper but for now it is considered crucial to
point out this update and update the tables especially since the main inferences of the paper remain
unchanged.

\begin{table*}[t]
\centering
%\scriptsize
%\footnotesize
\small
\caption{Tables 1,3,4,5 in the paper list what was thought to be $z_g$ but as pointed out here, the listed
values are of $z_{in}$.  So the heading in the tables in the paper should be exclusively `$z_{in}$'.  The gravitational
redshifts $z_g$ estimated from $z_{in}$ using Equation \ref{zg} are listed here for the same objects listed in the tables.
$z_{in}$ which is already tabulated in the tables in the paper is also included.   }
\begin{tabular}{l|c|c|l|c|c|l|c|c|l|c|c}
\hline
\multicolumn{3}{c|}{\bf Table 1} & \multicolumn{3}{c|}{\bf Table 3} & \multicolumn{3}{c|}{\bf Table 4} &
\multicolumn{3}{c}{\bf Table 5}\\
\hline
{\bf Quasar} & $\bf z_{in}$ & $\bf z_g$ & {\bf Quasar} & $\bf z_{in}$ & $\bf z_g$  & {\bf BL Lac} & $\bf z_{in}$ &
$\bf z_g$ & {\bf GRB} & $\bf z_{in}$ &  $\bf z_g$ \\
\hline
Q0013-004 & 1.133 & 0.531   &   Q048+163(em)&  0.844 & 0.457  & 0100-337 & 0.115 & 0.103     &  010222 &  0.285&  0.221 \\
Q0014+818 & 1.073 & 0.517   &   Q048+163 & 0.5493 & 0.354     & 0238+1636&  0.272 & 0.213    &  020405 & 0.151 & 0.131\\
Q0058+019 & 0.835 & 0.455   &   Q048+163 & 0.5562 & 0.357     & 0241+0043 & 0.219 & 0.179    &  020813 & 0.014 & 0.0138 \\
Q0119-046 & 0.772 & 0.435   &   Q048+163 & 0.5574 & 0.357     & 0334-4008 & 0.130 & 0.115    &  021004 & 0.398 & 0.284 \\
Q0150-203 & 1.265 & 0.558   &   Q048+163 & 0.8383 & 0.456     & 0423-0120 & 0.172 & 0.146    &  050505 & 0.957 & 0.489 \\
Q0207-003 & 0.883 & 0.468   &   Q0014+818(em) & 1.073 & 0.517 & 0428-3756 & 0.353 & 0.260    &  050820 & 1.136 & 0.531 \\
Q0229+131 & 1.235 & 0.552   &   Q0014+818 & 0.6548 & 0.395    & 0457-2324 & 0.057 & 0.0539   &  050908 & 0.707 & 0.414 \\
Q0348+061 & 1.186 & 0.542   &   Q0014+818 & 0.7992 & 0.444    & 0538-4405 & 0.130 & 0.115    &  051111 & 0.164 & 0.140 \\
Q0440-168 & 0.833 & 0.454   &   Q0014+818 & 0.8004 & 0.444    & 0745-0044 & 0.109 & 0.0982   &  060418 & 0.553 & 0.356 \\
Q0450-132 & 1.177 & 0.540   &   Q0014+818 & 1.0022 & 0.500    & 0909+0121 & 0.316 & 0.240    &  030323 & 0.814 & 0.448 \\
Q0528-250 & 0.937 & 0.483   &   Q0837+109(em) & 0.7561 & 0.430& 0942-0047 & 0.299 & 0.230    & &  \\
Q0837+109 & 0.756 & 0.430   &   Q0837+109 & 0.3869 & 0.278    & 0948+0839 & 0.199 & 0.165     && \\
Q0848+163 & 0.844 & 0.457   &   Q0837+109 & 0.6058 & 0.377    & 1147-3812 & 0.49 & 0.328     && \\
Q0852+197 & 1.276 & 0.560   &   Q0837+109 & 0.6817 & 0.405    & 1408-0752 & 0.099 & 0.0900    &  & \\
Q0958+551 & 1.216 & 0.548   &   54452-2824-554(em) & 0.1531 & 0.132 &   1410+0203 & 0.067 & 0.0627    & & \\
Q1222+228 & 0.822 & 0.451   &   54452-2824-554 & 0.0011 & 0.001  &   1427-4206 & 0.234 & 0.189    & & \\
Q1329+412 & 0.955 & 0.488   &   54452-2824-554 & 0.02 & 0.0196& 1522-2730 & 0.004 & 0.0039  & & \\
Q1331+170 & 0.768 & 0.434   &   54452-2824-554 & 0.0419 & 0.040  &      1743-0350 & 0.639 & 0.389   & & \\
Q1517+239 & 0.667 & 0.400   &   54452-2824-554 & 0.0965 & 0.0883 &      1956-3225 & 0.381 & 0.275    & & \\
Q1548+093 & 1.118 & 0.527   &   52178-0702-503(em) & 1.0163 & 0.504  &  2031+1219 & 0.047 & 0.0442    & & \\
Q1623+269 & 0.868 & 0.464   &   52178-0702-503 & 0.7305 & 0.422      &  2134-0153 & 0.017 & 0.016  & & \\
Q1715+535 & 1.142 & 0.533   &   52178-0702-503 & 0.7386 & 0.424      &  2225-0457 & 0.302 & 0.231    & & \\
Q2206-199 & 1.031 & 0.507   &   52178-0702-503 & 0.7629 & 0.432      &  0221+3556 & 0.154 & 0.133    & & \\
Q2342+089 & 1.196 & 0.544   &   52178-0702-503 & 0.9220 & 0.479     & &     & & & \\
Q2343+125 & 1.03 & 0.507    &   52618-1059-146(em) & 0.6783 & 0.404 &&    & & & \\
Q2344+125 & 0.839 & 0.456   &   52618-1059-146 & 0.4402 & 0.305 &&    & & & \\
Q2145+067 & 0.112 & 0.100   &   52618-1059-146 & 0.4769 & 0.322 &&    & &  &\\
          &       &            &   52618-1059-146 & 0.5432 & 0.351 &&    & & & \\
\hline
\end{tabular}
\label{tab1}
\end{table*}

Another edit is related to Equation 10 in the paper which implicitly uses $z_g$.
Figure 9 in the paper plots the gravitational linewidth for different $z_g$ and the expected fractional
thickness of the line forming zone.  The corrected figure is shown in
Figure \ref{fsrq}(a) for gravitational redshifts of 0.5 and 0.25 and should replace Figure 9 in the paper.
Since the shifted wavelength of a line, given a gravitational redshift can be estimated as
$\rm \lambda_{obs,g} = \frac{\lambda_{rest}}{(1-z_g)}$,
the displaced wavelength of the 1548 A line will be 3096 A for $z_g=0.5$ and will
be 2064 A for $z_g=0.25$.

Figure 13 in the paper shows the distribution of the redshift of the flat spectrum radio quasars
(FSRQ) and the estimated intrinsic redshift.  We show the
distribution of gravitational redshifts estimated from $z_{in}$ as detailed above in Figure \ref{fsrq}(b) here.
{\it (November 2017)}

\item
Alongwith the update in point 5, we also show a few results derived from datasets listing
Mg II absorption redshifts detected towards quasars and which also include
higher redshift quasars than were discussed in the paper.
We find that this analysis also supports a non-trivial contribution to the observed redshift
of spectral lines in quasar spectra by an intrinsic process in the quasar.
The sample of the Mg II absorption line redshifts in the range 0.35 to 2.3
in \citet{2013ApJ...779..161S,2016MNRAS.463.2640R} and the samples of quasars
which list Mg II absorption in the redshift range of 2 to 7
from \citet{2012ApJ...761..112M,2016arXiv161202829C} were analysed.
The results are shown in Figures \ref{zg4},\ref{zg3} and the estimated $z_g$ for the high redshift
quasars ($z > 3.5$) from \citet{2012ApJ...761..112M,2016arXiv161202829C} are listed in Table \ref{tab2}.
The gravitational redshift ($z_g$) component in the emission line redshift ($z=z_{em}$) is estimated using the
procedure suggested in the paper and also includes the last step specified in Equation \ref{eqn4} in point 5.
The important inputs to the procedure are emission line redshift of the quasar $z$ and $z_c$
which is the lowest redshift at which Mg II absorption is detected in a quasar spectrum -
the difference redshift is $z_{in}$ which is a measure of $z_g$.
Since the emission line shows the largest redshift in a quasar spectrum, this method estimates the
largest $z_g$ in the spectrum.  If the absorption line redshifts are used instead of $z_{em}$, then
the gravitational redshift component in those redshifts can be estimated in the same way.

\begin{figure}
\centering
\includegraphics[width=8cm]{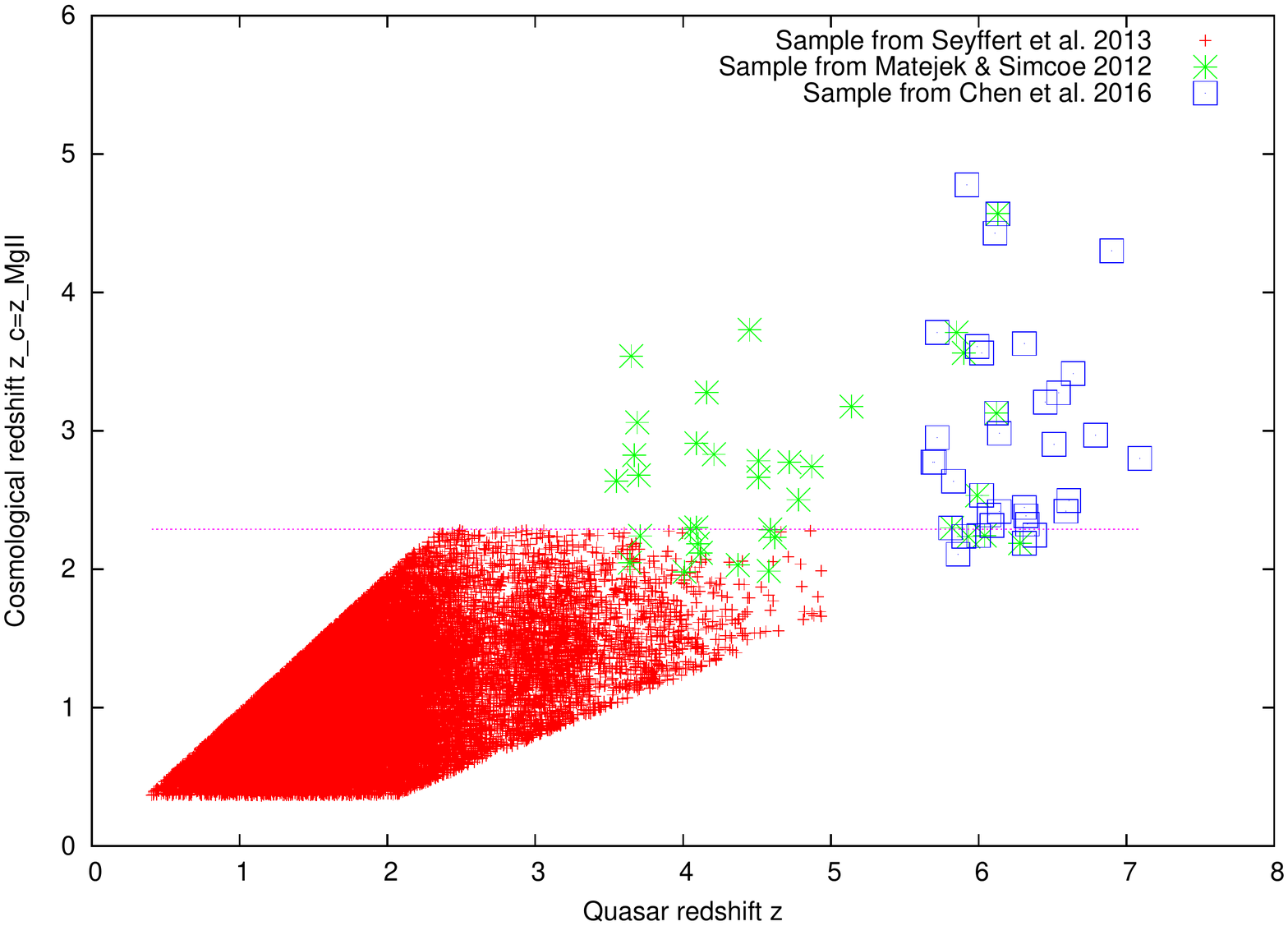}(a)
\includegraphics[width=8cm]{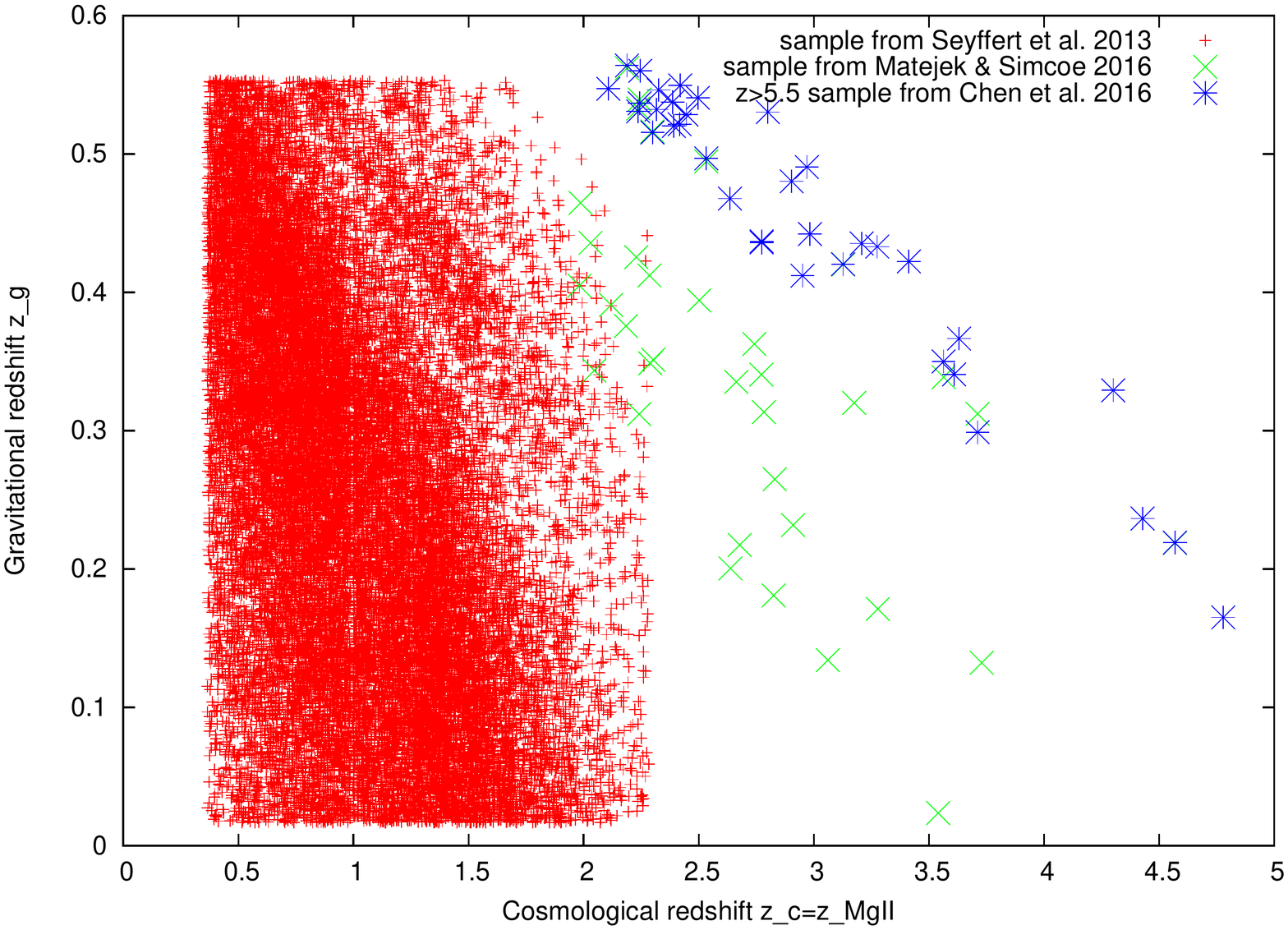}
\caption{(a) The lowest redshift of the Mg II absorption feature in the quasar spectrum taken from
\citet{2013ApJ...779..161S}
(red cross), \citet{2012ApJ...761..112M} (green star) and for the $z>5.5$ sample of quasars from
\citet{2016arXiv161202829C} (blue box).  The horizontal line is drawn at a redshift of 2.29.
(b) Figure shows the distribution of $z_g$ estimated using Equation \ref{eqn2}
against $z_c=z_{MgII}$ for the sample of quasars
in \citet{2013ApJ...779..161S, 2012ApJ...761..112M, 2016arXiv161202829C}.  Note that the quasars
with redshift $<1.5$ from the sample in \citet{2013ApJ...779..161S} show the entire range of $z_g$
(from 0.0166 to 0.5535) and the largest gravitational redshift observed appears to decrease as $z_c$ increases
for $z_c > 1.5$.  However quasars with high $z_c$ from \citet{2016arXiv161202829C} does show large
values of $z_g$ indicating decreasing $z_g$ with $z_c$ might also a result of observational bias
since the number of high redshift quasars and high $z_{MgII}$ searches are limited.  }
\label{zg4}
\end{figure}

\begin{figure}[h]
\includegraphics[width=8cm]{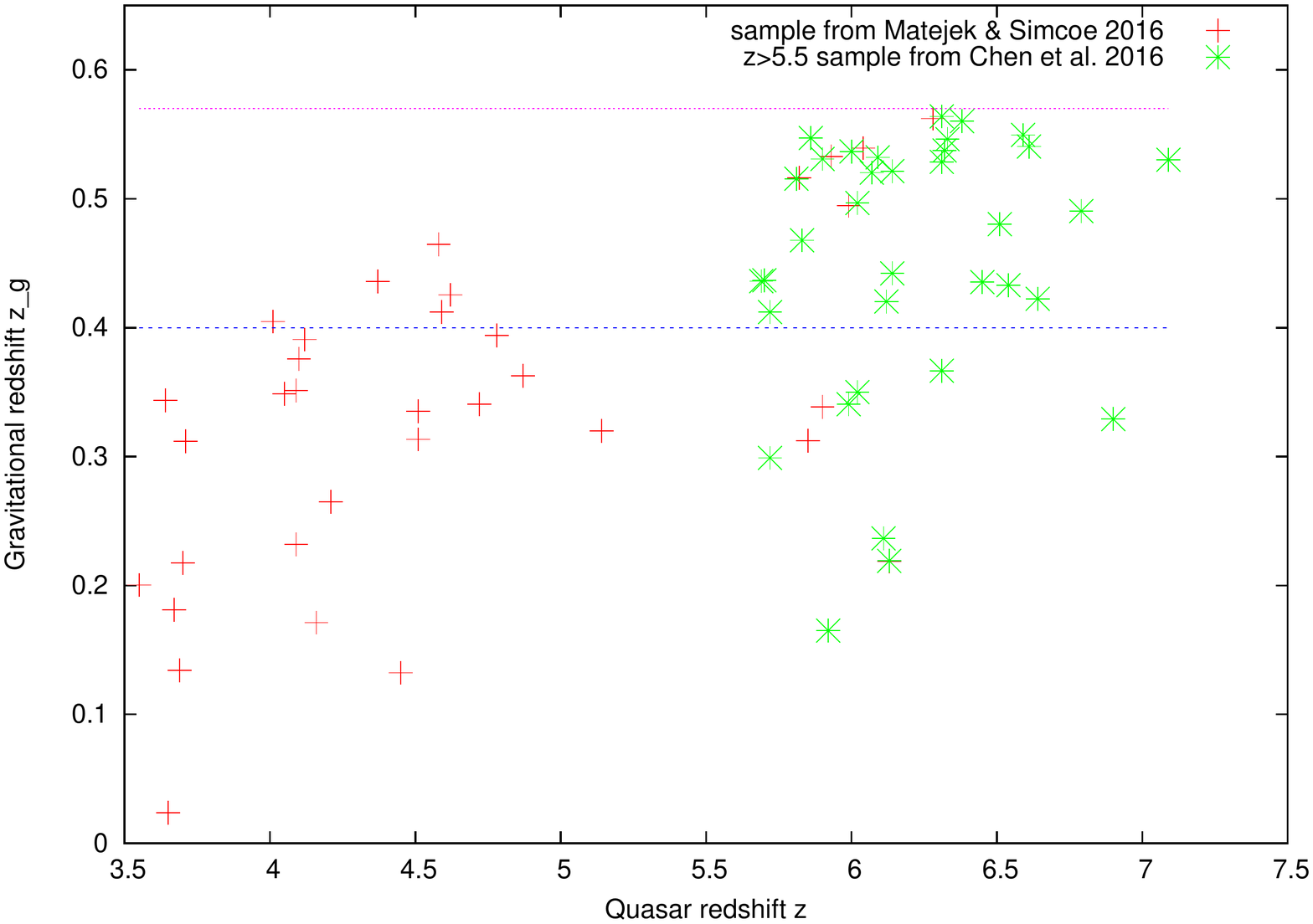}(a)
\includegraphics[width=8cm]{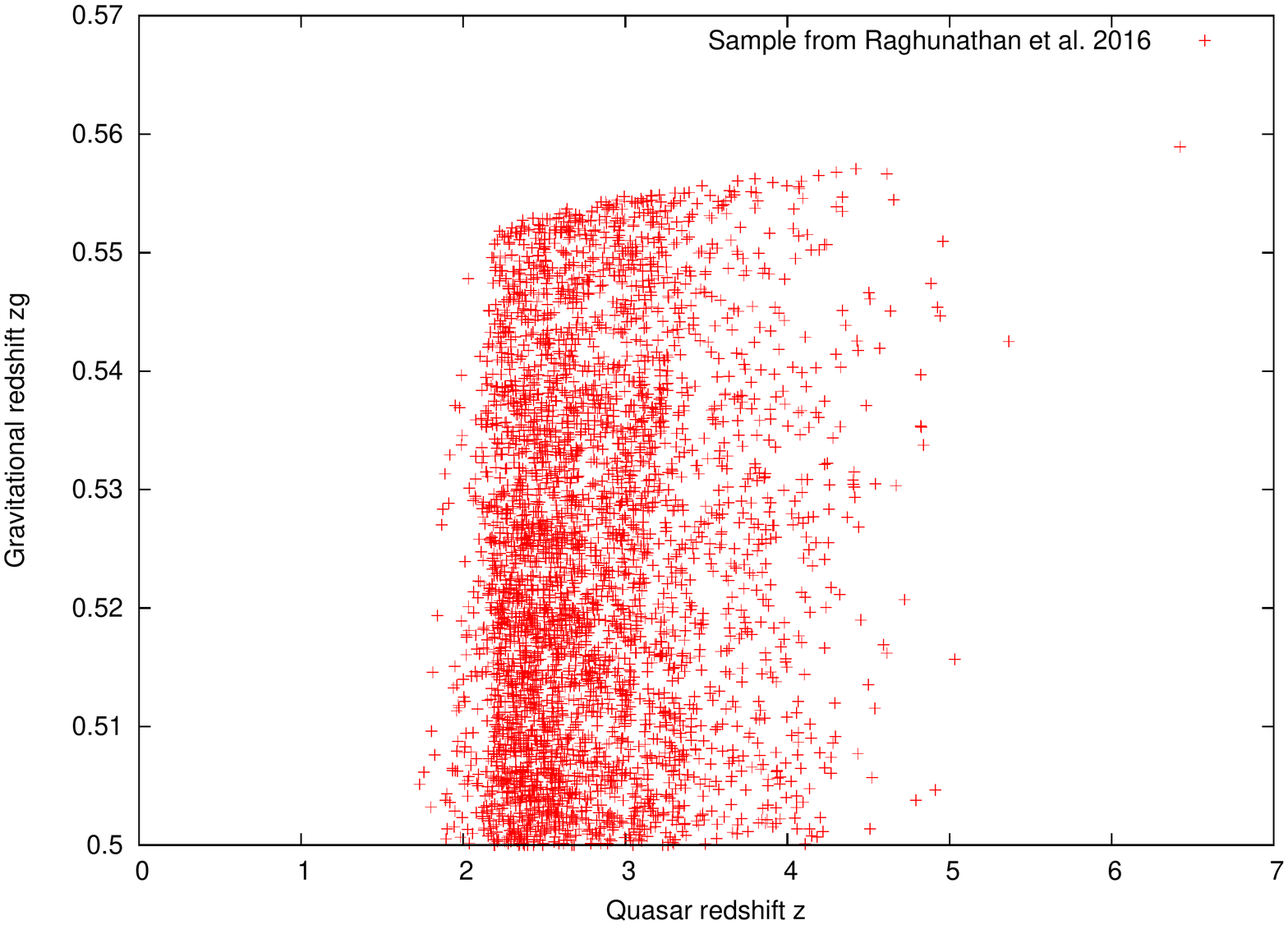}(b)
\caption{
(a) Distribution of gravitational redshifts estimated for the sample with $z>3.5$ in \citet{2012ApJ...761..112M}
and $z>5.5$ in \citet{2016arXiv161202829C}.  Horizontal lines are drawn at 0.4 and 0.57.
Notice the predominance of larger gravitational redshift for the larger quasar redshifts i.e.
the fraction of quasars with $z>5.5$ having
maximum $z_g$ between 0.4 and 0.57 is larger than compared to quasars with $z<5.5$.  This is expected
if the increasing quasar redshift is due to increasing contribution from both $z_c$ and $z_g$.
%In the sample, a quasar of $z=6.28$ has maximum $z_g=0.562$, $z=6.13$ has maximum $z_g=0.219$,
%$z=6.12$ has maximum $z_g=0.420$, $z=6.04$ has maximum $z_g=0.539$.
(b) A zoom-in into the $z_g$ range of 0.5 to 0.57 for the sample of quasars in
\citet{2016MNRAS.463.2640R} shows the same trend as in (a) ie an increasing largest $z_g$ with
$z$ in the quasar redshift range of 2 to 5 and which continues to the largest $z=6.423$ in the
sample for which $z_g= 0.5589$ has been estimated.}
\label{zg3}
\end{figure}

Examining the distribution of the lowest redshift of the Mg II absorption (range 0.3 to 2.5) detected
along a quasar sightline shown in Figure \ref{zg4}a, we note that
the sample of quasars from \citet{2013ApJ...779..161S} show a trapezium-shaped distribution in the
$z\rightarrow z_{MgII}$ plane.  The higher redshift sample from \citet{2012ApJ...761..112M,2016arXiv161202829C}
are also distributed in a similar fashion.
We focus on the observed distribution of $z_{MgII}$ in the larger sample from \citet{2013ApJ...779..161S}.
The upper ($z_{MgII}\sim 2.3$) and lower ($z_{MgII}\sim0.35$) bounds are
due to the redshift limits of the Mg II catalogue.  The left slanting side indicates the largest value of
$z_{MgII}=z_c$ i.e. the maximum $z_c$ detected in the sample for a given quasar redshift.
The right slanting side of the trapezium indicates the minimum $z_{MgII}=z_c$
that is detected in the sample for a given quasar redshift.
The largest and smallest $z_{MgII}=z_c$ at any given $z$
is the range of $z_{MgII}=z_c$ observed for that quasar redshift within the catalogue bounds.
While upto $z\sim2$, the smallest $z_{MgII}$ detected in the sample continues to be $\sim 0.35$ which is
the lower limit of the range of explored redshifts in the catalogue,
for $z\ge 2.2$, there are no quasars for which $z_{MgII}$
is $\sim 0.35$.  The smallest redshift at which Mg II absorption is detected for
a given quasar redshift keeps increasing with $z$ beyond $z\sim 2.2$.   For example, none of the
quasars with $z>3$ have recorded Mg II absorption with $z_{MgII}<0.7$.
We examine the behaviour of the detected Mg II redshifts,
in the context of the two types of origin of these features namely external, in the intervening clouds
along the sightline and intrinsic to the quasar.
In the intervening cloud origin,
we would have expected the smallest $z_{MgII}$ detected at all quasar redshifts to be
constant and in this case equal to the lower bound of the range of explored redshifts i.e. 0.35, since at least a few high
redshift quasars should have sightlines passing through the nearby Mg II clouds.   Since there is
not a single detection of the nearby clouds in the high redshift quasars (see Figure \ref{zg4}a) i.e.
low $z_{MgII}$ features are missing in high $z$ ($z>2.2$) quasar spectra, it would mean that the high redshift
quasars necessarily lie along sightlines where the
low redshift intervening clouds (detected in the spectra of quasars with $z<2$)
are systematically missing which, if true, would be a cosmic conspiracy.  A simpler
explanation is that the intervening cloud scenario for the origin of the Mg II absorption lines in the quasar
spectrum is not the correct interpretation.   In fact, the observed nature of the distribution of $z_{MgII}$
(lowest redshift detected in a quasar spectrum) shown
in Figure \ref{zg4}a seems to rule out an intervening origin for the Mg II features.

We now investigate the intrinsic origin scenario for the Mg II lines in which the lowest redshift Mg II absorption lines 
arise within the quasar system and their redshift is the cosmological redshift of the quasar as suggested in the paper.  
This would necessarily imply that all the spectral lines in a quasar spectrum 
arise within the quasar and the multiple redshifts contain a contribution from gravitational redshifts as discussed above.
Since the gravitational redshift shown by a line photon is a function of the separation of the line forming
region from the event horizon of the black hole, the multiple redshifts observed in the quasar spectrum are 
to be expected as long as photons arise in a region of finite extent located close to the black hole. 
%This explanation, it is found, expects the trapezium-shaped distribution of Figure \ref{zg4}a.
As the observed $z$ of the quasar increases, Figure \ref{zg4}a shows that
%the minimum and maximum of the range of $z_{MgII}=z_c$ that are detected also increase
$z_{MgII}$ systematically increases with $z$ and the lowest redshift Mg II features detected in low
redshift quasars are never detected in high redshift quasars.   In the intrinsic origin,
$z$ will increase either because both $z_c$ and $z_g$ have increased or if one of them has increased. 
Figure \ref{zg4}a shows that $z_c$ of quasars increases with $z$ and $z_c < z$ so that the difference redshift
$z_{in}$ is due to the effect of an intrinsic physical process i.e. gravitational redshift. 
%When the latter increases
%the difference between $z$ and $z_c$ will increase; moreover since $z_g$ is well-constrained,
%it means that as $z$ increases, $z_c$ has to increase as noted in Figure \ref{zg4}a.
Thus, qualitatively, the inclusion of an intrinsic redshift can explain the distribution in
Figure \ref{zg4}a.  We now quantify $z_g$ from $z$ and $z_c$.
Figure \ref{zg4}b shows the distribution of $z_g$ with $z_c=z_{MgII}$ for the same three
catalogues shown in Figure \ref{zg4}a.  Interesting results also emerge from this plot. 
The distribution of the data \citep{2013ApJ...779..161S} in the $z_c \rightarrow z_g$ plane
is rectangular with the upper right corner being sparsely populated.
The $z_g$ estimated for quasars at all $z_c$ i.e. 0.35 to 2.3 is between 0.0166 and $0.5535$ and
lines which show $z_g\ge0.5$ have to arise from matter inside the erogsphere of the rotating black hole in the quasar.  
We use these limits to estimate the expected quasar redshifts.
The quasar with $z_c=0.35$ and $z_g=0.5535$ will be detected at $z=2.02$ whereas a quasar
with $z_c=0.35$ and $z_g=0.0166$ will be detected at $z=0.373$.  The quasar with $z_c=2.3$ and $z_g=0.5535$ will
be detected at $z=6.39$ and if $z_g=0.0166$, it will be $z=2.355$.
The lower limit of $z_g=0.0166$ would be shown by lines forming
at a radial separation $\sim 30$ Schwarzchild radius from the black hole whereas the
upper limit of $0.5535$ would be shown by lines forming in matter located at $0.903$ Schwarzchild radius. 
In the sample of \citet{2013ApJ...779..161S} shown in 
Figure \ref{zg4}b,  the line emitting matter in all the detected quasars is located between these
two extremes.  
The upper limit to $z_g$ could be indicative of the combination of the rapidly increasing gravitational potential
and extreme physics that photons would experience as they emerge from within the
ergosphere of the black hole and observational constraints such as observable line widths or redshifts.
The paucity of quasars in the top right corner of the rectangular distribution, which is populated
by quasars with high $z_c$ and high $z_g$ and hence high $z$ can be explained by
the limited number of quasars detected at high $z$ (e.g. a quasar with $z_c=2.3$ and $z_g=0.5535$  
which would appear in the top right corner of the distribution will be detected at a redshift $z=6.39$).
Since the sample of quasars in \citet{2013ApJ...779..161S} is upto about $z\sim5$, the empty top right
hand corner of the rectangular distribution in Figure \ref{zg4}b is easily accounted for by the absence
of the high redshift quasars in the catalogue.  Considering that the sample of \citet{2016arXiv161202829C}
wherein quasars with $z>5.5$ have been targetted does show the presence of a bunch of quasars
with $z_g \sim 0.55$  supports this interpretation and indicates that
quasars show the entire range of $z_g$ at all $z_c$.  We can, hence, infer that $z_g$ is independent of $z_c$ as 
it should be since $z_g$ is intrinsic to the quasar and only dependent on black hole physics.  
A linear decrease in $z_g$ with increase in $z_c$ is noticeable in all the three
datasets, more so in the higher redshift catalogues.   This behaviour is also due to the limited
number of high $z$ quasars in the existing catalogues.  The high $z_c$ quasars (e.g. $z_c \sim 3$) 
detected within the high $z$ sample (e.g. $z \ge 5.5$), will necessarily be sensitive to quasars with low $z_g$
(e.g. $\sim 0.38$).  For detecting the high $z_g$ sample within the
high $z_c$ sample, even higher redshift quasars will have to be targetted.
Excepting the lowest $z_{MgII}$ in their sample, the catalogues of \citet{2012ApJ...761..112M,2016arXiv161202829C} detect 
relatively higher redshift Mg II absorption in a catalogue of higher redshift quasars and hence are sensitive to  
the low $z_g$ sample.  Thus, the distributions in Figure \ref{zg4}a, b are
expected if the contribution by gravitational redshift is included.
%Figure \ref{zg4} rules out any strong evolutionary effect on the distribution of matter around black holes
%which is encouraging since it should be dictated only by black hole physics and observational constraints.
The discussion so far supports the results in the paper and strongly favours the existence of an intrinsic
redshift component in addition to a cosmological redshift component in the observed quasar spectra.  This model satisfactorily explains
observational results from large datasets on quasars without having to resort to any contrived explanations
and we believe is the correct interpretation of the quasar spectra.

In Figure \ref{zg3}, the distribution of $z_g$ with $z$ of the quasar has been shown.
In Figure \ref{zg3}a, the high redshift sample in \citet{2012ApJ...761..112M,2016arXiv161202829C} indicate
that the number of quasars in which lines are shifted by a larger $z_g$ increase with $z$.
Thus, the high $z$ of quasars is due to a larger $z_c$ or $z_g$ or both as would be expected.
The largest $z_g$ in this sample is 0.563 estimated for a quasar with $z=6.31$ and $z_c=2.188$.
This $z_g$ will be shown by lines forming in matter located about 0.888
Schwarzchild radius from the black hole.  In Figure \ref{zg3}b, the distribution of $z_g$ between 0.5
and 0.57 using the quasar data between $z$ of 0.5 and 6.423 in \citet{2016MNRAS.463.2640R} has been shown
and their appears to be a trend of a gentle increase in the highest
detected $z_g$ with $z$.  The largest $z_g$ in this sample is 0.5589 estimated for a quasar
with $z=6.423$ and $z_c=2.274$.  It is not immediately obvious why higher $z_g$ quasars
are not detected at lower $z_c$.  However we note that data from \citet{2013ApJ...779..161S} does not show
a trend of increasing $z_g$ with $z$ and
hence it would be useful to investigate this further before any concrete inferences are drawn.

The above results mean that quasars are not as distant as customarily believed.  
The highest cosmological redshift detected in the sample of quasar redshifts upto 7.1 \citep{2016arXiv161202829C} 
is 4.778. 
The proximity of quasars then brings down the enormous luminosities inferred for quasars and will have
important implications on several other parameters and models which place quasars at extremely
large distances. 
An interesting point to note is that we can keep detecting quasars at ever increasing emission line redshift
for small increases in the gravitational redshift.   For example, for quasars at $z_c=2$, the
emission line redshifts for $z_{in}=1$ (corresponds to $z_g\sim0.5$) and for $z_{in}=2$ (corresponds
to $z_g\sim0.67$) would be $z=5$ and $z=8$ respectively.
$z_{in}$ and hence $z$ can keep increasing till $z_g$ approaches one for the maximally rotating black holes even if
there is no increase in $z_c$ of the quasar.
At this juncture, it is instructive to mention that $z_{in}$ was found well constrained
in the paper, which played a major role in convincing the author of a significant contribution by an intrinsic
process,  because the $z_g$ of detected quasars was well constrained to values less or around 0.5.  
We note that $z_{in}$ can take on a large range of values; for example if $z_g=0.5$ then $z_{in}=1$,
if $z_g=0.9$, then $z_{in} = 9$ and if $z_g=0.95$ then $z_{in}=19$!
The quasars with
$z_g>0.5$ that are present in the sample led to $z_{in}$ being estimated to be $>1$ and since the numbers were
small, the excess redshift above $z_{in}=1$ was incorrectly attributed to the contribution by another
unknown intrinsic factor such as Doppler flows.  With the correction detailed in point 5, this problem has been
removed.  

This important edit and examination of more data has only made the case for the intrinsic origin
of the entire quasar spectrum and the gravitational redshift origin for the multi-redshifted spectral
lines in the quasar spectrum stronger.  The observed redshift of the quasar is a combination of
cosmological redshift and gravitational redshift of the emission line.

\begin{table}[t]
\centering
\small
\caption{The observed emission line redshift of the quasar, the observed lowest redshift of the Mg II absorption and
the estimated gravitational redshifts are noted in columns 2,3,4 respectively for the sample in
\citet{2012ApJ...761..112M} and in columns 6,7,8 for the sample in \citet{2016arXiv161202829C}.
The intrinsic redshift ($z_{in}$) is estimated using the quasar redshift ($z_{em}$) and the lowest
redshift at which Mg II absorption ($z_{MgII}=z_c$) is detected in the quasar spectrum and
the gravitational redshift ($z_g$) is determined using Equation \ref{zg}.
%These are plotted in Figures \ref{zg4},\ref{zg3}.
There are upto 8 common quasars in the two catalogues listed here. }
\begin{tabular}{l|c|c|c|l|c|c|c}
\hline
\multicolumn{3}{c}{\citet{2012ApJ...761..112M}} & & \multicolumn{3}{c}{\citet{2016arXiv161202829C}} & \\
{\bf Quasar} & $\bf z_{em}$ & $\bf z_{MgII}$ & $\bf z_g=z_{in}/(1+z_{in})$ & {\bf Quasar}& $\bf z_{em}$ & $\bf z_{MgII}$ & $\bf z_g=z_{in}/(1+z_{in})$ \\
\hline
Q0000-26 & 4.10  & 2.184 & 0.375    &   SDSSJ0100+2802  &  6.33 &   2.326  &  0.546  \\
BR0004-6224 & 4.51 & 2.663 & 0.335    & VIKJ0109-3047  &  6.79  &  2.969 &   0.490  \\
BR0016-3544 & 4.51 & 2.783 & 0.313    & PSOJ029-29   & 5.99  &  3.609  &  0.340  \\
SDSS0106+0048 & 4.45&  3.729&  0.132    &       ULASJ0203+0012 &   5.72 &   3.711 &   0.298  \\
SDSS0113-0935 & 3.67 & 2.825&  0.180    &       ATLASJ025-33  &  6.31  &  2.446  &  0.528 \\
SDSS0140-0839 & 3.71 & 2.241&  0.311    &       VIKJ0305-3150 &   6.61 &   2.496 &   0.540  \\
SDSS0203+0012 & 5.85 & 3.711&  0.312    &       PSOJ036+03  &  6.54  &  3.275  &  0.433  \\
BR0305-4957&  4.78 & 2.502&  0.394    & PSOJ071-02  &  5.70  &  2.773  &  0.436  \\
BR0322-2928&  4.62 & 2.229&  0.425    & DESJ0454-4448  &  6.09 &   2.317 &   0.532  \\
SDSS0332-0654 & 3.69 & 3.061&  0.134    &       PSO065-26  &  6.14  &  2.983  &  0.442  \\
BR0353-3820&  4.58 & 1.987 & 0.464    & PSOJ071-02 &   5.69 &   2.773 &   0.436  \\
BR0418-5723 & 4.37 & 2.030 & 0.435    & SDSSJ0818+1722 &   6.02 &   3.563 &   0.35  \\
SDSS0818+1722 & 5.90 & 3.563 & 0.338    &       SDSSJ0836+0054 &   5.81 &   2.299 &   0.515  \\
SDSS0836+0054 & 5.82 & 2.299 & 0.516    &       SDSSJ0842+1218 &   6.07 &   2.392 &   0.520  \\
SDSS0949+0335 & 4.05 & 2.289 & 0.348    &       SDSSJ1030+0524 &   6.31 &   2.188 &   0.563  \\
SDSS1020+0922 & 3.64 & 2.046 & 0.343    &       J1048-0109  &  6.64   & 3.413  &  0.422  \\
SDSS1030+0524 & 6.28 & 2.188 & 0.562    &       ULASJ1120+0641 &   7.09 &   2.800 &   0.530  \\
SDSS1110+0244 & 4.12 & 2.119 & 0.390    &       ULASJ1148+0702 &   6.32 &   2.386 &   0.537  \\
SDSS1305+0521 & 4.09 & 2.302 & 0.351    &       PSOJ183-12  &  5.86  &  2.107  &  0.547  \\
SDSS1306+0356 & 5.99 & 2.533 & 0.494    &       SDSSJ1306+0356 &   6.02  &  2.533 &   0.496  \\
ULAS1319+0950 & 6.13 & 4.569 & 0.218    &       ULAS1319+0950  &  6.13   & 4.568  &  0.219  \\
SDSS1402+0146 & 4.16 & 3.277 & 0.171    &       SDSSJ1411+1217 &   5.90  &  2.237 &   0.530 \\
SDSS1408+0205 & 4.01 & 1.982 & 0.404    &       PSOJ213-22  &  5.92  &  4.778  &  0.165  \\
SDSS1411+1217 & 5.93 & 2.237 & 0.532    &       CFQS1509-1749 &   6.12 &   3.127 &   0.420  \\
Q1422+2309 & 3.65&  3.540 & 0.023    &  PSOJ159-02  &  6.38  &  2.246  &  0.560  \\
SDSS1433+0227 & 4.72 & 2.772 & 0.340    &       PSOJ183+05  &  6.45  &  3.207  &  0.435  \\
CFQS1509-1749 & 6.12 & 3.128 & 0.420    &       PSOJ209-26  &  5.72  &  2.951  &  0.412  \\
SDSS1538+0855 & 3.55 & 2.638 & 0.200    &       PSOJ217-16  &  6.14  &  2.417  &  0.521  \\
SDSS1616+0501 & 4.87&  2.741 & 0.362    &       VIKJ2211-3206 &   6.31 &   3.630 &   0.366  \\
SDSS1620+0020  &4.09 & 2.910 & 0.231    &       PSOJ231-20  &  6.59  &  2.419  &  0.549  \\
SDSS1621-0042 & 3.7 & 2.678 & 0.217    &        SDSSJ2310+1855 &   6.00&    2.243&    0.536  \\
SDSS2147-0838&  4.59 & 2.286&  0.412    &       VIKJ2318-3113  &  6.51 &   2.903 &   0.480  \\
SDSS2228-0757&  5.14 & 3.175&  0.320    &       VIKJ2348-3054  &  6.90 &   4.300 &   0.329  \\
SDSS2310+1855&  6.04 & 2.243&  0.539    &       PSOJ239-07  &  6.11 &   4.428  &  0.236  \\
BR2346-3729 & 4.21 & 2.830  &0.264    & PSOJ242-12  &  5.83 &   2.635  &  0.467  \\
\hline
\end{tabular}
\label{tab2}
\end{table}

{\it (November 2017)}

\end{enumerate}

\end{document}